\documentclass{aa}
\usepackage{graphicx}
\usepackage[varg]{txfonts}
\usepackage[colorlinks,citecolor=blue,linkcolor=blue,urlcolor=blue]{hyperref}
\usepackage{amsmath}
\usepackage[dvipsnames]{xcolor}
\usepackage{comment}
\usepackage{multirow}
\usepackage{url}
\usepackage[normalem]{ulem}

\newcommand{\udef}{\stackrel{\mathrm{def}}{=}}

\DeclareRobustCommand{\Eqref}[1]{Eq.~\ref{#1}}
\DeclareRobustCommand{\Figref}[1]{Fig.~\ref{#1}}
\DeclareRobustCommand{\Tabref}[1]{Tab.~\ref{#1}}
\DeclareRobustCommand{\Secref}[1]{Sec.~\ref{#1}}

\definecolor{Orange}{HTML}{FF9900}

\begin{document}
\title{A Systematic Survey of the Effects of
  Wind Mass Loss Algorithms on the
  Evolution of Single Massive Stars}
\authorrunning{Renzo et al.}

\author{M.~Renzo\inst{1,2,3,4}, C.~D.~Ott\inst{3},
  S.~N.~Shore\inst{2}, \and S.~E.~de~Mink\inst{1}}
\institute{{Astronomical Institute Anton Pannekoek, University of Amsterdam, 1098 XH Amsterdam, The Netherlands}\and{Dipartimento di Fisica `Enrico Fermi,' University of Pisa, I-56127 Pisa, Italy}\and
  {TAPIR, Walter Burke Institute for Theoretical Physics, MC 350-17,
    California Institute of Technology, Pasadena, CA 91125,
    USA}\and{Kavli Institute for Theoretical Physics, University of
    California, Santa Barbara, CA 93106, USA}}
\offprints{\href{mailto:m.renzo@uva.nl}{m.renzo@uva.nl}}
\date{}

\abstract {Mass loss processes are a key uncertainty in the evolution
  of massive stars. They determine the amount of mass and angular
  momentum retained by the star, thus influencing its evolution and
  presupernova structure. Because of the high complexity of the
  physical processes driving mass loss, stellar evolution calculations
  must employ parametric algorithms, and usually only include wind
  mass loss. We carry out an extensive parameter study of wind mass
  loss and its effects on massive star evolution using the open-source
  stellar evolution code MESA. We provide a systematic comparison of
  wind mass loss algorithms for solar-metallicity, nonrotating, single
  stars in the initial mass range of $15\,M_\odot$ to $35 M_\odot$. We
  consider combinations drawn from two hot phase (i.e.\ roughly the
  main sequence) algorithms, three cool phase
  (i.e.\ post-main-sequence) algorithms, and two Wolf-Rayet mass loss
  algorithms. We discuss separately the effects of mass loss in each
  of these phases. In addition, we consider linear wind efficiency
  scale factors of $1$, $0.33$, and $0.1$ to account for suggested
  reductions in mass loss rates due to wind inhomogeneities.  We find
  that the initial to final mass mapping for each zero-age
  main-sequence (ZAMS) mass has a $\sim 50\%$ uncertainty if all
  algorithm combinations and wind efficiencies are considered. The
  ad-hoc efficiency scale factor dominates this uncertainty. While the
  final total mass and internal structure of our models vary
  tremendously with mass loss treatment, final luminosity and
  effective temperature are much less sensitive for stars with ZAMS
  mass $\lesssim 30\,M_\odot$.  This indicates that uncertainty in
  wind mass loss does not negatively affect estimates of the ZAMS mass
  of most single-star supernova progenitors from pre-explosion
  observations.  Our results furthermore show that the internal
  structure of presupernova stars is sensitive to variations in both
  main sequence and post main-sequence mass loss. The compactness
  parameter $\xi \propto \mathcal{M}/R(\mathcal{M})$ has been
  identified as a proxy for the ``explodability'' of a given
  presupernova model. We find that $\xi$ varies by as much as $30\%$
  for models of the same ZAMS mass evolved with different wind
  efficiencies and mass loss algorithm combinations. This suggests
  that the details of the mass loss treatment might bias the outcome
  of detailed core-collapse supernova calculations and the predictions
  for neutron star and black hole formation.
   
}
    
\maketitle{}

\section{Introduction}
\label{sec:intro}
Mass loss is a key phenomenon for the co-evolution of massive stars
($M\gtrsim 8\,M_\odot$) and their environment, yet it is poorly
understood. It plays an important role throughout
the stellar evolution and it may have a deciding influence on the
outcome of core collapse. Mass loss is responsible for a large part of
the chemical enrichment of the interstellar medium, and its momentum
input can trigger star formation, but it can also sweep away gas from
stellar clusters, preventing further star formation.

In the standard picture of single massive star evolution, mass loss
influences the duration of different evolutionary phases \citep[e.g.,
][]{meynet:15}, especially the amount of time spent on the red
supergiant (RSG) branch.  It has been suggested to be important for
the solution of the so-called ``red supergiant problem''
\citep{smartt:09,smith:11}, i.e.\ the discrepancy between the observed
maximum mass for type IIP supernovae (SNe) and the theoretical
predictions for the core collapse of RSG stars. This discrepancy
indicates an incomplete theoretical understanding of the evolution
and explosion of massive stars, especially in
the mass range $\sim$$16-30\,M_\odot$.

Mass loss also plays an essential role in the formation of Wolf-Rayet
(WR) stars. Two competing scenarios exist for the removal of (most of)
the hydrogen-rich envelope of stars: the single star picture
\citep[so-called ``Conti scenario'',][]{conti:75,maeder:94,lamers:13}
and the binary formation channel
\citep[e.g.,][]{woosley:95,wellstein:99, smith:15b,
  shara:17}. Understanding mass loss phenomena is necessary to
discriminate between these two scenarios. Mass loss is invoked to
explain the variety of core-collapse SNe \citep[e.g.,][]{eldridge:04,
  smith:11, georgy:12, groh:13,smith:14}, because it can modify the
surface composition of the pre-SN star, possibly removing the
hydrogen-rich envelope (and perhaps some of the helium-rich shell) and
leading to type Ib/c SNe. Observations of SNe IIn and superluminous
SNe may perhaps be explained by the strong interaction between the SN
shock and shells of material ejected from the star in late stages of
its evolution, \citep[e.g.,][]{smith:11,smith:14,shiode:14}. It has
also been proposed that the coupling of mass loss and rotation might
prevent pair instability SNe for very massive, low metallicity stars
\citep[][]{ekstrom:08,woosley:16b}.

Finally, mass loss plays a key role in shaping the (hydrostatic)
internal structure of massive stars at the pre-SN stage, which can
influence the expected SN outcome \citep[see,
  e.g.,][]{belczynski:10,oconnor:11,oconnor:13,ugliano:12,sukhbold:16}:
will the star successfully explode and leave a neutron star (NS)
remnant? Will the explosion be highly asymmetric or weak, leading to
fallback accretion and black hole (BH) formation? Or will the
explosion fail completely, leaving a BH with little
\citep[][]{lovegrove:13} or no electromagnetic counterpart?
 
Depending on the amount and geometry of the ejecta, which are governed
by the mass loss during the stellar lifetime and the SN energetics,
also the SN kick can vary 
\citep[][]{zwicky:57,blaauw:61,boersma:61,janka:13,janka:16}. The kick
can change the post-explosion orbital parameters if the star is in a
binary system and, therefore, our incomplete understanding of massive
star mass loss affects the predicted populations of
sources for gravitational wave astronomy \citep[][]{LVC:16b}.

There exist three main channels of
mass loss in the evolution of massive stars:

\begin{enumerate}
\item Steady-state
  winds. These are radiatively driven in hot stars (i.e., on the main
  sequence; see \citealt{puls:08} for a review). In the supergiant
  phase, the driving mechanism is uncertain. Winds could be
  radiatively driven via lines or dust, or
  by other mechanisms, e.g., wave energy deposition 
  (see, e.g., \citealt{bennett:10} for a brief review).  
\item Impulsive, pulsational and/or eruptive mass
  loss, e.g.\ disk shedding at criticial
    rotation or giant eruptions such as those of Luminous Blue
    Variables (LBVs) \citep[e.g.,][]{puls:08,smith:14,puls:15};
\item Roche lobe overflow (RLOF) and possibly common
  envelope ejection in binary systems, which can result in mass loss
  from the donor star and also mass loss from the system as a whole in
  non-conservative cases.
\end{enumerate}

Which of these processes dominates in terms of the total mass shed has
been a matter of debate in the literature \citep[see][and references
therein] {smith:14}.

Mass loss is an intrinsically dynamical phenomenon which involves bulk
acceleration of matter for escaping the star. Because of the
dynamical nature of mass loss, it is
difficult to include in stellar evolution codes: most simulations
focus on single stars, and are carried out with hydrostatic codes
which cannot account for dynamical or
impulsive events in a
physical way. Even hydrodynamical codes do not permit a
self-consistent development of impulsive events, since the physical
processes triggering them are currently poorly understood or even
unknown. Impulsive outbursts of mass loss can, however, be included using
physically plausible prescriptions \citep[e.g.,][]{morozova:15}.

Most massive stars are found in binary systems
\citep[e.g.,][]{mason:09,sana:11,sana:12,kiminki:12,chini:12,kobulnicky:14,almeida:16},
where mass loss also determines the angular momentum losses, thus the
orbital evolution, and ultimately the binary evolution path and its
end point (merger, disruption of the binary system, double compact
object binary, etc.).

\medskip
In this paper, our focus is on steady, radiatively-driven wind mass
loss \citep[][]{lucy:70} and our goal is to understand how different
treatments of this process affect massive star
evolution and pre-SN structure. We consider single massive stars
and do not address the problem of mass loss in binaries directly.

Most evolutionary calculations of massive stars only include wind mass
loss, which can be treated in the steady state approximation,
although, strictly speaking, a wind is dynamical as well.

The wind is in fact driven by a radiative acceleration which formally
enters into the momentum equation for the stellar plasma
\citep[][]{castor:75}, the stellar
structure responds secularly, since wind mass loss rates are low and
change only slowly compared to impulsive mass loss events. At solar metallicity, the wind mass loss rate is $10^{-10}\,M_\odot\,\mathrm{yr^{-1}}
\lesssim \dot{M}_\mathrm{wind} \lesssim
10^{-5}\,M_\odot\,\mathrm{yr^{-1}}$, while RLOF and LBV eruptions
yield
$10^{-6}\,M_\odot\,\mathrm{yr^{-1}}\lesssim\dot{M}_\mathrm{RLOF}
\lesssim \dot{M}_{LBV} \lesssim 10^{-2}\,M_\odot\,\mathrm{yr^{-1}}$,
\citep[e.g.,][]{dejager:88,vink:01,vanloon:05,smith:06,crowther:07,puls:08,langer:12,smith:14}. These
mass loss rates correspond to timescales $\tau_\mathrm{wind} =
M/\dot{M}_\mathrm{wind} \gg \tau_\mathrm{RLOF},\tau_\mathrm{LBV}$.
Therefore, wind mass loss can be included in stellar evolution
calculations using parametric algorithms\footnote{These are often
  called ``recipes'' in jargon. However, we prefer the term
    ``algorithm'' \citep[][]{Fibonacci} because it underlines that
    these are mathematical representations of physical phenomena
    relying on specific sets of assumptions.}.

Stellar winds are radiatively driven by the interaction of photons
with metallic ions (line-driven mass loss) or dust grains (dust-driven
mass loss). Therefore, they depend on the opacity and thus
chemical composition, ionization state, and density stratification, so
indirectly also on the equation of state of the outermost stellar
layers. Metals effectively provide all the opacity in stellar
atmospheres because of their large number of lines. Photons come out of the photosphere with well defined
direction, interact with metallic atoms/ions via bound-bound processes
(absorption and line scattering) and cede their momentum to the
atoms/ions. When these de-excite, they emit photons isotropically.
The momentum of de-excitation photons averages out, and the result is
a net gain of momentum in the direction of the initial photon
\citep[see, e.g.,][]{puls:08}.

Note that the radiation field in the stellar atmosphere, i.e.\ above the
photosphere, is \emph{not} isotropic. If there were not a net radial
flux of photons, also the momentum of the incoming photons would
average out.  In a nutshell, the incident photons push metals outward, and metals drag hydrogen and
helium through collisional Coulomb coupling \citep[see][and references therein]{puls:08,vink:14}.

This simple theoretical picture of line-driven stellar winds is
complicated by two phenomena: the high nonlinearity of the driving
mechanism, and the possible presence of inhomogeneities (so called
``clumpiness'') in the stellar atmosphere. The high nonlinearity
arises because the outflow of mass is driven by the rate of
interaction of photons with metals, but this in turn depends on the
local opacity, and therefore on the outflow properties, such as
density, and velocity (which can Doppler shift the lines), see
\cite{lamers:99, puls:08} and references therein. The presence of
inhomogeneities in the stellar atmosphere is both theoretically
expected \citep[see, e.g.,][]{owocki:84,owocki:88,feldmeier:95,
  owocki:99,dessart:05,puls:08,smith:14} and observed comparing
diagnostic spectral lines sensitive to the density $\rho$
(e.g.,\ lines with P Cygni profiles) and $\rho^2$
(e.g.,\ recombination lines, such as H$\alpha$) in the same stellar
wind \citep[see e.g.][]{fullerton:06,bouret:05,evans:04}. This
comparison shows that the averaged
$\langle\rho^2\rangle>\langle\rho\rangle^2$ (for reviews see
\citealt{puls:08} and \citealt{smith:14}). Therefore, the presence of
over-dense clumps in the wind causes an overestimation of the density
inferred from observed spectral features sensitive to $\rho^2$. This
is not taken into account in most wind mass loss algorithms in the
literature. The overestimation of the density directly results in an
overestimated mass loss rate. Work by
\cite{crowther:02,hillier:03,bouret:05,fullerton:06,puls:08,smith:14,LVC:16b},
suggests that the algorithms used in stellar evolution calculations
may yield mass loss rates that are a factor of 2 to 10 too
high. \cite{puls:08} and \cite{smith:14} suggest a factor of 3 as the
most realistic overestimate. We refer the interested reader to
\cite{puls:08}, \cite{smith:14}, and \cite{thesis} for more details.

To date, there has been no systematic comparison of
the various wind mass loss
algorithms, with varying corrections for clumpiness, and their
combinations and effects throughout the evolution and on the final
structure of massive stars. However, \cite{eldridge:04} compared
different combinations of semi-empirical mass loss rates to find the
threshold in mass between type Ibc and type II SNe. \cite{yoon:10b}
discussed consequences of the revision of mass loss rates because of
the clumpiness during the WR phase.

In this study, we employ the open-source stellar evolution
code MESA
\citep[][]{paxton:11,paxton:13,paxton:15} to compare various
combinations of mass loss algorithms, using different efficiency factors
to account for the inhomogeneities in the wind (albeit in an
\emph{ad-hoc} fashion). Our aim is to understand the systematics of
massive star evolution and pre-SN structure caused by variations in
the treatment of wind mass loss, focusing on the differences in the
evolution and pre-SN structure (effective temperature, total
  mass, core masses, and interior structure).

The remainder of this paper is structured as follows. In
\Secref{sec:methods}, we discuss some general aspects of the
implementation of wind mass loss in stellar evolution codes and give a
very brief overview of the physical bases of the wind mass loss
algorithms we compare. A longer review of these, including the
limitations and the formulae implemented in MESA, can be found in
Appendix~\ref{app:schemes_review}.  The more technical points not
relevant to the physics of stellar winds are discussed in
Appendix~\ref{app:MESA_technical}, with the explicit aim of making our
result reproducible. We compare our models when their final pre-SN
mass is determined in \Secref{sec:end_ML}. We discuss the impact of
winds during the hot, cool, and WR phases (if reached) separately in
\Secref{sec:hot}, \ref{sec:cool}, and \ref{sec:WRs}, respectively.  We
compare a subset of our models at oxygen depletion in
\Secref{sec:O_depl_res}. The evolution and pre-SN structure of the
core is also sensitive to the mass loss history of the stellar model,
including the early mass loss during the main sequence, as we show in
\Secref{sec:xi_O_depl}. In \Secref{sec:onset_cc}, we discuss the
evolution of a subset of our models from oxygen depletion to the onset
of core collapse. We discuss the implications of
uncertainties in wind mass loss and potential
observational constraints in
\Secref{sec:discussion}, before concluding in \Secref{sec:conclusion}.

\section{Methods}
\label{sec:methods}

\subsection{Overview of the mass loss algorithms}
\label{sec:presc}

Stellar evolution codes do not explicitly compute the acceleration of
the gas unbound in mass loss processes. The usual approach for
including wind mass loss is to use parametric algorithms prescribing a
mass loss rate averaged over each timestep. The time averaging is
needed to compute each timestep with a constant mass loss
rate. Homogeneity of the wind is implicit in the standard formalism, which is known to be a poor approximation and could
cause a significant overestimate of the mass loss rate. Stellar
wind algorithms are either parametric fits to observed mass loss rates, or theoretically
derived models with free parameters chosen empirically or
heuristically. Each algorithm gives a formula for the mass loss rate
as a function of some quantities characterizing the star
$\dot{M}\equiv\dot{M}(L,T_\mathrm{eff}, Z,\dots)$. The precise set of
variables assumed to be independent varies between the
algorithms. Most algorithms do not include an explicit metallicity
dependence, because they either assume a specific chemical composition
of the stellar atmosphere or are based on observed samples with a
specific metallicity $Z$. It is common practice to impose a smooth
scaling with $Z$ such as
\begin{equation}\label{eq:scaling}
  \dot{M} \propto Z^a \ \ ,
\end{equation}
with $a \simeq 0.5$
\citep[e.g.,][]{vink:00,woosley:02}. \Eqref{eq:scaling} is in
reasonable agreement with more sophisticated mass loss rate
determinations \citep[see, e.g.,][]{vink:01}, but deviations should be
expected both at very low $Z$ (because of the lack of metal lines to
drive the wind) and very high $Z$ (because of line saturation
preventing further driving of the wind). In this study, we only
consider solar metallicity. Note also that stellar evolution codes
usually neglect the errors on the coefficients of the algorithms
obtained as fits to observations.

Since mass loss rates have large
uncertainties, it is common practice to employ a linear efficiency
factor $\eta$ to rescale rates to account for various physical uncertainties. For example, $\eta\lesssim
1$ can be used to account for  wind clumpiness
\citep[e.g.,][]{hamann:98,woosley:02, woosley:07}.  Mass loss channels other
than winds exist (e.g.,\ LBV eruptions, binary interactions, etc.) and
some authors \citep[e.g.,][]{dessart:13,meynet:15} use $\eta>1$ to
explore the effects of increased
total mass loss. However, an averaged steady wind approximation might
not properly capture the readjustment of the stellar structure to
non-wind mass loss events, which may be sensitive to mass loss timing,
and/or the physical process triggering them, and may not be
radiatively driven.

Most wind mass loss algorithms are tailored to a specific evolutionary
stage. To carry out a simulation of the entire evolution of the star,
several mass loss algorithms are commonly combined using computational
definitions of the evolutionary phases. This may introduce somewhat
arbitrary switching points in the evolution. Below, we list the
physical basis and the abbreviations for the two hot phase, three cool
phase, and two WR phase wind mass loss algorithms that we combine and
compare here.  We define each phase of the evolution in
\Secref{sec:comb}.

\begin{description}
\item {\bf \cite{vink:00,vink:01} (V):} This wind scheme is a
  theoretical algorithm obtained with numerical simulation of the
  line-driven process. It explicitly includes the metallicity
  dependence and applies to OB stars during their hot evolutionary
  phase. It also includes a detailed
  treatment of the so-called ``bistability jump''. This corresponds to
  a non-monotonic behavior of the mass loss rate $\dot{M}$ as a
  function of effective temperature $T_\mathrm{eff}$ in certain
  temperature ranges (e.g.,
  $T_\mathrm{eff}^\mathrm{jump}\simeq25\,000\,\mathrm{K}$) because of
  the recombination of certain ions, which provides more lines in the
  spectral domain relevant to drive the wind,
  cf. Appendix~\ref{sec:vink}.

\item {\bf \cite{kudritzki:89} (K):} This analytical mass loss rate is
  obtained using the \cite{castor:75} model for line-driven
  acceleration. It assumes that the wind is stationary, isothermal,
  spherically symmetric, and without viscosity and heat
  conduction. The analytical solution is obtained assuming a velocity
  structure $v\equiv v(r)$ as a function of the radius  of the wind,
  and solving self-consistently for density and radiative
  acceleration, cf. Appendix~\ref{sec:kudr}.

\item {\bf \cite{dejager:88} (dJ):} This empirical mass loss rate describes the
``averaged statistical behavior'' of stars (excluding WR and Be
stars) in the HR diagram. It is commonly used for the cool (giant)
phase of the evolution of massive stars, cf. also Appendix~\ref{sec:dejager}.

\item {\bf \cite{nieuwenhuijzen:90} (NJ):} This algorithm is also an empirical mass loss rate drawn from the same sample of stars used by
\cite{dejager:88}. The two algorithms differ in the physical
quantities the mass loss rate is assumed to depend on: NJ used
pre-computed stellar models to add a
dependence on the total mass ($\dot{M}\equiv\dot{M}(M)$), which is not
a directly observable quantity for single stars. It is also usually
adopted for the cool phase of stellar evolution. See
Appendix~\ref{sec:nieuw} for more details.
 
\item {\bf \cite{vanloon:05} (vL):} This empirical mass loss
rate is derived from a sample of oxygen rich asymptotic giant branch
(AGB) and red supergiants (RSG) stars in the
Large Magellanic Cloud. It assumes a dust-driven wind, i.e.\
mass loss is driven by photons impinging on dust grains instead of
metallic ions. Note, however, that the presence of dust and its role as
a wind driving agent in supergiant stars is still debated in the
literature \citep[][]{vanloon:05,ferrarotti:06}, cf. Appendix~\ref{sec:van_loon}.

\item {\bf \cite{nugis:00} (NL):} This empirical mass loss rate is for
  WR stars. Fitting the data for two populations of WR stars (one of
  known distance and one for which they carry out a distance
  determination), they provide an algorithm, which depends strongly on
  the surface chemical composition of the star, cf.
  Appendix~\ref{sec:nl}.

\item {\bf \cite{hamann:82,hamann:95} (H):} This is a theoretical mass
  loss rate for WR stars. It is derived assuming a spherically
  symmetric, homogeneous, and stationary wind. They avoid solving for
  the dynamics with a complicated radiative acceleration term by
  imposing a velocity structure $v\equiv v(r)$. In this way, they are
  able to produce synthetic spectra, which they then fit to observed
  WR stars to infer $\dot{M}$. \cite{hamann:98} suggest to reduce the
  mass loss rate by a factor between 2 and 3 to account for wind
  clumpiness, cf. Appendix~\ref{sec:hamann}.
\end{description}

We report in \Tabref{tab:scalingLT} an approximate scaling of the mass
loss rate with the luminosity ($L$), which quantifies the amount of
photons available to drive the wind (neglecting the frequency
dependence of the line transitions), and the effective temperature
($T_\mathrm{eff}$), which can be considered as a rough parametrization
of the ionization state at the base of the wind and therefore the
opacity. We strongly recommend to consult \Tabref{tab:scaling} for the
scaling of $\dot{M}$ with physical stellar quantities for each of
these algorithms for anything beyond simple order of magnitude
estimates.

\begin{table}[hbp]
\setlength{\abovecaptionskip}{-0.1cm}
\setlength{\belowcaptionskip}{-0.2cm}
\begin{center}
  \caption{Approximate scaling of the mass loss rate
    with luminosity and effective
    temperature, $\log_{10}(\dot{M})\propto
    \alpha\log_{10}(L)+\beta\log_{10}(T_\mathrm{eff})$, predicted by
    the considered mass loss
    algorithms. Appendix~\ref{app:schemes_review}
    provides the full functional form of the algorithms implemented in
    MESA and \Tabref{tab:scaling} lists the scaling with all physical
    quantities. \label{tab:scalingLT}}
\begin{tabular}{ll|cr}\hline\hline
 & \multicolumn{1}{l|}{ID} & $\alpha$  & \multicolumn{1}{c}{$\beta$}       \\ \hline\hline
\multirow{2}{*}{\begin{sideways}Hot\end{sideways}} & V                       & $2.2$ & $1.0$     \\
 & K                       & $1.2$ & $0.6$     \\
  \hline
\multirow{3}{*}{\begin{sideways}Cool\end{sideways}} & dJ                      & $1.8$ & $-1.7$    \\
 & NJ                      & $1.6$ & $-1.6$    \\
 & vL                      & $1.0$ & $-6.3$    \\
  \hline
\multirow{3}{*}{\begin{sideways}WR\end{sideways}} & NL                      & $1.3$ &  $0.0$           \\
 & H  ($L>4.5L_\odot$)   & $1.5$ &  $0.0$           \\  
 & H  ($L\leq4.5L_\odot$)& $6.8$ &  $0.0$           \\  
\end{tabular}
\end{center}
\end{table}

\subsection{Combination of mass loss algorithms}
\label{sec:comb}

\begin{table*}[tbp]
  \begin{center}
  \caption{Combinations of wind mass loss algorithms
    employed in this study.  The
    temperature threshold separating the hot phase and the cool phase
    is $T_\mathrm{th}=15\,000$ (10\,000) K when using the Kudritzki
    \emph{et al.} (Vink \emph{et al.})  algorithm. The WR phase is
    defined using the surface (outermost computational cell) hydrogen
    mass fraction $X_s$, without constraints on $T_\mathrm{eff}$. In
    the text, we do not mention the WR phase algorithm if the model
    discussed does not enter this phase. For a description of the
    algorithms, see \Secref{sec:presc} and the appendices listed in this
    table. We discuss the definition of each evolutionary phase in
    \Secref{sec:comb}.  \label{tab:comb}}
    \begin{tabular}{l|lr|lr|lr}\hline\hline
      \multicolumn{1}{c|}{\multirow{2}{*}{ID}} & \multicolumn{2}{c|}{Hot phase} & \multicolumn{2}{c|}{Cool phase} & \multicolumn{2}{c}{WR phase}\\
       & \multicolumn{2}{c|}{$T_\mathrm{eff} \gtrsim T_\mathrm{th}$} & \multicolumn{2}{c|}{$T_\mathrm{eff} \lesssim T_\mathrm{th} $} & \multicolumn{2}{c}{$X_s <0.4$} \\ \hline
      \hline
      V-dJ-NL & Vink \emph{et al.} & \ref{sec:vink} & de Jager \emph{et al.} & \ref{sec:dejager} & Nugis \& Lamers      & \ref{sec:nl}\\
      V-dJ-H  & Vink \emph{et al.} & \ref{sec:vink} & de Jager \emph{et al.} & \ref{sec:dejager} & Hamann \emph{et al.} & \ref{sec:hamann} \\
      V-NJ-NL & Vink \emph{et al.} & \ref{sec:vink} & Nieuwenhuijzen \& de Jager & \ref{sec:nieuw} & Nugis \& Lamers  & \ref{sec:nl} \\
      V-NJ-H  & Vink \emph{et al.} & \ref{sec:vink} & Nieuwenhuijzen \& de Jager &\ref{sec:nieuw} & Hamann \emph{et al.} &\ref{sec:hamann}\\
      V-vL-H  & Vink \emph{et al.} & \ref{sec:vink} & van Loon \emph{et al.} &\ref{sec:van_loon} & Hamann \emph{et al.}    &\ref{sec:hamann} \\
      V-vL-NL & Vink \emph{et al.} & \ref{sec:vink} & van Loon \emph{et al.} &\ref{sec:van_loon} & Nugis \& Lamers & \ref{sec:nl}\\
      K-dJ-NL & Kudritzki \emph{et al.} & \ref{sec:kudr} & de Jager \emph{et al.} & \ref{sec:dejager} & Nugis \& Lamers & \ref{sec:nl}\\
      K-dJ-H  & Kudritzki \emph{et al.} & \ref{sec:kudr} & de Jager \emph{et al.} &\ref{sec:dejager} & Hamann \emph{et al.} & \ref{sec:hamann}\\
      K-NJ-NL & Kudritzki \emph{et al.} & \ref{sec:kudr} & Nieuwenhuijzen \& de Jager &\ref{sec:nieuw} & Nugis \& Lamers & \ref{sec:nl}\\
      K-NJ-H  & Kudritzki \emph{et al.} & \ref{sec:kudr} & Nieuwenhuijzen \& de Jager & \ref{sec:nieuw} & Hamann \emph{et al.} & \ref{sec:hamann} \\
      K-vL-NL & Kudritzki \emph{et al.} & \ref{sec:kudr} & van Loon \emph{et al.} & \ref{sec:van_loon} & Nugis \& Lamers & \ref{sec:nl}\\
      K-vL-H  & Kudritzki \emph{et al.} & \ref{sec:kudr} & van Loon \emph{et al.} & \ref{sec:van_loon} & Hamann \emph{et al.} & \ref{sec:hamann} \\
    \end{tabular}
  \end{center}
\end{table*}

The wind mass loss of massive stars is usually divided into three
separate phases, whose definition is somewhat arbitrary. When using
the algorithm K for the hot phase, we adopt the
following thresholds based on the effective temperature
$T_\mathrm{eff}$ and surface (i.e.,\ outermost computational cell)
hydrogen mass fraction $X_s$:

\begin{itemize}
\item[$\bullet$] {\bf Hot phase}: $T_\mathrm{eff} \geq 15\,000 \ \mathrm{K}$;
\item[$\bullet$] {\bf Cool phase}: $T_\mathrm{eff} < 15\,000 \ \mathrm{K}$;
\item[$\bullet$] {\bf WR phase}: $X_s < 0.4 \ \mathrm{regardless\ of} \ T_\mathrm{eff}$.
\end{itemize}

To follow the suggestions of \cite{glebbeek:09}, and to have a smoother transition between the hot and
cool phase wind algorithm, we use a slightly different definition of
the cool and hot evolutionary stages when using the V mass loss
algorithm for the hot phase:
\begin{itemize}
\item[$\bullet$] {\bf Hot phase}: $T_\mathrm{eff}  \geq 11\,000 \ \mathrm{K}$;
\item[$\bullet$] {\bf Cool phase}: $T_\mathrm{eff} \leq 10\,000 \ \mathrm{K}$;
\end{itemize}
and we use a linear interpolation between the hot phase wind and cool
phase wind in between. We choose this different threshold when
  using V to
  match how the cool and hot phases are defined
for the ``\texttt{Dutch}'' wind scheme in the MESA code. Note
that the interval from $T_\mathrm{eff}\simeq 10,\mathrm{kK}$ to $\sim
15\,\mathrm{kK}$ is covered during a fraction of the Hertzsprung gap,
in a very short time. 

The threshold dividing the cool and hot
phases is qualitatively justified as follows: the radiation pressure is
determined by the product of opacity 
and flux, which peaks between $10000$--$15000 \ \mathrm{K}$ because of iron
recombination. Therefore, an effective temperature (i.e.,\ in other
words, a radius, for each given luminosity) in this range is a
physically meaningful threshold to switch between wind mass loss algorithms.

The third phase is the WR phase and our
criterion $X_s < 0.4$ just requires a hydrogen-poor stellar
surface. This has very little in common with the
observational definition of what a WR star is, which is based on
spectral features that are not tracked by stellar evolution
codes. Specifically, WR stars are identified by their surface hydrogen
depletion \citep[][]{schmutz:99} \emph{and} the presence of broad
emission lines \citep[][]{vanderhutch:01,marchenko:10}, indicating the
presence of a wind with a steep density and velocity gradient.
Moreover, typical WR stars have high $T_\mathrm{eff}$,
and our definition might, in principle, produce unrealistically cold
(and red) WR stars. However, in absence of strong mixing processes
(e.g., due to rotation), the required surface
hydrogen depletion can only be reached by removing mass from the
surface and revealing deep and hot stellar layers. We do not find in
our calculations cool but hydrogen depleted models. WR stars are further subdivided into classes (WNH, WN, WC, WO, etc.)
based on the relative flux of specific lines. Here, we do not attempt
to distinguish between different WR sub-classes, because our
simulations do not produce the stellar spectra that would be necessary
to distinguish these sub-classes \citep[see however][and references
therein]{meynet:03,groh:14}.

Our definitions of the evolutionary phases are commonly used
in the literature
\citep[see, e.g.,][]{limongi:06,eggenberger:07,woosley:07}. We list in
\Tabref{tab:comb} the algorithm combinations explored here: each
combination is labeled by combining the abbreviations for each wind
algorithm introduced above.

To study the effects of the possible overestimate of the mass loss
rate caused, for example, by wind
inhomogeneities (i.e., ``clumpiness''), we use three different values
of the wind efficiency $\eta = 1.0,\, 0.33,\, 0.1$. These values span
the range of observational estimates of the volume filling factor of
clumps (see \citealt{smith:14} for a review). For simplicity, we use
the same $\eta$ during the entire evolution of
a given model. 

\subsection{The Grid of Stellar Models}
\label{sec:models}

We employ release version 7624 of the open-source stellar evolution
code MESA \citep[][]{paxton:11,paxton:13,paxton:15} and compute a grid
of nonrotating solar metallicity models. We choose $Z_\odot=0.019$ to
match precisely the value adopted in \cite{vink:01}.
We consider initial masses of
$15$, $20$, $25$, $30$, and $35\,$ $M_\odot$.  Higher initial mass
models are more strongly affected by numerical (and possibly physical)
instabilities (see Appendix~\ref{app:MESA_technical} and references therein), which makes them less reliable for our
purpose. Appendix~\ref{app:MESA_technical} describes the details of
our MESA simulations that are not directly related to mass loss. Here,
we only mention that we use a 45-isotope nuclear reaction network
(\texttt{mesa\_45.net}) until oxygen depletion, and a mixing length
parameter $\alpha_\mathrm{mlt}=2.0$ with exponential overshooting.

We run our grid of models in three steps and at each checkpoint we
make a selection of models to run at the next step\footnote{The models
  saved at each checkpoint are available at
  \url{https://zenodo.org/record/292924\#.WK_eENWi60i} . The
  input files and customized routines are available at 
  \url{https://stellarcollapse.org/renzo2017}.}. This selection is
necessary to reduce the total computational cost of our grid of
models. In the first step, we evolve our models from the zero age main
sequence (ZAMS) to when the temperature in the central computational
cell rises above $T_c\geq10^9\,\mathrm{K}$ (``end of the mass loss
phase''). At this point, the star has little time left to live (a few
years) and lose mass through winds (for a typical mass loss rate the
star loses less than
$10^{-4}\,M_\odot$). Also, when
$T_c\geq10^9\,\mathrm{K}$ is reached, MESA
starts to artificially damp the mass loss rate for stability
reasons. Mass loss is completely shut off when
$T_c\geq2\times10^9\,\mathrm{K}$.

Note, however, that the
observation of SN impostors, type IIn SNe, and intra-night flash
spectroscopy of normal type II SNe suggests that non-wind
mass loss phenomena (neglected here) may occur in the very late phase
of the stellar life \citep[see, e.g.,][]{quataert:12,smith:14b,quataert:16,khazov:16}.

In the second step, we restart a subset of MESA models and run to
oxygen depletion (see \Secref{sec:O_depl_res}). This is defined as the
first time when the mass fraction of $^{16}\mathrm{O}$ in the central
cell drops below 0.04 and the mass fraction of $^{28}\mathrm{Si}$ is
higher than 0.01
\citep[indicating that some oxygen burning has already occurred,
  following][]{sukhbold:14}. Note that these thresholds are an
artificial choice, since the evolution of the star is 
continuous throughout its lifetime.

Finally, we restart a select subset of stars at oxygen depletion and
run to the onset of core collapse, defined by
\begin{equation}\label{eq:onset_cc}
\mathrm{max}\{ |v| \} \geq 10^3 \ \mathrm{km \ s^{-1}} \ \ ,
\end{equation}
where $v$ is the radial infall velocity \citep[see, e.g.,][]{heger:05,
  sukhbold:14}. For this last phase of the evolution, we switch from
the 45-isotopes nuclear reaction network to a customized 203-isotopes
nuclear reaction network (see also Appendix~\ref{app:MESA_technical}),
to capture the details of the weak interaction physics (electron
captures and $\beta$-decays) occurring during silicon burning and
before collapse. This physics determines the final number of electrons
per nucleon $Y_e$, thus the effective Chandrasekhar mass of the
stellar core, and ultimately the core structure at the onset of core
collapse. Experiments with a single-zone model for silicon burning
show that at least $\sim 200$ isotopes are required to obtain a
converged final value of $Y_e$ in the core \citep[i.e., one that is
  independent of the size of the nuclear reaction network; see
  also][]{farmer:16}.  Since MESA solves the fully coupled set of
equations for the chemical composition and structure of the star, the
increase in the number of isotopes forces us to reduce the number of
spatial mesh points in each model because of memory limitations (see
Appendix~\ref{app:re-meshing}).

\section{Results}
\label{sec:res}

\begin{figure*}[!htp] \centering
  \includegraphics[width=\textwidth]{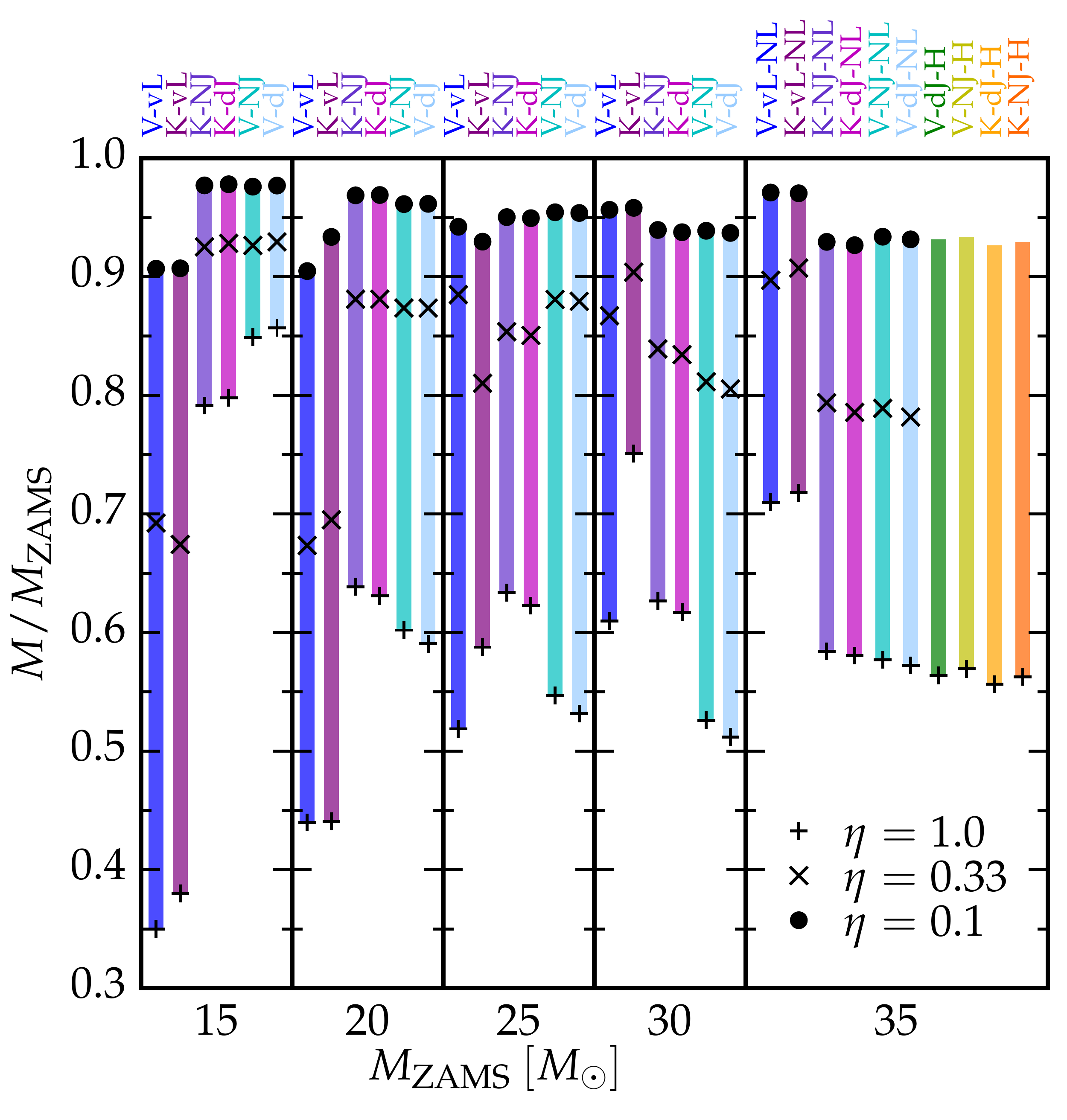} 
  \caption{Uncertainty in the mapping between
    $M_\mathrm{ZAMS}$ and the relative final mass
    $M/M_\mathrm{ZAMS}$ due to wind mass loss.  Each colored bar
    corresponds to a specific wind algorithm combination defined in
    \Tabref{tab:comb}.  The pluses, crosses, and circles correspond
    to $\eta=1.0,\,0.33,\,0.1$, respectively. We employ the vertical
    bars to emphasize the spread.  The uncertainty in wind mass loss
    limits the predictive
    power of stellar
    evolution studies for the
    final mass of stars of given initial mass. Only models with
    $M_\mathrm{ZAMS} = 35\,M_\odot$ and wind efficiency $\eta = 1$
    reach the WR stage (cf.~\Tabref{tab:colors} and
    \Secref{sec:WRs}). They are shown in the rightmost panel and we
    list the WR mass loss algorithm only for them. Note that the
    maximum relative mass for the four 
    algorithm combinations using the H WR
    mass loss algorithm in the rightmost panel are the
    results obtained using the corresponding hot and cool
    mass loss combination with
    $\eta=0.1$ (not reaching the WR phase).\label{fig:mrel_mzams}}
\end{figure*}

\subsection{Overview at the end of the mass loss phase, $T_c\geq10^9\,\mathrm{K}$}
\label{sec:end_ML}


\begin{table*}[!htbp]
  \begin{center}
    \caption{Model summary at the end of the
        mass loss phase (when $T_c\geq10^9\,\mathrm{K}$, roughly
        corresponding to neon core ignition). We provide total mass
      $M$, helium core mass $M_\mathrm{He}$, and carbon-oxygen (CO)
      core mass $M_\mathrm{CO}$. We
      omit models differing only by the WR wind scheme if they do not
      reach the WR stage (i.e.\ $X_s>0.4$ at all times). We define the
      edge of the CO core as first location going inward where
      $X(^{4}\mathrm{He})<0.01$, without requiring a minimum mass
      fraction of carbon or oxygen. Analogously, we define the outer
      edge of the helium core as the first location going inward where
      $X(^1\mathrm{H})<0.01$.
    \vspace{-0.3cm}}
 \label{tab:res}
     \small
\begin{tabular}{clccc||clccc}\hline \hline
  \multicolumn{10}{c}{End of mass loss phase: $T_c \geq 10^9 \,\mathrm{K}$}\\\hline\hline
  \multicolumn{5}{c||}{$M_\mathrm{ZAMS}=15\,M_\odot$} & \multicolumn{5}{c}{$M_\mathrm{ZAMS}=20\,M_\odot$}\\
  $\eta$ & \multicolumn{1}{l}{ID}  & $M \ [M_\odot]$ &
  $M_\mathrm{He} \ [M_\odot]$ & $M_\mathrm{CO} \ [M_\odot]$ & $\eta$ & \multicolumn{1}{l}{ID} & $M \ [M_\odot]$ & $M_\mathrm{He} \ [M_\odot]$ & $M_\mathrm{CO} \ [M_\odot]$ \\ \hline
  \hline
\multirow{6}{*}{0.1}  & V-dJ  & 14.66 & \phantom{0}4.99 & \phantom{0}3.20                                & \multirow{6}{*}{0.1}  & V-dJ             & 19.23 & \phantom{0}7.04 & 4.91   \\            
                      & V-NJ  & 14.64 & \phantom{0}4.99 & \phantom{0}3.20                                &                       & V-NJ             & 19.23 & \phantom{0}7.04 & 4.91    \\   
                      & V-vL  & 13.60 & \phantom{0}4.98 & \phantom{0}3.20                                &                       & V-vL             & 18.10 & \phantom{0}7.03 & 4.91    \\   
                      & K-dJ  & 14.67 & \phantom{0}5.00 & \phantom{0}3.22                                &                       & K-dJ             & 19.38 & \phantom{0}7.01 & 4.90    \\   
                      & K-NJ  & 14.66 & \phantom{0}5.00 & \phantom{0}3.22                                &                       & K-NJ             & 19.37 & \phantom{0}7.01 & 4.90    \\   
                      & K-vL  & 13.61 & \phantom{0}5.00 & \phantom{0}3.21                                &                       & K-vL             & 18.67 & \phantom{0}7.01 & 4.90    \\   
\cline{1-6}\cline{7-10}
\multirow{6}{*}{0.33} & V-dJ  & 13.94 & \phantom{0}4.93 & \phantom{0}3.16                                & \multirow{6}{*}{0.33} & V-dJ             & 17.47 & \phantom{0}7.01 & 4.88    \\   
                      & V-NJ  & 13.90 & \phantom{0}4.93 & \phantom{0}3.16                                &                       & V-NJ             & 17.48 & \phantom{0}7.01 & 4.88    \\   
                      & V-vL  & 10.39 & \phantom{0}4.93 & \phantom{0}3.15                                &                       & V-vL             & 13.47 & \phantom{0}6.99 & 4.87    \\   
                      & K-dJ  & 13.92 & \phantom{0}4.98 & \phantom{0}3.19                                &                       & K-dJ             & 17.62 & \phantom{0}7.00 & 4.87    \\   
                      & K-NJ  & 13.88 & \phantom{0}4.98 & \phantom{0}3.19                                &                       & K-NJ             & 17.62 & \phantom{0}7.00 & 4.88    \\   
                      & K-vL  & 10.11 & \phantom{0}4.97 & \phantom{0}3.19                                &                       & K-vL             & 13.90 & \phantom{0}6.98 & 4.87    \\   
\cline{1-6}\cline{7-10} 
\multirow{6}{*}{1.0}  & V-dJ  & 12.86 & \phantom{0}4.65 & \phantom{0}2.95                                & \multirow{6}{*}{1.0}  & V-dJ             & 11.81 & \phantom{0}7.06 & 4.92    \\   
                      & V-NJ  & 12.74 & \phantom{0}4.65 & \phantom{0}2.95                                &                       & V-NJ             & 12.04 & \phantom{0}7.06 & 4.92    \\   
                      & V-vL  & \phantom{0}5.25 & \phantom{0}4.64 & \phantom{0}2.94                      &                       & V-vL             & \phantom{0}8.80 & \phantom{0}7.02 & 4.89    \\      
                      & K-dJ  & 11.97 & \phantom{0}4.92 & \phantom{0}3.15                                &                       & K-dJ             & 12.62 & \phantom{0}6.95 & 4.84    \\   
                      & K-NJ  & 11.87 & \phantom{0}4.92 & \phantom{0}3.14                                &                       & K-NJ             & 12.77 & \phantom{0}6.95 & 4.84    \\   
                      & K-vL  & \phantom{0}5.70 & \phantom{0}4.90 & \phantom{0}3.13                      &                       & K-vL             & \phantom{0}8.81 & \phantom{0}6.90 & 4.80\\
\hline\hline
\multicolumn{5}{c||}{$M_\mathrm{ZAMS}=25\,M_\odot$} & \multicolumn{5}{c}{$M_\mathrm{ZAMS}=30\,M_\odot$}\\
  $\eta$ & \multicolumn{1}{l}{ID}  & $M \ [M_\odot]$ &
  $M_\mathrm{He} \ [M_\odot]$ & $M_\mathrm{CO} \ [M_\odot]$ & $\eta$ &
  \multicolumn{1}{l}{ID} &  $M \ [M_\odot]$ & $M_\mathrm{He} \ [M_\odot]$ & $M_\mathrm{CO} \ [M_\odot]$ \\ \hline
  \hline
\multirow{6}{*}{0.1}  & V-dJ  & 23.85 & \phantom{0}9.14 & \phantom{0}6.58          & \multirow{6}{*}{0.1}  & V-dJ            & 28.11 & 10.97 & 7.97    \\                                                                       
                      & V-NJ  & 23.86 & \phantom{0}9.14 & \phantom{0}6.59          &                       & V-NJ            & 28.17 & 10.97 & 7.94    \\ 
                      & V-vL  & 23.56 & \phantom{0}9.14 & \phantom{0}6.59          &                       & V-vL            & 28.70 & 10.96 & 7.95    \\ 
                      & K-dJ  & 23.74 & \phantom{0}9.23 & \phantom{0}6.66          &                       & K-dJ            & 28.13 & 10.91 & 7.90    \\ 
                      & K-NJ  & 23.76 & \phantom{0}9.24 & \phantom{0}6.67          &                       & K-NJ            & 28.19 & 10.91 & 7.86    \\ 
                      & K-vL  & 23.24 & \phantom{0}9.23 & \phantom{0}6.67          &                       & K-vL            & 28.75 & 10.90 & 7.91    \\ 
\cline{1-6}\cline{7-10}                                                                                                                          
\multirow{6}{*}{0.33} & V-dJ  & 21.98 & \phantom{0}8.87 & \phantom{0}6.37          & \multirow{6}{*}{0.33} & V-dJ            & 24.16 & 11.11 & 8.13    \\ 
                      & V-NJ  & 22.02 & \phantom{0}8.87 & \phantom{0}6.38          &                       & V-NJ            & 24.34 & 11.11 & 8.13    \\ 
                      & V-vL  & 22.13 & \phantom{0}8.87 & \phantom{0}6.36          &                       & V-vL            & 26.02 & 11.10 & 8.12    \\ 
                      & K-dJ  & 21.27 & \phantom{0}9.11 & \phantom{0}6.57          &                       & K-dJ            & 25.03 & 10.89 & 7.88    \\ 
                      & K-NJ  & 21.34 & \phantom{0}9.11 & \phantom{0}6.57          &                       & K-NJ            & 25.18 & 10.89 & 7.87    \\ 
                      & K-vL  & 20.26 & \phantom{0}9.11 & \phantom{0}6.55          &                       & K-vL            & 27.12 & 10.87 & 7.80    \\ 
\cline{1-6}\cline{7-10}                                                                                                                         
\multirow{6}{*}{1.0}  & V-dJ  & 13.29 & \phantom{0}9.05 & \phantom{0}6.50          & \multirow{6}{*}{1.0}  & V-dJ            & 15.36 & 11.28 & 8.23    \\ 
                      & V-NJ  & 13.67 & \phantom{0}9.05 & \phantom{0}6.48          &                       & V-NJ            & 15.78 & 11.29 & 8.26    \\ 
                      & V-vL  & 12.97 & \phantom{0}9.04 & \phantom{0}6.51          &                       & V-vL            & 18.29 & 11.31 & 8.27    \\ 
                      & K-dJ  & 15.57 & \phantom{0}8.89 & \phantom{0}6.38          &                       & K-dJ            & 18.51 & 10.92 & 7.96    \\ 
                      & K-NJ  & 15.84 & \phantom{0}8.89 & \phantom{0}6.37          &                       & K-NJ            & 18.80 & 10.92 & 7.94    \\ 
                      & K-vL  & 14.69 & \phantom{0}8.90 & \phantom{0}6.38          &                       & K-vL            & 22.53 & 10.89 & 7.94    \\ 
\hline\hline
\multicolumn{5}{c||}{$M_\mathrm{ZAMS}=35\,M_\odot$}                                        & \multicolumn{5}{}{\multirow{22}{}{a}} \\ 
  $\eta$ & \multicolumn{1}{l}{ID} & $M \ [M_\odot]$ &
  $M_\mathrm{He} \ [M_\odot]$ & $M_\mathrm{CO} \ [M_\odot]$ & \multicolumn{5}{}{} \\
\cline{1-5}\cline{1-5}
\multirow{6}{*}{0.1}  & V-dJ-NL  & 32.61 & 10.25 & \phantom{0}7.19      & & & & \\
                      & V-NJ-NL  & 32.69 & 10.24 & \phantom{0}7.19      & & & & \\
                      & V-vL-NL  & 33.99 & 10.03 & \phantom{0}7.03      & & & & \\
                      & K-dJ-NL  & 32.43 & 11.87 & \phantom{0}8.54      & & & & \\
                      & K-NJ-NL  & 32.53 & 11.87 & \phantom{0}8.56      & & & & \\
                      & K-vL-NL  & 33.97 & 11.83 & \phantom{0}8.51      & & & & \\
\cline{1-5}
\multirow{6}{*}{0.33} & V-dJ-NL  & 27.36 & 12.72 & \phantom{0}9.32     & & & & \\
                      & V-NJ-NL  & 27.62 & 12.72 & \phantom{0}9.32     & & & & \\
                      & V-vL-NL  & 31.40 & 12.67 & \phantom{0}9.23     & & & & \\
                      & K-dJ-NL  & 27.50 & 12.68 & \phantom{0}9.21     & & & & \\
                      & K-NJ-NL  & 27.78 & 12.68 & \phantom{0}9.28     & & & & \\
                      & K-vL-NL  & 31.76 & 12.63 & \phantom{0}9.21     & & & & \\
\cline{1-5}
\multirow{10}{*}{1.0} & V-dJ-NL  & 20.03 & 13.30 & \phantom{0}9.81     & & & & \\
                      & V-NJ-NL  & 20.20 & 13.30 & \phantom{0}9.94     & & & & \\
                      & V-vL-NL  & 24.85 & 13.30 & \phantom{0}9.97     & & & & \\
                      & K-dJ-NL  & 20.32 & 13.82 & 10.33               & & & & \\
                      & K-NJ-NL  & 20.45 & 13.82 & 10.36               & & & & \\
                      & K-vL-NL  & 25.13 & 13.82 & 10.41               & & & & \\
                      & V-dJ-H   & 19.73 & 13.30 & \phantom{0}9.93     & & & & \\
                      & V-NJ-H   & 19.93 & 13.30 & \phantom{0}9.99     & & & & \\
                      & K-dJ-H   & 19.48 & 13.81 & 10.41               & & & & \\
                      & K-NJ-H   & 19.69 & 13.82 & 10.41               & & & & \\

\end{tabular}
\end{center}
\end{table*}


\begin{table*}[!htbp]
  \begin{center}
\caption{Final color and surface properties of the computed
  models. We follow the definitions of
  \cite{georgy:12} to classify models as RSGs, YSGs, or BSGs
    (see also text).
  WR stars have $X_s\leq0.4$, regardless
  of their surface temperature. The first and
  second columns indicate the wind efficiency and the mass loss
  algorithm combination,
  respectively. The WR stars are computed twice, once with the NL
  mass loss algorithm (see \Secref{sec:nl}) and once with the H
  algorithm (see \Secref{sec:hamann}).  \label{tab:colors}} \small
\begin{tabular}{clcccl||clcccl}\hline \hline
  \multicolumn{12}{c}{End of mass loss phase: $T_c \geq 10^9 \,\mathrm{K}$}\\\hline\hline
  \multicolumn{6}{c||}{$M_\mathrm{ZAMS}=15\,M_\odot$} & \multicolumn{6}{c}{$M_\mathrm{ZAMS}=20\,M_\odot$}\\
  $\eta$ & \multicolumn{1}{l}{ID}  & $R \ [R_\odot]$ & $\log_{10}(L/L_\odot)$ &
 $\log_\mathrm{10}(T_\mathrm{eff}/\mathrm{[K]})$& color & $\eta$ & \multicolumn{1}{c}{ID}  & $R \ [R_\odot]$ & $\log_{10}(L/L_\odot)$ &$\log_\mathrm{10}(T_\mathrm{eff}/\mathrm{[K]})$& color \\ \hline
  \hline
\multirow{6}{*}{0.1}      & V-dJ & 911 & 5.06 & 3.55 & \textcolor{red}{\bf RSG}                         & \multirow{6}{*}{0.1}   & V-dJ & 992 & 5.27 & 3.58 & \textcolor{red}{\bf RSG}    \\            
                          & V-NJ & 911 & 5.06 & 3.55 & \textcolor{red}{\bf RSG}                         &                        & V-NJ & 992 & 5.27 & 3.58 & \textcolor{red}{\bf RSG}    \\   
                          & V-vL & 927 & 5.06 & 3.54 & \textcolor{red}{\bf RSG}                         &                        & V-vL & 996 & 5.27 & 3.58 & \textcolor{red}{\bf RSG}    \\   
                          & K-dJ & 914 & 5.06 & 3.55 & \textcolor{red}{\bf RSG}                         &                        & K-dJ & 987 & 5.27 & 3.58 & \textcolor{red}{\bf RSG}    \\   
                          & K-NJ & 914 & 5.06 & 3.55 & \textcolor{red}{\bf RSG}                         &                        & K-NJ & 989 & 5.27 & 3.58 & \textcolor{red}{\bf RSG}    \\   
                          & K-vL & 930 & 5.06 & 3.54 & \textcolor{red}{\bf RSG}                         &                        & K-vL & 991 & 5.27 & 3.58 & \textcolor{red}{\bf RSG}    \\   
\cline{1-6}\cline{7-12}                                                                                                                                                                  
\multirow{6}{*}{0.33}     & V-dJ & 916 & 5.05 & 3.54 & \textcolor{red}{\bf RSG}                         & \multirow{6}{*}{0.33}  & V-dJ & 998 & 5.27 & 3.58 & \textcolor{red}{\bf RSG}    \\   
                          & V-NJ & 916 & 5.05 & 3.54 & \textcolor{red}{\bf RSG}                         &                        & V-NJ & 999 & 5.27 & 3.58 & \textcolor{red}{\bf RSG}    \\   
                          & V-vL & 968 & 5.05 & 3.53 & \textcolor{red}{\bf RSG}                         &                        & V-vL & 985 & 5.27 & 3.58 & \textcolor{red}{\bf RSG}    \\   
                          & K-dJ & 920 & 5.06 & 3.54 & \textcolor{red}{\bf RSG}                         &                        & K-dJ & 998 & 5.27 & 3.58 & \textcolor{red}{\bf RSG}    \\   
                          & K-NJ & 921 & 5.06 & 3.54 & \textcolor{red}{\bf RSG}                         &                        & K-NJ & 998 & 5.27 & 3.58 & \textcolor{red}{\bf RSG}    \\   
                          & K-vL & 959 & 5.06 & 3.54 & \textcolor{red}{\bf RSG}                         &                        & K-vL & 991 & 5.27 & 3.58 & \textcolor{red}{\bf RSG}    \\   
\cline{1-6}\cline{7-12}                                                                                                                                                                  
\multirow{6}{*}{1.0}      & V-dJ & 895 & 5.02 & 3.54 & \textcolor{red}{\bf RSG}                         & \multirow{6}{*}{1.0}   & V-dJ & 950 & 5.28 & 3.59 & \textcolor{red}{\bf RSG}    \\   
                          & V-NJ & 896 & 5.02 & 3.54 & \textcolor{red}{\bf RSG}                         &                        & V-NJ & 956 & 5.28 & 3.59 & \textcolor{red}{\bf RSG}    \\   
                          & V-vL & 630 & 5.02 & 3.62 & \textcolor{Orange}{\bf YSG}                      &                        & V-vL & 672 & 5.28 & 3.67 & \textcolor{Orange}{\bf YSG} \\      
                          & K-dJ & 947 & 5.05 & 3.54 & \textcolor{red}{\bf RSG}                         &                        & K-dJ & 972 & 5.27 & 3.59 & \textcolor{red}{\bf RSG}    \\   
                          & K-NJ & 949 & 5.05 & 3.54 & \textcolor{red}{\bf RSG}                         &                        & K-NJ & 976 & 5.27 & 3.58 & \textcolor{red}{\bf RSG}    \\   
                          & K-vL & 644 & 5.05 & 3.62 & \textcolor{Orange}{\bf YSG}                      &                        & K-vL & 712 & 5.27 & 3.65 & \textcolor{Orange}{\bf YSG} \\
\hline\hline
\multicolumn{6}{c||}{$M_\mathrm{ZAMS}=25\,M_\odot$} & \multicolumn{6}{c}{$M_\mathrm{ZAMS}=30\,M_\odot$}\\
  $\eta$ & \multicolumn{1}{l}{ID}  & $R \ [R_\odot]$ & $\log_{10}(L/L_\odot)$ &
 $\log_\mathrm{10}(T_\mathrm{eff}/\mathrm{[K]})$& color & $\eta$ & \multicolumn{1}{l}{ID}  & $R \ [R_\odot]$ & $\log_{10}(L/L_\odot)$ &$\log_\mathrm{10}(T_\mathrm{eff}/\mathrm{[K]})$& color \\ \hline
  \hline
\multirow{6}{*}{0.1}   & V-dJ & 899 & 5.41 & 3.64 & \textcolor{Orange}{\bf YSG}         & \multirow{6}{*}{0.1}    & V-dJ & 698 & 5.50 & 3.72 & \textcolor{Orange}{\bf YSG}         \\                                                                       
                       & V-NJ & 900 & 5.42 & 3.64 & \textcolor{Orange}{\bf YSG}         &                         & V-NJ & 697 & 5.51 & 3.72 & \textcolor{Orange}{\bf YSG}         \\ 
                       & V-vL & 902 & 5.42 & 3.64 & \textcolor{Orange}{\bf YSG}         &                         & V-vL & 696 & 5.51 & 3.72 & \textcolor{Orange}{\bf YSG}         \\ 
                       & K-dJ & 891 & 5.43 & 3.64 & \textcolor{Orange}{\bf YSG}         &                         & K-dJ & 705 & 5.53 & 3.72 & \textcolor{Orange}{\bf YSG}         \\ 
                       & K-NJ & 891 & 5.44 & 3.65 & \textcolor{Orange}{\bf YSG}         &                         & K-NJ & 704 & 5.53 & 3.72 & \textcolor{Orange}{\bf YSG}         \\ 
                       & K-vL & 893 & 5.43 & 3.64 & \textcolor{Orange}{\bf YSG}         &                         & K-vL & 708 & 5.52 & 3.72 & \textcolor{Orange}{\bf YSG}         \\ 
\cline{1-6}\cline{7-12}                                                                                                                                                       
\multirow{6}{*}{0.33}  & V-dJ & 926 & 5.42 & 3.63 & \textcolor{Orange}{\bf YSG}         & \multirow{6}{*}{0.33}   & V-dJ & 670 & 5.54 & 3.73 & \textcolor{Orange}{\bf YSG}         \\ 
                       & V-NJ & 924 & 5.41 & 3.63 & \textcolor{Orange}{\bf YSG}         &                         & V-NJ & 670 & 5.54 & 3.73 & \textcolor{Orange}{\bf YSG}         \\ 
                       & V-vL & 932 & 5.42 & 3.63 & \textcolor{Orange}{\bf YSG}         &                         & V-vL & 675 & 5.55 & 3.73 & \textcolor{Orange}{\bf YSG}         \\ 
                       & K-dJ & 896 & 5.42 & 3.64 & \textcolor{Orange}{\bf YSG}         &                         & K-dJ & 719 & 5.53 & 3.72 & \textcolor{Orange}{\bf YSG}         \\ 
                       & K-NJ & 897 & 5.42 & 3.64 & \textcolor{Orange}{\bf YSG}         &                         & K-NJ & 717 & 5.52 & 3.71 & \textcolor{Orange}{\bf YSG}         \\ 
                       & K-vL & 899 & 5.43 & 3.64 & \textcolor{Orange}{\bf YSG}         &                         & K-vL & 711 & 5.53 & 3.72 & \textcolor{Orange}{\bf YSG}         \\ 
\cline{1-6}\cline{7-12}                                                                                                                                                       
\multirow{6}{*}{1.0}   & V-dJ & 797 & 5.43 & 3.67 & \textcolor{Orange}{\bf YSG}         & \multirow{6}{*}{1.0}    & V-dJ & 614 & 5.56 & 3.76 & \textcolor{Orange}{\bf YSG}         \\ 
                       & V-NJ & 811 & 5.43 & 3.67 & \textcolor{Orange}{\bf YSG}         &                         & V-NJ & 628 & 5.56 & 3.75 & \textcolor{Orange}{\bf YSG}         \\ 
                       & V-vL & 782 & 5.43 & 3.67 & \textcolor{Orange}{\bf YSG}         &                         & V-vL & 610 & 5.55 & 3.76 & \textcolor{Orange}{\bf YSG}         \\ 
                       & K-dJ & 874 & 5.42 & 3.65 & \textcolor{Orange}{\bf YSG}         &                         & K-dJ & 755 & 5.54 & 3.71 & \textcolor{Orange}{\bf YSG}         \\ 
                       & K-NJ & 878 & 5.42 & 3.65 & \textcolor{Orange}{\bf YSG}         &                         & K-NJ & 758 & 5.55 & 3.71 & \textcolor{Orange}{\bf YSG}         \\ 
                       & K-vL & 857 & 5.42 & 3.65 & \textcolor{Orange}{\bf YSG}         &                         & K-vL & 723 & 5.53 & 3.72 & \textcolor{Orange}{\bf YSG}         \\ 
\hline\hline
\multicolumn{6}{c||}{$M_\mathrm{ZAMS}=35\,M_\odot$}                                        & \multicolumn{6}{}{\multirow{22}{}{a}} \\ 
  $\eta$ & \multicolumn{1}{l}{ID}  & $R \ [R_\odot]$ & $\log_{10}(L/L_\odot)$ &
 $\log_\mathrm{10}(T_\mathrm{eff}/\mathrm{[K]})$& color & \multicolumn{6}{}{} \\
\cline{1-6}\cline{1-6}
\multirow{6}{*}{0.1}     & V-dJ & 860 & 5.53 & 3.68 & \textcolor{Orange}{\bf YSG}   & & & & \\
                         & V-NJ & 861 & 5.53 & 3.68 & \textcolor{Orange}{\bf YSG}   & & & & \\
                         & V-vL & 920 & 5.52 & 3.66 & \textcolor{Orange}{\bf YSG}   & & & & \\
                         & K-dJ & 726 & 5.59 & 3.73 & \textcolor{Orange}{\bf YSG}   & & & & \\
                         & K-NJ & 730 & 5.59 & 3.73 & \textcolor{Orange}{\bf YSG}   & & & & \\
                         & K-vL & 755 & 5.58 & 3.72 & \textcolor{Orange}{\bf YSG}   & & & & \\
\cline{1-6}
\multirow{6}{*}{0.33}    & V-dJ & 529 & 5.62 & 3.80 & \textcolor{blue}{\bf BSG}   & & & & \\
                         & V-NJ & 531 & 5.62 & 3.80 & \textcolor{blue}{\bf BSG}   & & & & \\
                         & V-vL & 569 & 5.61 & 3.79 & \textcolor{Orange}{\bf YSG} & & & & \\  
                         & K-dJ & 539 & 5.62 & 3.80 & \textcolor{blue}{\bf BSG}   & & & & \\
                         & K-NJ & 541 & 5.62 & 3.80 & \textcolor{blue}{\bf BSG}   & & & & \\
                         & K-vL & 595 & 5.62 & 3.78 & \textcolor{Orange}{\bf YSG} & & & & \\  
\cline{1-6}
\multirow{10}{*}{1.0}    & V-dJ-NL & 258 & 5.53 & 3.94 & {\bf WR}                 & & & & \\                                                    
                         & V-NJ-NL & 255 & 5.66 & 3.97 & {\bf WR}                  & & & & \\                                                   
                         & V-vL & 398 & 5.66 & 3.88 & \textcolor{blue}{\bf BSG}    & & & & \\                                                   
                         & K-dJ-NL & 167 & 5.75 & 4.09 & {\bf WR}                  & & & & \\                                                   
                         & K-NJ-NL & 168 & 5.73 & 4.08 & {\bf WR}                  & & & & \\                                                   
                         & K-vL & 313 & 5.67 & 3.93 & \textcolor{blue}{\bf BSG}    & & & & \\                                                   
                         & V-dJ-H & 253 & 5.60 & 3.96 & {\bf WR}                   & & & & \\                                                   
                         & V-NJ-H & 260 & 5.65 & 3.97 & {\bf WR}                   & & & & \\                                                   
                         & K-dJ-H & 157 & 5.72 & 4.09 & {\bf WR}                   & & & & \\                                                   
                         & K-NJ-H & 158 & 5.72 & 4.09 & {\bf WR} & & & &
                       
\end{tabular}
\end{center}
\end{table*}

\begin{table*}[!htbp]
\setlength{\belowcaptionskip}{0.5cm}
  \begin{center}
    \caption{Maximum spread in
      total mass, core masses, and radius at the end of the mass loss
      phase, for models differing in mass loss algorithm
      combination and efficiency $\eta$. We
      also list the maximum and minimum total mass for each
      $M_\mathrm{ZAMS}$.}
    \label{tab:max_spreads}
\begin{tabular}{c|cccccc}\hline\hline
$M_\mathrm{ZAMS}$ & $\max\Delta M$  & $\max M$    & $\min M$          & $\max\Delta M_\mathrm{He}$ &  $\max\Delta M_\mathrm{CO}$ & $\max\Delta R$ \\
$[M_\odot]$       &  $[M_\odot]$    & $[M_\odot]$ & $[M_\odot]$       & $[M_\odot]$                &  $[M_\odot]$                & $[R_\odot]$ \\ \hline\hline
15                & \phantom{0}9.42 & 14.66       & \phantom{0}5.25   & 0.36                       &     0.28                    & 338         \\
20                & 10.58           & 19.38       & \phantom{0}8.80   & 0.16	                   &     0.12                    & 327         \\
25                & 10.89           & 23.86       & 12.97             & 0.37	                   &     0.31                    & 150         \\
30                & 13.39           & 28.75       & 15.36             & 0.44	                   &     0.47                    & 148         \\
35                & 14.51	    & 33.99       & 19.48             & 3.79	                   &     3.38                    & 763         
\end{tabular}
\end{center}
\end{table*}

The lifetime remaining for the star after $T_c\geq10^9\,\mathrm{K}$ is
short ($\sim15$ years for $15\,M_\odot$, $\sim4$ years for
$30\,M_\odot$), and the photosphere of the star remains frozen at the
same effective temperature $T_\mathrm{eff}$ and luminosity $L$ as
long as no impulsive mass loss events or other instabilities take
place. The subsequent evolution will lead to changes
  of the internal structure only. Therefore, we discuss the final
mass and pre-SN appearance of our models already at this stage.

Figure~\ref{fig:mrel_mzams} shows the overall impact of the
uncertainty in stellar wind mass loss rates on the final mass. We plot
the final mass relative to the initial mass for each considered $M_\mathrm{ZAMS}$ and the large vertical spread illustrates the uncertainty. As an
example, we can consider a star of $M_\mathrm{ZAMS} = 20\,M_\odot$. It
can, in principle, reach the onset of collapse with $M=19.38\,M_\odot$
if evolved with the V-NJ mass loss combination with reduced efficiency
$\eta=0.1$. But it might also evolve to $M = 8.81\,M_\odot$ with
the K-vL combination and efficiency $\eta=1.0$. If these are truly
limiting cases and anything in between is unconstrained, then the
uncertainty in the final mass is greater than $50\%$. For more quantitative results, see \Tabref{tab:max_spreads}
where we report the maximum spreads of the ZAMS to pre-SN mass mapping
for our entire grid of models.  \Tabref{tab:res} and
\Tabref{tab:colors} list the main physical quantities for all the
stars in our grid at the end of the mass loss phase.

It is important to note that the vertical spread in the ZAMS to pre-SN
mass mapping shown in \Figref{fig:mrel_mzams} is dominated by the 
highly uncertain wind efficiency $\eta$. At fixed $\eta$, variations
due to different wind algorithm combinations are minor. \emph{This
  makes $\eta$ the most important free parameter for wind mass loss in
  stellar evolution calculations.}

In \Figref{fig:mrel_mzams}, the spread in $M_\mathrm{ZAMS}$ to pre-SN
mass decreases for higher $M_\mathrm{ZAMS}$. This is, however, only
because we show the relative final mass, i.e.\ at higher
$M_\mathrm{ZAMS}$ we divide the final mass by a larger number. In
absolute numbers, the uncertainty in the final to initial mass
relation increases for more massive stars. We summarize this in
\Tabref{tab:res}. As an example, the maximum spread between the final
masses of $35\,M_\odot$ models is $\max{\Delta M} \equiv
33.99\,M_\odot-19.48\,M_\odot=14.51\,M_\odot$. In the $15\,M_\odot$
case, it is $\max{\Delta M}\equiv 14.66\,M_\odot - 5.25\,M_\odot =
9.42\,M_\odot$.  As expected, all wind mass loss algorithm
combinations yield a higher mass loss rate for more massive (and thus
more luminous) stars. For stars with
$M_\mathrm{ZAMS}\lesssim20\,M_\odot$, the models using the vL
algorithm (dust driven mass loss in the cool evolutionary phase)
produce much higher mass loss than all other algorithms (see also
\Secref{sec:cool}): for example, the $15\,M_\odot$ model using the
V-vL combination with full efficiency $\eta=1.0$ results in a pre-SN
mass of only $5.25\,M_\odot$, while using the combination V-dJ or
V-NJ, we obtain a final mass of $\sim12.7\,M_\odot$. We discuss this
effect in more detail in \Secref{sec:cool}.

From \Figref{fig:mrel_mzams}, we also note that for
$M_\mathrm{ZAMS}\gtrsim20\,M_\odot$ we obtain a range $1.0\lesssim M
/M_\mathrm{ZAMS} \lesssim 0.5$. For any given $\eta$, the range
  is smaller. In any case, the size of this interval does not
decrease going to higher masses, indicating
that the different mass loss prescriptions and efficiencies do not
appear to be converging with $M_\mathrm{ZAMS}$ in the 
mass range we consider here.

\begin{figure}[!tb]
 \centering
 \resizebox{\hsize}{!}{\includegraphics{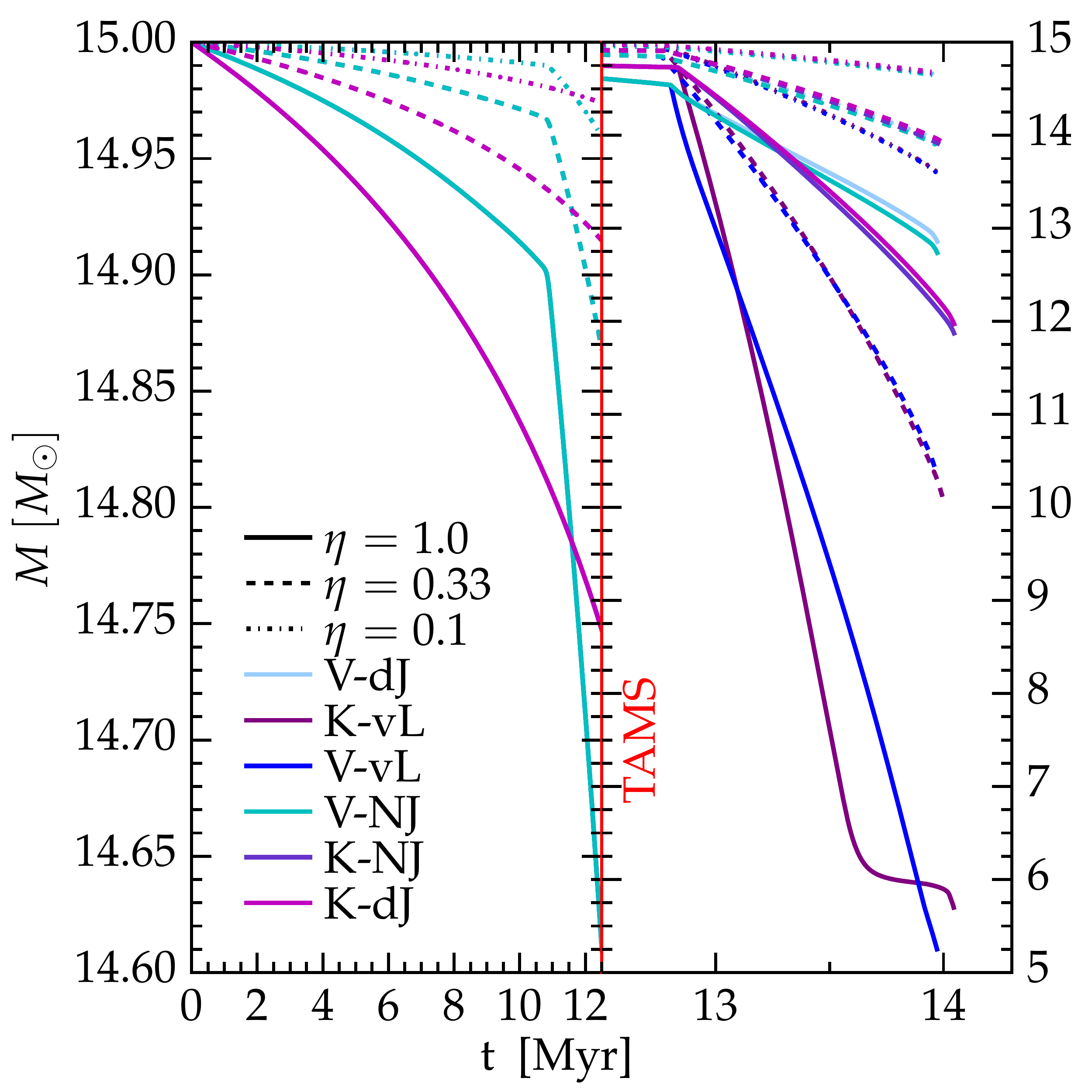}}
 \caption{Total stellar mass as a function of time for the $15M_\odot$
   models. Each color corresponds to a given combination of wind
   algorithms (see \Tabref{tab:comb}). None of these $15\,M_\odot$
   models reach the WR phase.  The solid, dashed, and dot-dashed
   curves correspond to efficiency $\eta = 0.1,\,0.33,\,1.0$,
   respectively, regardless of color. The red vertical line
   indicates roughly the terminal age main sequence (TAMS,
   i.e. $X_c<0.01$). The enhanced mass loss of models using
   the \cite{vink:00,vink:01} (V) algorithm close to TAMS in the
   left panel is due to the bistability jump,
   see \Secref{sec:hot}. Note that the left panel has a different
   vertical scale to magnify the differences in the tracks during the
   main sequence evolution. 
 \label{fig:m_t}}
\end{figure}

In \Figref{fig:m_t}, we show, as a representative example, the time
evolution of the total mass for our $15\,M_\odot$ models. The overall
qualitative behavior is the same for all the other considered
$M_\mathrm{ZAMS}$.  \Figref{fig:m_t} shows that the amount of mass
lost during the main sequence (i.e.\ prior to the vertical dot-dashed
line) is relatively small, only a few percent of the total
mass for $M_\mathrm{ZAMS}=15\,M_\odot$, even when using $\eta=1.0$. Most of the mass is lost after hydrogen
core burning.  Both the mass loss rate and the spread between the
predictions of different algorithms increases dramatically after the
end of the main sequence. Hence, uncertainties in the
post-main-sequence cool phase mass loss rates have
by far the greatest effect on the final mass of the star.


Using our nonrotating single-star models at the end of the mass loss
phase and ignoring any potential subsequent non-wind mass loss events, we can
attempt to classify their pre-SN color.  We follow \cite{georgy:12} by
assuming $\log_{10}(T_\mathrm{eff}/\mathrm{[K]})\lesssim3.6$ for RSG,
$3.6\lesssim\log_{10}(T_\mathrm{eff}/\mathrm{[K]})\lesssim3.8$ for
YSG, $\log_{10}(T_\mathrm{eff}/\mathrm{[K]})\gtrsim3.8$ for BSG, and
we only require the surface abundance of hydrogen to be $X_s<0.4$
without temperature thresholds for WR stars. We list the outcome of
this classification in \Tabref{tab:colors}.
Almost all of our $15\,M_\odot$
and $20\,M_\odot$ models end their evolution as
RSGs. The exceptions are those using the vL cool mass loss
algorithm and $\eta=1.0$, which end their lives as YSGs. All
of our $25\,M_\odot$ and $30\,M_\odot$ models
end as YSGs. The $35\,M_\odot$ models computed with $\eta=0.1$ also
end as YSGs. With $\eta=0.33$ they instead become BSGs, because of the
higher mass loss rate, except when using the vL cool mass loss rate
(cf.\ \Secref{sec:cool}), which produces YSGs. With $\eta=1.0$ we find
WR pre-SN models, unless vL is used during the cool phase, in which
case our $35\,M_\odot$ models would explode as BSGs.  However, these
results are highly dependent on the somewhat arbitrary temperature
thresholds assumed to divide the categories. For example, assuming
$\log_{10}(T_\mathrm{eff}/\mathrm{[K]})\leq3.68$ as the threshold
dividing RSG and YSG, all models
with $M_\mathrm{ZAMS}\leq25\,M_\odot$ would be RSGs.

\begin{figure}[!tbp]
 \centering \resizebox{\hsize}{!}{\includegraphics{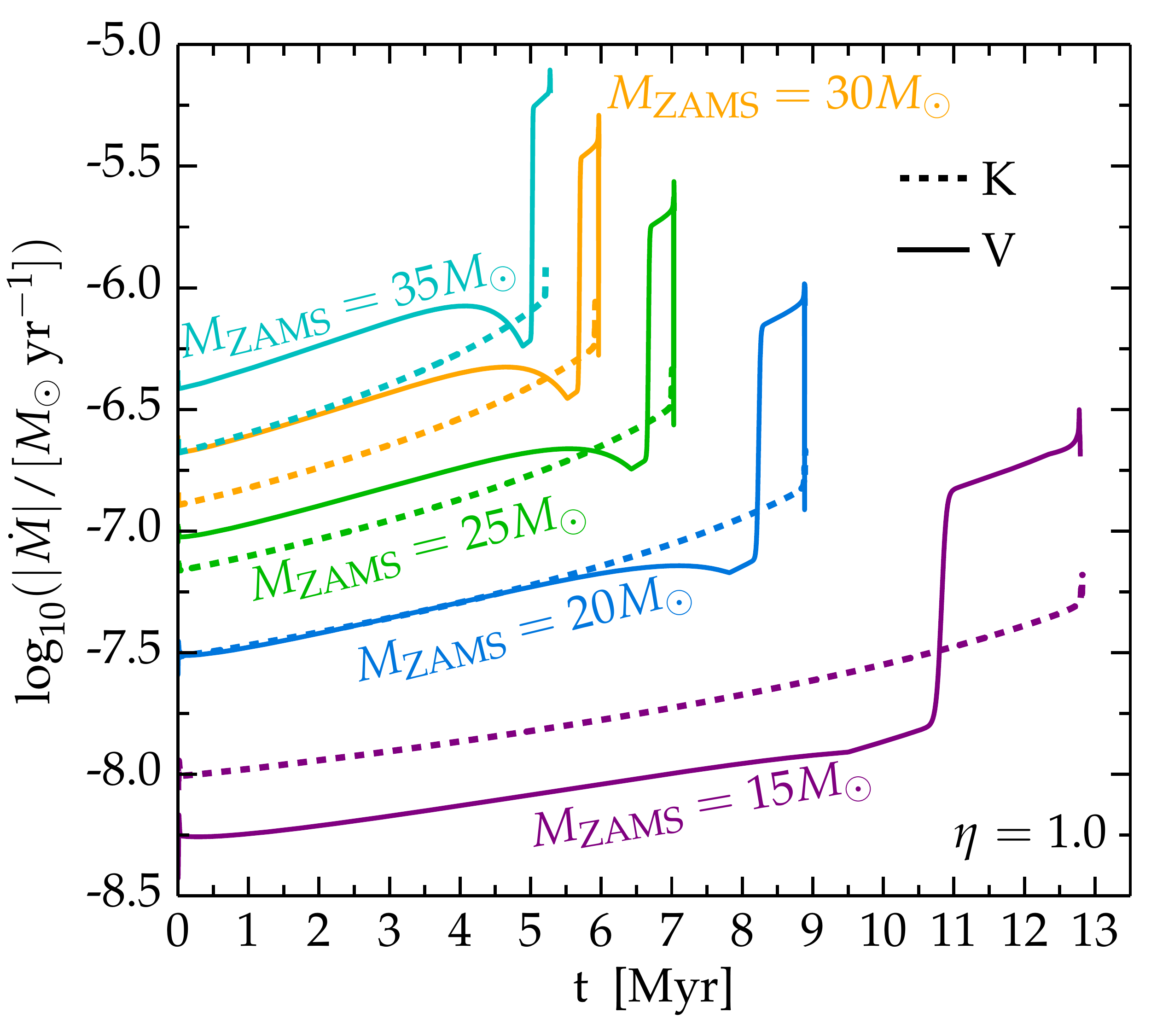}}
  \caption{Mass loss rate during the hot evolutionary phase (including
    the main sequence, see \Secref{sec:comb}) as a function of time
    for all computed
    $M_\mathrm{ZAMS}$ and wind efficiency
    $\eta = 1$. The solid and dashed curves are computed using the
    V \citep[][]{vink:00,vink:01} and K \citep[][]{kudritzki:89} mass
    loss algorithm,
    respectively. The rapid rise of the solid curves is due to the
    inclusion of the bistability jump (see
    \Secref{sec:hot}). The qualitative behavior of the
    curves shown does not change substantially when varying
    $\eta$. The curves end when
    $T_\mathrm{eff}=15\,000\,\mathrm{K}$.\label{fig:hot_mdot}}
\end{figure}

\subsection{The ``hot phase'' mass loss}\label{sec:hot}

\begin{table*}[!htbp] \setlength{\textfloatsep}{-2.5cm}
\setlength{\abovecaptionskip}{-0.4cm}
\setlength{\belowcaptionskip}{0.3cm}
  \begin{center}
    \caption{ Stellar properties at the end of
        hot phase, when $T_\mathrm{eff}$ decreases below
        $15\,000\,\mathrm{K}$
        for the first
        time. For the descriptions of the V and K
        algorithms see \Secref{sec:vink} and \Secref{sec:kudr},
        respectively. \label{tab:hot_end_res}}
    \begin{tabular}{ccc|cccccc}\hline\hline
\renewcommand{\arraystretch}{1.2} {Hot wind algorithm} & $M_\mathrm{ZAMS} \ [M_\odot]$ & $\eta$ & $R \ [R_\odot]$ & $L \ [10^4L_\odot]$ & $M \ [M_\odot]$ & $M_\mathrm{He} \ [M_\odot]$ & age [Myr] \\\hline\hline 
V & 15 & 1.0\phantom{0} & \phantom{0}34.85 & \phantom{0}5.44 & 14.54 & \phantom{0}3.69  & 12.7917 \\
K & 15 & 1.0\phantom{0} & \phantom{0}35.44 & \phantom{0}5.64 & 14.73 & \phantom{0}3.71  & 12.8239 \\
\hline                                                                                 
V & 15 & 0.33           & \phantom{0}35.91 & \phantom{0}5.76 & 14.85 & \phantom{0}3.76  & 12.7503 \\
K & 15 & 0.33           & \phantom{0}35.97 & \phantom{0}5.82 & 14.91 & \phantom{0}3.77  & 12.7612 \\
\hline                                                                                 
V & 15 & 0.1\phantom{0} & \phantom{0}36.27 & \phantom{0}5.85 & 14.95 & \phantom{0}3.78  & 12.7363 \\
K & 15 & 0.1\phantom{0} & \phantom{0}35.96 & \phantom{0}5.83 & 14.97 & \phantom{0}3.78  & 12.7359 \\
\hline \hline                                                                          
V & 20 & 1.0\phantom{0} & \phantom{0}51.97 & 12.20           & 19.06 & \phantom{0}5.83  & \phantom{0}8.8937 \\ 
K & 20 & 1.0\phantom{0} & \phantom{0}52.75 & 12.32           & 19.42 & \phantom{0}5.87  & \phantom{0}8.8947 \\
\hline                                                                                 
V & 20 & 0.33           & \phantom{0}53.10 & 12.67           & 19.69 & \phantom{0}5.95  & \phantom{0}8.8373 \\
K & 20 & 0.33           & \phantom{0}53.35 & 12.76           & 19.81 & \phantom{0}5.95  & \phantom{0}8.8374 \\
\hline                                                                                 
V & 20 & 0.1\phantom{0} & \phantom{0}53.46 & 12.80           & 19.91 & \phantom{0}5.99  & \phantom{0}8.8176 \\
K & 20 & 0.1\phantom{0} & \phantom{0}54.39 & 13.10           & 19.94 & \phantom{0}5.98  & \phantom{0}8.8183 \\
\hline \hline                                                                          
V & 25 & 1.0\phantom{0} & \phantom{0}67.58 & 20.63           & 23.24 & \phantom{0}8.03  & \phantom{0}7.0367 \\
K & 25 & 1.0\phantom{0} & \phantom{0}68.70 & 21.41           & 24.00 & \phantom{0}7.91  & \phantom{0}7.0094 \\
\hline                                                                                 
V & 25 & 0.33           & \phantom{0}70.59 & 22.55           & 24.44 & \phantom{0}7.97  & \phantom{0}6.9653 \\
K & 25 & 0.33           & \phantom{0}70.05 & 22.04           & 24.66 & \phantom{0}8.07  & \phantom{0}6.9546 \\
\hline                                                                                 
V & 25 & 0.1\phantom{0} & \phantom{0}70.88 & 22.39           & 24.84 & \phantom{0}8.10  & \phantom{0}6.9390 \\
K & 25 & 0.1\phantom{0} & \phantom{0}70.50 & 22.19           & 24.90 & \phantom{0}8.13  & \phantom{0}6.9355 \\
\hline \hline                                                                          
V & 30 & 1.0\phantom{0} & \phantom{0}80.82 & 29.52           & 27.00 &           10.18  & \phantom{0}5.9700 \\
K & 30 & 1.0\phantom{0} & \phantom{0}84.32 & 31.82           & 28.46 &           10.00  & \phantom{0}5.9215 \\
\hline                                                                                 
V & 30 & 0.33           & \phantom{0}84.64 & 32.11           & 29.10 &           10.16  & \phantom{0}5.8853 \\
K & 30 & 0.33           & \phantom{0}85.14 & 32.83           & 29.48 &           10.11  & \phantom{0}5.8689 \\
\hline                                                                                 
V & 30 & 0.1\phantom{0} & \phantom{0}85.58 & 32.85           & 29.77 & \phantom{0}9.96  & \phantom{0}5.8558 \\
K & 30 & 0.1\phantom{0} & \phantom{0}85.34 & 32.93           & 29.84 & \phantom{0}9.89  & \phantom{0}5.8503 \\
\hline \hline                                                                          
V & 35 & 1.0\phantom{0} & \phantom{0}95.27 & 40.28           & 30.32 &           12.29  & \phantom{0}5.2807 \\
K & 35 & 1.0\phantom{0} & \phantom{0}97.93 & 41.83           & 32.83 &           12.51  & \phantom{0}5.2164 \\
\hline                                                                                 
V & 35 & 0.33           & \phantom{0}98.15 & 43.52           & 33.62 &           12.19  & \phantom{0}5.1879 \\ 
K & 35 & 0.33           & \phantom{0}99.84 & 44.24           & 34.27 &           12.09  & \phantom{0}5.1661 \\
\hline                                                                                 
V & 35 & 0.1\phantom{0} & 100.81           & 45.59           & 34.63 &           11.58  & \phantom{0}5.1559 \\
K & 35 & 0.1\phantom{0} & 100.35           & 45.03           & 34.78 &           11.78  & \phantom{0}5.1484

\end{tabular}
\end{center}
\end{table*}

``Hot phase'' evolution,
i.e.\ $T_\mathrm{eff}\gtrsim15\,000\,\mathrm{K}$ if using the
\cite{kudritzki:89} (K) mass loss algorithm, or $T_\mathrm{eff}\gtrsim11\,000\,\mathrm{K}$ if using
\cite{vink:00,vink:01} (V), roughly covers the main sequence
evolution, the subsequent
overall contraction caused by hydrogen depletion (known as the main
sequence ``hook''), and the initial part of the Hertzsprung gap.

In \Tabref{tab:hot_end_res}, we summarize key
quantitative results for our models at the end of the hot phase.
While this is the longest phase of the evolution, covering $\gtrsim
90\%$ of the stellar lifetime, the amount of mass lost during this
phase is relatively small regardless of the algorithm used. For example, initially
$35\,M_\odot$ stars computed with $\eta=1.0$ lose $\sim15\%$ of their
mass, while initially $15\,M_\odot$ stars only lose a few percent

We can also infer from
\Tabref{tab:hot_end_res} that the amount of mass lost during the
hot phase is always higher with
V than with K, and the difference increases with increasing $\eta$ and
$M_\mathrm{ZAMS}$: it is only $\sim 0.02\,M_\odot$ for $\eta=0.1$ and
$M_\mathrm{ZAMS}=15\,M_\odot$, and grows to $\sim 0.2\,M_\odot$ for
$\eta=1.0$ and the same initial mass. For the $35\,M_\odot$ models and
$\eta=1.0$, the difference between the total mass shed using V or K
reaches $\sim 0.5\,M_\odot$ with $\eta=1.0$.

In \Figref{fig:hot_mdot}, we plot the mass loss rates $\dot{M}$ given
by V and K as a function of time in the hot phase for models with
$\eta = 1$. The reason for the higher total mass lost with V is that
this algorithm includes a detailed treatment of the bistability jump,
which is an increase in the cross section for photon interactions
caused by the recombination of ions driving the mass loss. This
enhancement of the cross section happens when the effective
temperature drops below
$T_\mathrm{eff}^\mathrm{jump}\simeq25\,000\,\mathrm{K}$,
\cite{vink:00}.  This is what causes the sudden tremendous increase in
the mass loss rate in V models seen in \Figref{fig:hot_mdot}. The
subsequent drop in models with masses higher than $20\,M_\odot$
happens because these cross the bistability jump region twice during
the contraction following core hydrogen
  depletion. Overall,
the average mass loss rate $\langle |\dot{M}_\mathrm{hot}|\rangle$ of
V models is driven up and surpasses that of K models on the main
sequence, resulting in higher mass loss in V models.  Note that the
use of two different thresholds to separate the hot and cool phase of
evolution for V and K does not significantly influence the duration of
the hot phase: the gap between the two thresholds is covered in a
fraction of the Herzsprung gap duration.

Different values of $\eta$ produce small ($\lesssim 2\%$) age
differences: models remaining more massive (i.e.\ computed with lower
$\eta$) evolve slightly faster. However, these differences are too
small to potentially be used as observational tests for the wind
efficiency.

Figure~\ref{fig:hot_mdot} also shows that at any point in time, the
mass loss rate of more massive stars is higher, because they produce a
higher photon flux to drive the wind. A factor of $\sim 2.3
\left(\simeq 35/15\right)$ in initial mass translates to a difference
of almost two orders of magnitude in the mass loss rate. The
difference between the V and K rates is a non-monotonic function of
the mass, because of the different functional dependencies of the two
algorithms: for $20\,M_\odot$ models they are roughly equal before the
surface cools enough for the bistability jump to occur; for
$15\,M_\odot$ models, the K algorithm gives an initially higher mass
loss rate, and, for $30\,M_\odot$ models, K mass loss is instead
lower.

Although only a small amount of mass is lost, the hot phase mass loss
can significantly influence the core evolution. This is because no
shell sources decouple the surface from the convective core during
most of this phase. The use of different algorithms during the hot
phase can thus create small (seed) differences in the core structure,
which may then be amplified by the subsequent evolution of the star
and contraction of the core. These differences are small at the end
of the hot phase, and we will discuss them at
later stages in the evolution in \Secref{sec:xi_O_depl} and
\ref{sec:onset_cc}.

In \Tabref{tab:hot_end_res}, we also list the helium core masses
$M_\mathrm{He}$ of our models at the end of the hot phase.  The
effects of hot phase wind mass loss on $M_\mathrm{He}$ are less
straightforward to interpret than its influence on the total mass.
First, the value of $M_\mathrm{He}$ depends on the definition of the
helium core. We define the outer edge of the helium core as the
first location going inward where $X(^1\mathrm{H})<0.01$. Second, where this
interface is at the end of the hot phase is very sensitive to mixing:
depending on the mass and metallicity of the star (and the
convective stability
criterion adopted), deep convective shells can develop at the
beginning of the Hertzsprung gap, and they shape the chemical
composition profile and determine $M_\mathrm{He}$.
\Tabref{tab:hot_end_res} shows that the maximum spread in
$M_\mathrm{He}$ increases with $M_\mathrm{ZAMS}$, starting from
$\max(\Delta M_\mathrm{He})\simeq 0.1\,M_\odot$ for $15\,M_\odot$
models, up to $\max(\Delta M_\mathrm{He})\simeq 0.9\,M_\odot$ for
$35\,M_\odot$ models. The spread in these values is almost entirely
due to variations in $\eta$. The difference in $M_\mathrm{He}$ between
models of same mass and $\eta$ (thus differing only in the use of V or
K) is also increasing with increasing $M_\mathrm{ZAMS}$ but remains
below $\sim 0.2\,M_\odot$. This trend directly reflects the larger
uncertainties in the modeling of winds from more massive stars.

\subsection{The ``cool phase'' mass loss}\label{sec:cool}

Regardless of efficiency $\eta$ and mass loss algorithm, most of
the mass loss through stellar winds happens during the cool phase of
the evolution. \Figref{fig:m_t} demonstrates this clearly for the
$15\,M_\odot$ models, which lose $2\%-60\%$ of their total initial
mass during this phase, depending on the algorithm combination and
wind efficiency. The increase in the mass loss rate from the hot phase
can be understood in terms of the effective gravity of the star
(although we stress that the algorithms compared here do not depend
explicitly on it): for any given luminosity of a massive star, if the
stellar surface is cool, necessarily its radius must be large, and
thus it will be easier for matter to leave the gravitational potential
well of the star. Moreover, at lower temperature the opacity tends to
be higher because of recombination of ions and possibly dust
formation, thus enhancing the wind driving.

Table~\ref{tab:res} summarizes key quantitative results of our models
at the end of the mass loss phase, including the total mass and the
core masses. One striking result is that the dust-driven
\cite{vanloon:05} (vL) mass loss algorithm results in significantly
different total masses and core masses than the \cite{dejager:88} (dJ)
and the \cite{nieuwenhuijzen:90} (NJ) algorithms. This can also be
seen in \Figref{fig:mrel_mzams}, where wind combinations using vL for
the cool phase produce different vertical spreads. These differences
are strongest for $\eta=1.0$.  The most extreme example are
$15\,M_\odot$ models computed with $\eta=1.0$: regardless of the hot
phase mass loss algorithm, they end their evolution with masses of
$\sim5-6\,M_\odot$ with vL in the cool phase, while they remain
as massive as $\sim11-12\,M_\odot$ with dJ or NJ. We find
the opposite trend at the upper end of our mass range: a
$30\,M_\odot$ star computed with $\eta=1.0$ and V during the hot phase
reaches the end of the mass loss phase with a total mass of $\sim
15M_\odot$ if using either dJ or NJ in the cool phase, while it ends
its life with $\sim 18\,M_\odot$ with vL.

On the one hand, the similarities between the dJ and the NJ rates are
expected \citep[see also][]{mauron:11,eldridge:04}: both are
semi-empirical rates derived from the same sample of observed
stars. They differ only in the choice of the stellar variables used to
parametrize $\dot{M}$. On the other hand, the vL algorithm is also
semi-empirical, but based on the analysis of a different sample of
stars assuming a dust-driven model of the wind, i.e.\ wind mass loss
is not driven by photons impinging on metallic ions, but rather on
dust particles. If a dust-driven (instead of line-driven) model of the
wind is assumed, the resulting mass loss rate is generally higher, and
much more $T_\mathrm{eff}$-dependent (see Tabs.~\ref{tab:scalingLT}
and \ref{tab:scaling}).

The very steep dependence of the vL rate on $T_\mathrm{eff}$ causes
the different evolution of models above and below $\sim
25\,M_\odot$. During the early stage of the cool phase, the vL rate is
always much higher than the others, and the stellar wind described by
this algorithm reveals the deeper and hotter layers of the star (see
also \Tabref{tab:colors}).  As $T_\mathrm{eff}$ increases, the vL mass
loss rate decreases rapidly ($\propto T^{-6.3}$, see
Tabs.~\ref{tab:scalingLT} and \ref{tab:scaling}), which is attributed
to the temperature sensitivity of the microscopic dust formation
processes \citep{wachter:02} that we of course do not track explicitly
in our calculations. Also, for any mass loss process at a given
luminosity, higher $T_\mathrm{eff}$ correspond to smaller radii, and
thus higher effective gravity at the stellar surface. For
$M_\mathrm{ZAMS}\gtrsim25\,M_\odot$, the steep
$T_\mathrm{eff}$-dependence leads to a self regulation of the vL rate.
For lower initial masses, the vL rate is also initially higher than
the dJ or NJ rate, but not high enough to reach the self-regulating
regime: the vL rate remains higher than the dJ and NJ rates for the
whole evolution and produces pre-SN structures of a much lower final
mass than when dJ or NJ are used. This is summarized in
\Tabref{tab:res}.

The comparison of $M_\mathrm{He}$ listed in \Tabref{tab:res} for
models with the same $\eta$ and hot phase mass loss (either V or K)
reveals that the effect of the cool phase mass loss on the helium core
mass is very small and almost negligible. We
find the only appreciable differences when using $\eta=1.0$ and the vL
cool mass loss rate, and they are only of order $\sim0.01\,M_\odot$.

\subsection{Models reaching the WR stage}
\label{sec:WRs}


Out of the 94 models computed to $T_c\geq10^9\,\mathrm{K}$, only 8
reach the conditions to switch to the WR wind scheme. These are all
$35\,M_\odot$ models computed with $\eta=1.0$, and none use the vL
algorithm in the cool phase that precedes the
WR phase (cf.~\Tabref{tab:colors}). The lack of WR models using vL is
explained by the self-damping of this mass loss scheme for more
massive and thus more luminous stars (see \Secref{sec:cool}).

The typical duration of the WR phase is
$\sim0.02-0.05\,\mathrm{Myr}$. The
differences in duration of the WR phase with the \cite{nugis:00} (NL)
and \cite{hamann:82,hamann:95,hamann:98} (H) algorithms are
negligible. 
However, models computed with the dJ algorithm in the cool
phase have WR phases that are systematically longer by a few ten
thousand years than the corresponding models computed with the NJ
algorithm. 
Moreover, the
duration of the WR phase of models computed with the K algorithm in
the hot phase is about a factor of $\sim2.3$ longer than in models
using the V algorithm. This is because $M_\mathrm{ZAMS} = 35\,M_\odot$
models computed with $\eta=1.0$ and the V algorithm reach the end of
the hot phase with a helium core that is $0.22\,M_\odot$ less massive
than models computed with the K algorithm
(cf.~\Tabref{tab:hot_end_res}). Therefore, the subsequent evolution is
slowed down, and the WR phase is reached slightly later.

The NL WR phase algorithm produces higher final masses than the H
algorithm: the difference is $\sim 0.3$ $(\sim 0.8)\,M_\odot$ for models
using the V (K) hot phase algorithm and does not depend strongly on
cool phase mass loss (cf.~\Tabref{tab:res}). Note that these
differences are very small fractions of the initial mass
$M_\mathrm{ZAMS}=35\,M_\odot$ of these models
(cf.~\Figref{fig:mrel_mzams}).

The WR wind does not have a strong influence on $M_\mathrm{He}$ at the
end of the mass loss phase. For stars with
$M_\mathrm{ZAMS}\lesssim40M_\odot$, the He core mass is determined
well before the beginning of the WR phase, and the wind mass loss is
not strong enough to dig into the He core directly. More massive stars
with stronger winds, may become hydrogen depleted already during the
core hydrogen burning phase.  In that case, the WR mass loss rate
might have an impact on $M_\mathrm{He}$ through the quasi-static
response of the convective core to mass loss
\citep[][]{meynet:94,dekoter:97,crowther:10,bestenlehner:11}.  Also, in even
more massive stars, the entire hydrogen-rich envelope can be lost to
winds, making $M_\mathrm{He}$ the total mass of the star, and winds
can then further reduce it \citep[e.g.][]{woosley:16b}.

While we find no systematic effect of WR mass loss on $M_\mathrm{He}$,
the situation is different for $M_\mathrm{CO}$ (cf.~\Tabref{tab:res}).
NL models yield $M_\mathrm{CO}$ that are systematically lower by $\sim
0.05-0.1\,M_\odot$ than H models and the largest differences are
between models that also use different hot phase mass loss
algorithms. We find that models differing only in the WR algorithm
have lower $M_\mathrm{CO}$ for higher final masses. For example, the
V-dJ-NL model has $M_\mathrm{CO}=9.81\,M_\odot$ and final mass
$M=20.03\,M_\odot$. In contrast, the V-dJ-H model has a higher
$M_\mathrm{CO}=9.93\,M_\odot$, but a lower final mass of
$M=19.73\,M_\odot$. However, we note that the
trend that lower final masses correspond to higher $M_\mathrm{CO}$
holds for most of our $35\,M_\odot$ models, independent of if they
become WR stars or not.

The differences in both total mass and $M_\mathrm{CO}$ found varying the
WR mass loss algorithm are more sensitive to the previously employed
hot phase mass loss algorithm than to the cool phase
algorithm. Moreover, $M_\mathrm{He}$ is also almost insensitive to the
cool phase (cf.~\Secref{sec:cool}) and WR phase mass loss. Therefore,
the differences found here are most likely related to the differences
in $M_\mathrm{He}$ at the end of
the hot phase (and in the total mass for a given $M_\mathrm{He}$), and
consequently the position of the hydrogen burning shell, which
indirectly influences the helium burning, the mixing processes shaping
the composition profile, and ultimately the resulting $M_\mathrm{CO}$
and the amount of mass lost during the WR phase.

\subsection{Models at oxygen depletion}
\label{sec:O_depl_res}

\begin{table*}[!htp]
    \centering
    \caption{Stellar properties at core
      oxygen depletion, i.e.\ $X_c(\mathrm{^{16}O})<0.04$ and
      $X_c(\mathrm{^{28}Si})>0.01$.  The last column shows the maximum
      difference in the compactness parameter
      $\xi_{2.5}^\mathrm{O\ depl}$ for each choice
      of $\eta$ and
        $M_\mathrm{ZAMS}$.  These runs are re-started from the
      corresponding MESA models saved at the end of the mass loss
      phase, when $T_c\geq10^9\,\mathrm{K}$.
      \label{tab:runs_to_O_depl}} 
    \begin{tabular}{c|cl|cccccc}\hline \hline
      \multicolumn{8}{c}{Oxygen depletion: $X_c(\mathrm{^{16}O})<0.04$ and $X_c(\mathrm{^{28}Si})>0.01$}\\\hline\hline
      $M_\mathrm{ZAMS} \ [M_\odot]$ & $\eta$ &  \multicolumn{1}{c|}{ID}  & $R \ [R_\odot]$ &
      $M_\mathrm{tot} \ [M_\odot]$ & $M_\mathrm{He} \ [M_\odot]$ &
      $M_\mathrm{CO} \ [M_\odot]$ & $\xi_{2.5}^\mathrm{O\ depl}$ & $\max \Delta
        \xi_{2.5}^\mathrm{O\ depl} $\\\hline\hline
      \multirow{9}{*}{15} & \multirow{3}{*}{0.1\phantom{3}} & \phantom{A}V-dJ & \phantom{0}908 & 14.66 & \phantom{0}4.99 & 3.17 & 0.155  & \multirow{3}{*}{0.001}\\ 
                          &                                 & \phantom{A}V-vL & \phantom{0}924 & 13.60 & \phantom{0}4.98 & 3.17 & 0.156  &                        \\ 
                          &                                 & \phantom{A}K-NJ & \phantom{0}911 & 14.66 & \phantom{0}5.00 & 3.17 & 0.156  &                        \\ 
      \cline{2-9}                               
                          & \multirow{2}{*}{0.33}           & \phantom{A}V-NJ & \phantom{0}914 & 13.90 & \phantom{0}4.93 & 3.13 & 0.153  & \multirow{2}{*}{0.001}\\ 
                          &                                 & \phantom{A}V-vL & \phantom{0}967 & 10.39 & \phantom{0}4.93 & 3.13 & 0.152  &                       \\ 
      \cline{2-9}                                                              
                          & \multirow{4}{*}{1.0\phantom{3}} & \phantom{A}V-NJ & \phantom{0}895 & 12.74 & \phantom{0}4.65 & 2.95 & 0.141  & \multirow{4}{*}{0.014} \\ 
                          &                                 & \phantom{A}V-vL & \phantom{0}629 & \phantom{0}5.25 & \phantom{0}4.64 & 2.94 & 0.139 &\\ 
                          &                                 & \phantom{A}K-NJ & \phantom{0}946 & 11.87 & \phantom{0}4.92 & 3.12 & 0.153 & \\ 
                          &                                 & \phantom{A}K-vL & \phantom{0}643 & \phantom{0}5.70 & \phantom{0}4.90 & 3.11 & 0.152 &\\ 
      \hline\hline       
      \multirow{12}{*}{20}& \multirow{4}{*}{0.1\phantom{3}} & \phantom{A}V-dJ & \phantom{0}994 & 19.23 & \phantom{0}7.04 & 4.83 & 0.182& \multirow{4}{*}{0.007}\\ 
                          &                                 & \phantom{A}V-vL & \phantom{0}998 & 18.10 & \phantom{0}7.03 & 4.83 & 0.178&\\ 
                          &                                 & \phantom{A}K-NJ & \phantom{0}991 & 19.37 & \phantom{0}7.01 & 4.82 & 0.179&\\ 
                          &                                 & \phantom{A}K-vL & \phantom{0}991 & 18.67 & \phantom{0}7.01 & 4.82 & 0.175&\\ 
                          \cline{2-9}                               
                          & \multirow{4}{*}{0.33}           & \phantom{A}V-NJ &   1001 & 17.48 & \phantom{0}7.01 & 4.81 & 0.161& \multirow{4}{*}{0.052}\\ 
                          &                                 & \phantom{A}V-vL & \phantom{0}987 & 13.47 & \phantom{0}6.99 & 4.80 & 0.175&\\ 
                          &                                 & \phantom{A}K-dJ & \phantom{0}999 & 17.62 & \phantom{0}7.00 & 4.80 & 0.213&\\ 
                          &                                 & \phantom{A}K-vL & \phantom{0}993 & 13.90 & \phantom{0}6.98 & 4.80 & 0.175&\\ 
                          \cline{2-9}                                  
                          & \multirow{4}{*}{1.0\phantom{3}}  & \phantom{A}V-dJ & \phantom{0}951 & 11.81 & \phantom{0}7.06 & 4.85 & 0.182& \multirow{4}{*}{0.010}\\ 
                          &                                 & \phantom{A}V-vL & \phantom{0}673 & \phantom{0}8.80 & \phantom{0}7.02 & 4.82 & 0.176&\\ 
                          &                                 & \phantom{A}K-NJ & \phantom{0}978 & 12.77 & \phantom{0}6.95 & 4.77 & 0.176&\\ 
                          &                                 & \phantom{A}K-vL & \phantom{0}712 & \phantom{0}8.81 & \phantom{0}6.90 & 4.73 & 0.172&\\ 
      \hline\hline                                 
      \multirow{11}{*}{25}& \multirow{4}{*}{0.1\phantom{3}} & \phantom{A}V-dJ & \phantom{0}898 & 23.85 & \phantom{0}9.14 & 6.46 & 0.180& \multirow{4}{*}{0.023}\\ 
                          &                                 & \phantom{A}V-vL & \phantom{0}899 & 23.56 & \phantom{0}9.14 & 6.44 & 0.179&\\ 
                          &                                 & \phantom{A}K-NJ & \phantom{0}888 & 23.76 & \phantom{0}9.24 & 6.52 & 0.159&\\ 
                          &                                 & \phantom{A}K-vL & \phantom{0}889 & 23.24 & \phantom{0}9.23 & 6.50 & 0.157&\\ 
                          \cline{2-9}                                                 
                          & \multirow{3}{*}{0.33}           & \phantom{A}V-NJ & \phantom{0}922 & 22.02 & \phantom{0}8.87 & 6.25 & 0.161& \multirow{3}{*}{0.049} \\ 
                          &                                 & \phantom{A}V-vL & \phantom{0}929 & 22.13 & \phantom{0}8.87 & 6.20 & 0.164&\\ 
                          &                                 & \phantom{A}K-dJ & \phantom{0}894 & 21.27 & \phantom{0}9.11 & 6.42 & 0.210&\\ 
                          \cline{2-9}                               
                          & \multirow{4}{*}{1.0\phantom{3}} & \phantom{A}V-NJ & \phantom{0}812 & 13.67 & \phantom{0}9.05 & 6.32 & 0.164& \multirow{4}{*}{0.050} \\ 
                          &                                 & \phantom{A}V-vL & \phantom{0}786 & 12.97 & \phantom{0}9.04 & 6.29 & 0.211&\\ 
                          &                                 & \phantom{A}K-dJ & \phantom{0}875 & 15.57 & \phantom{0}8.89 & 6.21 & 0.161&\\ 
                          &                                 & \phantom{A}K-vL & \phantom{0}860 & 14.69 & \phantom{0}8.90 & 6.23 & 0.200&\\ 
                          \hline\hline       
      \multirow{11}{*}{30}& \multirow{4}{*}{0.1\phantom{3}} & \phantom{A}V-dJ& \phantom{0}701 & 28.11 & 10.97 & 7.89 & 0.242& \multirow{4}{*}{0.001}\\ 
                          &                                 & \phantom{A}V-NJ& \phantom{0}697 & 28.17 & 10.97 & 7.90 & 0.243&\\ 
                          &                                 & \phantom{A}K-NJ& \phantom{0}706 & 28.19 & 10.91 & 7.83 & 0.242&\\ 
                          &                                 & \phantom{A}K-vL& \phantom{0}719 & 28.75 & 10.90 & 7.74 & 0.242&\\ 
                          \cline{2-9}                                            
                          & \multirow{4}{*}{0.33}           & \phantom{A}V-dJ& \phantom{0}668 & 24.16 & 11.11 & 7.99 & 0.243& \multirow{4}{*}{0.023} \\ 
                          &                                 & \phantom{A}V-vL& \phantom{0}723 & 26.02 & 11.10 & 8.00 & 0.243& \\ 
                          &                                 & \phantom{A}K-NJ& \phantom{0}716 & 25.18 & 10.89 & 7.85 & 0.228& \\ 
                          &                                 & \phantom{A}K-vL& \phantom{0}709 & 27.12 & 10.87 & 7.79 & 0.220& \\ 
                          \cline{2-9}                               
                          & \multirow{3}{*}{1.0\phantom{3}}  & \phantom{A}V-dJ& \phantom{0}610 & 15.36 & 11.28 & 8.20 
                                                                                                & 0.179 & \multirow{3}{*}{0.065}           \\ 
                          &                                 & \phantom{A}K-dJ& \phantom{0}756 & 18.51 & 10.92 & 7.85 & 0.244&\\ 
                          &                                 & \phantom{A}K-vL& \phantom{0}723 & 22.53 & 10.89 & 7.74 & 0.243&\\ 
    \end{tabular}
  \end{table*}

To reduce the computational cost of our model grid, we select a subset
of 44 stars to continue until oxygen depletion, which we define as the
time when $X_c(^{16}\mathrm{O})\leq 0.04$ and
$X_c(^{28}\mathrm{Si})\geq0.01$. These models span the range of
properties found at the end of the mass loss phase, and are listed in
\Tabref{tab:runs_to_O_depl} together with their properties at oxygen
depletion.  This selection allows us to avoid running multiple models
that have very similar evolutionary paths from the end of the mass
loss phase onward. The duration of the evolution between the end of
the mass loss phase ($T_c\geq10^9\,\mathrm{K}$) and oxygen depletion
is of order years to decades, depending on the total mass and core
masses at the end of the mass loss phase.

In the very short time to oxygen depletion,
neither total mass nor helium core mass change appreciably. This can
be inferred by comparing the entries of \Tabref{tab:res} and
\Tabref{tab:runs_to_O_depl}, which also reveals that the stellar
radii vary only within $\pm3\,R_\odot$ in most models. 

The CO core masses at oxygen depletion summarized in
\Tabref{tab:runs_to_O_depl} are systematically a few percent lower
than those listed in \Tabref{tab:res} at the end of the mass loss
phase. This seems counter-intuitive, since one would expect the CO
core to grow in mass because of the ashes of helium shell
burning. However, the boundary location for the CO core is determined
by convective mixing within and above the He burning shell, which
brings helium-rich material inward and moves the CO core boundary to a
smaller mass coordinate. \emph{This implies that the core mass is very
  sensitive to the mixing parameters.} The maximum spread in
$M_\mathrm{CO}$ obtained varying the wind algorithm is of order $\sim
0.1\,M_\odot$ and it increases with $M_\mathrm{ZAMS}$ up to about
$\max(\Delta M_\mathrm{CO})\simeq0.5\,M_\odot$ for $30\,M_\odot$
models (see also \Tabref{tab:max_spreads}). Note that the
$15\,M_\odot$ models are outliers in that they show a larger spread of
CO core masses of up to $0.28\,M_\odot$ between models using the V and
K hot mass loss algorithm. This is a consequence of a combination of
(\emph{i}) the V algorithm leading to a lower total mass and a lower
He core mass at the end of the hot phase and (\emph{ii}) the
relatively low mass loss during the cool phase (compared to more
massive stars), which results in deeper convective episodes. Together,
these lead to small CO cores: the two $15 M_\odot$ models using the V
algorithm with $\eta = 1.0$ have $M_\mathrm{CO} \simeq 2.95 M_\odot$
at the end of the mass loss phase, which is about $0.2\,M_\odot$
smaller than the average for $15\,M_\odot$ models.

\subsection{The Compactness Parameter
  $\xi_{\mathcal{M}}$}
\label{sec:xi_O_depl}

Although the internal structure of our models is not yet final at
oxygen depletion, some quantities (e.g., core masses) are already
close to their final values, and it becomes possible to draw first
connections between internal structure and the potential final outcome
of core collapse \citep[][]{sukhbold:14}. For this, we include the
compactness parameter $\xi_{2.5}^\mathrm{O\,depl}$ in
\Tabref{tab:runs_to_O_depl}. \cite{oconnor:11} define the compactness
parameter as
\begin{equation}
  \label{eq:xi_def}
  \xi_{\mathcal{M}}\udef \frac{\mathcal{M} /
    M_\odot}{R(\mathcal{M})/1000\ \mathrm{km}} \ \ .
\end{equation}
This parameter provides a single measure for the complex inner core
structure (mass coordinate smaller than $\mathcal{M}$) of a star,
allowing for a simplified discussion of the differences
in the internal structure produced by the various
mass loss algorithms we compare.

We set $\mathcal{M}=2.5\,M_\odot$ because this is the typical mass above which the proto-NS that will
form during core collapse will become a BH
\citep[][]{oconnor:11}. This mass cut remains well outside of the
typical iron core mass and includes the layers of the star that the
shock will encounter after core bounce. These layers determine the
accretion ram pressure that the shock has to overcome for a successful
explosion. The qualitative behavior of the supernova dynamics and
outcome with $\xi_\mathcal{M}$ is known to be robust against different
choices of $\mathcal{M}$
\citep[][]{oconnor:11,oconnor:13,ugliano:12,sukhbold:14}.

One-dimensional parametric core-collapse SN explosion simulations
\citep[][]{oconnor:11,oconnor:13,ugliano:12,ertl:16,sukhbold:16} show
that the value of $\xi_{2.5}$ at the onset of core collapse indicates
the most probable remnant.  High values of $\xi_{2.5}$ indicate a more
compact pre-SN structure that is harder to explode and that will more
likely result in a BH remnant. Conversely, low values of $\xi_{2.5}$
indicate a steeper density gradient and an easier to explode
structure, suggesting that the remnant will more likely be a NS
\citep[][]{oconnor:11, ugliano:12, clausen:15}.  \cite{sukhbold:14}
suggest that the value of the compactness parameter evaluated at
oxygen depletion, $\xi_{2.5}^\mathrm{O\ depl}$, can already be used to
infer the most likely outcome of the core collapse event. The
evolution from oxygen depletion to core collapse tends to increase the
compactness and amplify the differences between different stellar
models, but the key features that determine the interpretation of
$\xi_{2.5}$ appear to be set already at oxygen depletion.  Other
parameters to relate the pre-SN structure to the most-likely remnant
can be defined in the context of neutrino-driven explosions
\citep[see][]{ertl:16}, but they rely on physical quantities that are
not at all set at earlier stages of the evolution (e.g., the entropy
profile throughout the silicon layer and iron core), and therefore
they are not useful diagnostics before the onset of core collapse.

\subsubsection{Evolution of $\xi_{2.5}$ until oxygen depletion}

\begin{figure}[!ht]
  \centering
  \resizebox{\hsize}{!}{\includegraphics{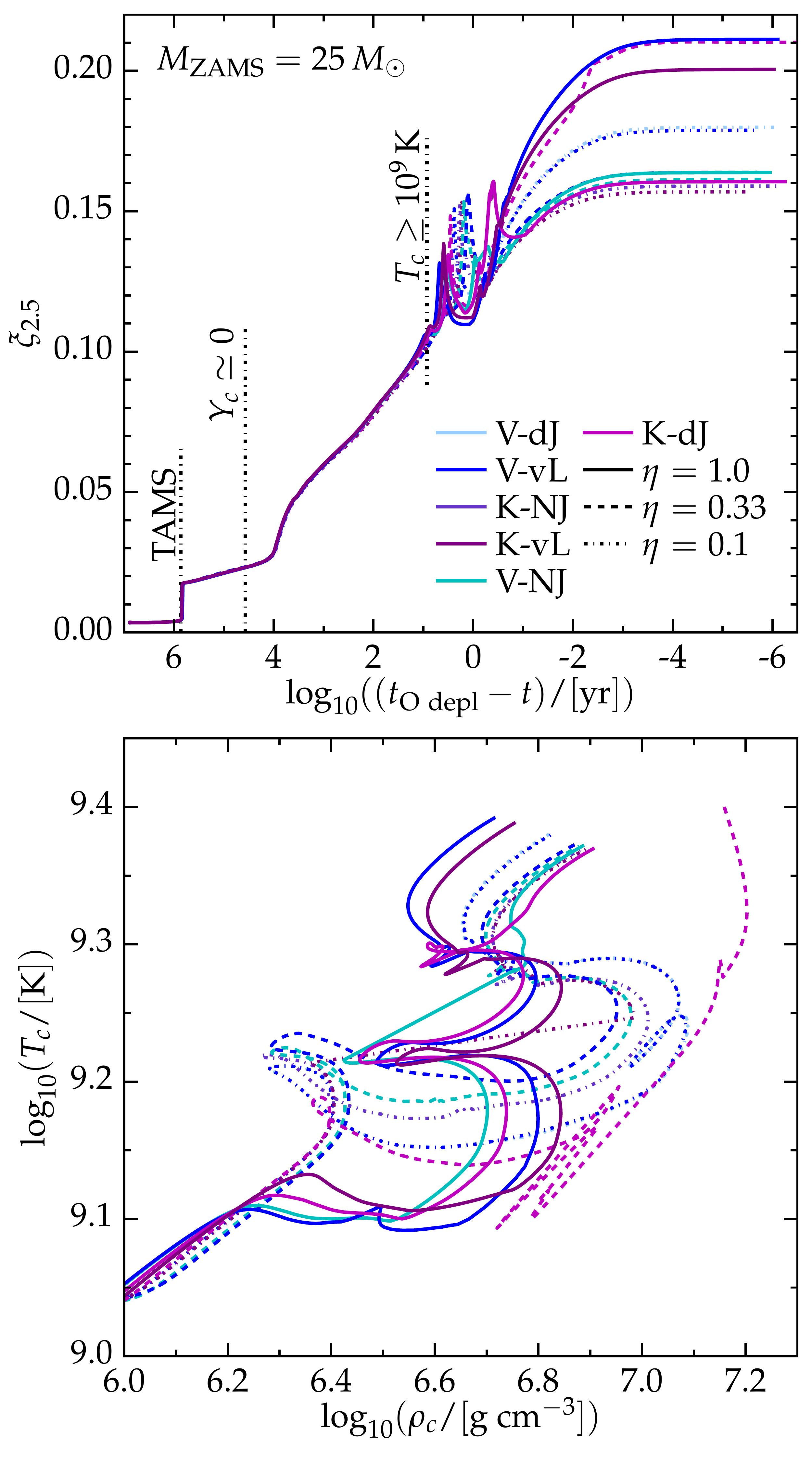}}
  \caption{Top panel: Time evolution of the compactness parameter
    $\xi_{2.5}$ for all $M_\mathrm{ZAMS}=25\,M_\odot$ models computed
    until oxygen depletion. Bottom panel: central temperature and
    density for the same models from when the central density
    increases above $10^6\,\mathrm{g\ cm^{-3}}$. All curves end at
    oxygen depletion ($X_c(^{16}\mathrm{O})\leq 0.04$ and
    $X_c(^{28}\mathrm{Si})\geq0.01$). The color indicates the wind
    algorithm combination, labeled according to
    \Tabref{tab:comb}. Dot-dashed, dashed, and solid curves are
    calculated using $\eta=0.1,\,0.33,\,1.0$, respectively. Note that
    we only show models listed in \Tabref{tab:runs_to_O_depl}.  The
    vertical dot-dashed lines in the top panel approximately indicate
    core hydrogen exhaustion (TAMS), core helium exhaustion
    ($Y_c\simeq0$), and end of the mass loss phase
    ($T_c\geq10^9\,\mathrm{K}$).  \label{fig:xi_t_25}}
\end{figure}

The compactness parameter is a function of time,
$\xi_{\mathcal{M}}\equiv\xi_{\mathcal{M}}(t)$, because of the changes
in the radius of a given mass coordinate $\mathcal{M}$. These can be
caused by contraction of the core, onset of partial electron
degeneracy within the mass coordinate $\mathcal{M}$, and by episodes
of convective mixing and shell burning \citep[][]{sukhbold:14}. The
top panel of \Figref{fig:xi_t_25} shows examples of the evolution of
$\xi_{2.5}$ until oxygen depletion in our $25\,M_\odot$ models. Note that
we use a reversed logarithmic scale on the x axis to
emphasize the late evolutionary stages.
 
Figure~\ref{fig:xi_t_25} shows that the
compactness parameter is constant during
the main sequence evolution. During this phase, it is also almost independent of
$M_\mathrm{ZAMS}$ and mass loss algorithm because all
stars considered here have convective main-sequence cores that are
always much larger than the mass coordinate at which we evaluate the compactness. After core
hydrogen exhaustion, $\xi_{2.5}$ increases because of the overall
contraction, reaching $\xi_{2.5}\simeq 0.02$ in our $25\,M_\odot$
models. Then it slowly continues to increase during the hydrogen-shell
burning and helium core burning phases. The increase speeds up
significantly during core carbon burning, reaching values of
$\xi_{2.5}\simeq 0.1$ in our $25\,M_\odot$ models. Neon core burning
ignition and the onset of carbon shell burning mark a critical point
in the evolution of $\xi_{2.5}$ at which the various curves in
Fig.~\ref{fig:xi_t_25} begin to diverge. \cite{sukhbold:14} find the
same and point out that the subsequent evolution of $\xi_{2.5}$ is
highly sensitive to the details of carbon shell burning (i.e. the
number, locations, and durations of shell burning episodes). It is
important to note from Fig.~\ref{fig:xi_t_25} that the effects of mass
loss on core structure (represented by $\xi_{2.5}$) are delayed: At
the time mass loss ends in our models (at $T_c > 10^9\,\mathrm{K}$),
differences in $\xi_{2.5}$ are minute. These seed differences grow and
become substantial only in the last decade before core collapse.

The bottom panel of \Figref{fig:xi_t_25} depicts central
density--temperature tracks for our $25\,M_\odot$ models that are
evolved to oxygen depletion. The tracks start roughly at neon core
ignition and show that the mass-loss history
(i.e.\ the choice of wind mass loss algorithm
combination) also influences the innermost core thermodynamics and
structure.  This is because the nuclear burning processes in the core
are regulated by the amount of mass that needs to be sustained by the
core itself, i.e.\ the mass below the innermost burning shell above
the core. This in turn depends on the location and luminosity of the
shell burning regions and therefore on the total mass of the star.

\begin{figure}[!tbp]
  \centering
  \resizebox{\hsize}{!}{\includegraphics{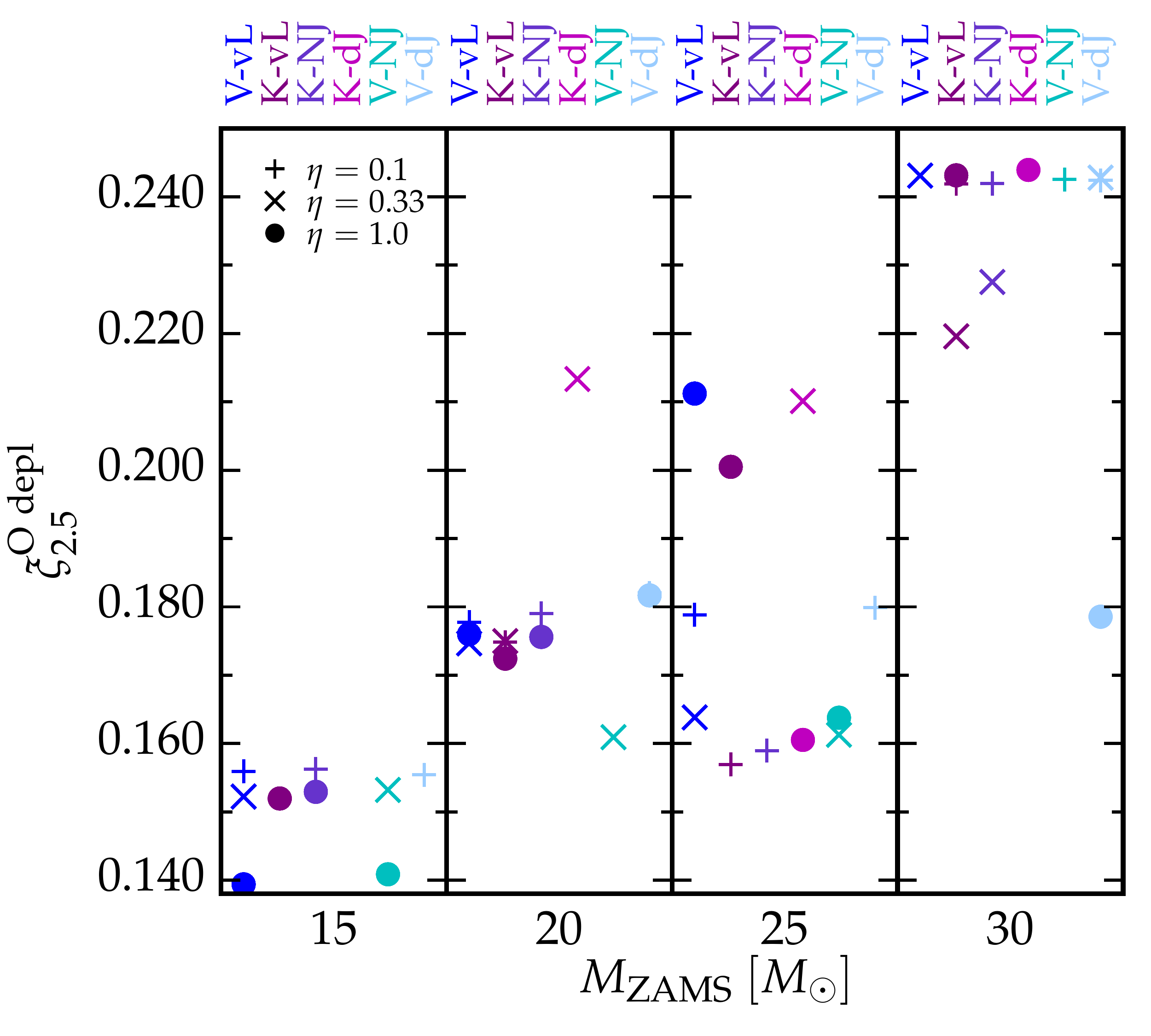}}
  \caption{Compactness parameter at oxygen depletion
    $X_c(^{16}\mathrm{O})\leq 0.04$ and $X_c(^{28}\mathrm{Si})\leq 0.01$. Each color corresponds to a wind
    algorithm combination (top axis), and
    each panel shows a different initial mass,
      indicated on the bottom. Crosses, pluses, and
      circles correspond to $\eta=0.1,\,0.33,\,1.0$, respectively. Note
      that we show only the models listed in
      \Tabref{tab:runs_to_O_depl}. The vertical spread indicates
      different core structures, which will evolve differently to the
      onset of core collapse, and possibly result in different SN
      outcomes.\label{fig:xi_O_depl}
      \vspace*{-0.3cm}
}
\end{figure}

In \Figref{fig:xi_O_depl}, we show the values of
$\xi_{2.5}^\mathrm{O\ depl}$ for all models that we run to oxygen
depletion. The spread in each panel is due to the different
algorithmic treatments of wind mass loss: for example, $25\,M_\odot$
models show values ranging between 0.210 and 0.157. Generally
speaking, the spread in $\xi_{2.5}^\mathrm{O\ depl}$ increases with
increasing $\eta$ and $M_\mathrm{ZAMS}$, i.e.\ the stronger the
stellar wind, the more it influences the core structure.  We emphasize
that a few percent variation of $\xi_{2.5}^\mathrm{O\ depl}$ can
result in important differences in the core structure at the onset of
core collapse: the subsequent contraction of the core and the details
of carbon, oxygen, and silicon shell burning, amplify the differences
between models that are still relatively similar at oxygen depletion
\citep[see \Secref{sec:onset_cc} and][]{sukhbold:14}.

\subsubsection{Effects of the wind efficiency on $\xi_{2.5}^\mathrm{O\
    depl}$} 
The effects of varying $\eta$ on $\xi_{2.5}^\mathrm{O \ depl}$ can be
inferred from the comparison of models in
  \Tabref{tab:runs_to_O_depl} of the same $M_\mathrm{ZAMS}$ using the
same wind algorithm combinations but different efficiencies. The
variations of $\xi_{2.5}^\mathrm{O \ depl}$ with $\eta$ are typically
non-monotonic: for example, the $25\,M_\odot$ model computed using the
V-vL combination reaches oxygen depletion with $\xi_{2.5}^\mathrm{O
  \ depl}=0.179$, 0.164, and 0.211 for $\eta=0.1$, 0.33, and 1.0,
respectively.  Within the framework of each wind algorithm, higher
values of $\eta$ correspond to higher mass loss rates and thus to a
progressive shift of the evolution toward that of lower initial
mass. However, the compactness parameter is known to be a highly
non-monotonic function of $M_\mathrm{ZAMS}$
\citep[][]{sukhbold:14}. Therefore, a higher mass loss rate can
sometimes result in a decrease, an increase, or even almost no
variation of $\xi_{2.5}^\mathrm{O \ depl}$.  For example, in
$15\,M_\odot$ models computed with V-vL, $\xi_{2.5}^\mathrm{O\ depl}$
decreases when going from $\eta=0.1$ to $\eta = 1.0$. In $25\,M_\odot$
models computed with V-vL, we instead find that
$\xi_{2.5}^\mathrm{O\ depl}$ increases with the same change in
$\eta$. And in the case of $20\,M_\odot$ models computed with V-dJ, we
find only tiny variations of $\xi_{2.5}^\mathrm{O\ depl}$ when $\eta$
is varied from $0.1$ to $1.0$.  See Tab.~\ref{tab:runs_to_O_depl} for
more examples and details.

\subsubsection{Effects of varying the mass loss algorithm on
  $\xi_{2.5}^\mathrm{O\ depl}$}

The last column of \Tabref{tab:runs_to_O_depl} shows that the effects
of different wind mass loss algorithms on $\xi_{2.5}$ are in most
cases small until oxygen depletion for $\eta = 0.1$ and $\eta = 0.33$.
For $\eta=0.1$, the V algorithm generally results in higher values of
$\xi_{2.5}^\mathrm{O \ depl}$ than the K algorithm. This holds for all
studied ZAMS masses with the exception of the $30\,M_\odot$
models. These do not exhibit this trend because all their burning
shells are outside the mass coordinate $\mathcal{M}=2.5\,M_\odot$, as
can be seen from the values of $M_\mathrm{CO}$ listed in
\Tabref{tab:runs_to_O_depl}.

The spread in $\xi_{2.5}^\mathrm{O \ depl}$ between different mass loss
algorithms increases for models with $\eta=0.33$. For example,
\Figref{fig:xi_O_depl} shows that the $25\,M_\odot$ model with the
K-dJ combination reaches $\xi_{2.5}^\mathrm{O \ depl}\simeq0.21$
(similar to its $20\,M_\odot$ counterpart,
cf.~\Tabref{tab:runs_to_O_depl}), which is $\sim 30\%$ higher than
with other algorithm combinations.

Models computed with $\eta=1.0$ are most suitable to discuss the
effect of different wind mass loss algorithm combinations. Both the
hot phase and the cool phase mass loss algorithms influence
$\xi_{2.5}^\mathrm{O\ depl}$, but their detailed effect varies with
$M_\mathrm{ZAMS}$ and is strongest in the $25\,M_\odot$ and
$30\,M_\odot$ models.  For example, from \Tabref{tab:runs_to_O_depl},
we find $\xi_{2.5}^\mathrm{O\ depl} \sim 0.16$, for the $\eta=1.0$, $25\,M_\odot$ models with the V-NJ and
K-dJ combinations, while models with V-vL and K-vL result in
$\xi_{2.5}^\mathrm{O\ depl} \sim 0.2$.  At lower $M_\mathrm{ZAMS}$,
even for $\eta = 1.0$, differences in $\xi_{2.5}^\mathrm{O\ depl}$ due
to the choice of mass loss algorithm combination are overall (with few
exceptions) rather small and typically at the few percent level. While
these differences will be 
amplified by the subsequent
evolution toward core collapse, they are small compared to the
tremendous differences in total mass resulting from the different
algorithm combinations (cf.~\Tabref{tab:runs_to_O_depl}).

An interesting question to address is the
relative importance of hot phase (i.e.\ main sequence) and cool
phase (i.e.\ post main sequence) mass loss for
$\xi_{2.5}^\mathrm{O\ depl}$.  Naively, one would think that by the
time the core and envelope are essentially decoupled, loss of envelope
mass should have limited impact on the subsequent evolution of the
core. Our results suggest that this is not generally the case.

The limited set of models run to oxygen depletion and listed in
\Tabref{tab:runs_to_O_depl} give a complex, but necessarily
incomplete picture of the relative importance of each mass
loss phase for the core structure. Which phase is most relevant depends on $M_\mathrm{ZAMS}$ and
$\eta$. For brevity, we focus here on the $\eta = 1$ case and
compare models evolved with the same hot phase mass loss algorithm
(V or K) and different cool phase algorithms.

For $15\,M_\odot$ models with $\eta = 1$, the tremendous mass loss
with the vL algorithm in the cool phase has little effect on He and
CO core masses and on $\xi_{2.5}^\mathrm{O\ depl}$.  For example,
the final masses of K-vL and K-NJ are $5.70\,M_\odot$ and
$11.87\,M_\odot$, respectively.  Yet their
$\xi_{2.5}^\mathrm{O\ depl}$ are very close to each other, $0.152$
and $0.153$, respectively. Qualitatively, the same is true for the
V-NJ and V-vL combinations. On the other hand there is a larger
spread between combinations using V and K, suggesting that the small
differences seeded by hot phase mass loss dominate in the
$15\,M_\odot$ $\eta = 1$ case. Similarly, we find for $30\,M_\odot$
models with $\eta = 1$ that K-dJ and K-vL lead to final masses of
$18.51\,M_\odot$ and $22.53\,M_\odot$, respectively, but their
$\xi_{2.5}^\mathrm{O\ depl}$ are $0.244$ and $0.243$. The situation
is more complicated for $25\,M_\odot$ and $20\,M_\odot$ models that
straddle the $M_\mathrm{ZAMS}$ range where $\xi_{2.5}$ varies
chaotically \citep{sukhbold:14}. From \Tabref{tab:runs_to_O_depl} we
see that for the $20\,M_\odot$, $\eta = 1$ models all considered K-
and V- combinations yield roughly the same
$\xi_{2.5}^\mathrm{O\ depl}$. In $25\,M_\odot$, $\eta = 1$ models, on
the other hand, cool phase mass loss has the dominant impact on
$\xi_{2.5}^\mathrm{O\ depl}$. For example, K-dJ and K-vL have
compactness of $0.161$ and $0.200$, respectively, although their
final masses differ by only $\sim 1\,M_\odot$ due to the
self-regulation of the vL algorithm in this mass range.

\subsubsection{Comparison with \cite{sukhbold:14}}

\cite{sukhbold:14} employed the NJ mass loss algorithm without any
efficiency scaling factor (i.e.\ at efficiency $\eta = 1.0$)
throughout the entire evolution of their models.  Comparing our
\Tabref{tab:runs_to_O_depl} with their Fig.~23, we find that the
compactness parameter values at oxygen depletion of our models lie in
the same range as theirs, with a tendency toward slightly higher
values.

Our $15\,M_\odot$ models produce values of $\xi_{2.5}^\mathrm{O
  \ depl}\simeq0.15$, which is slightly higher than their value of
$\sim 0.11-0.13$, especially for reduced wind mass loss rates
(i.e.\ $\eta<1.0$). For this initial mass, increasing $\eta$ decreases
the compactness of the core and reduces the difference between our
models and those of \cite{sukhbold:14}.

Most of our $20\,M_\odot$ models have $\xi_{2.5}^\mathrm{O
  \ depl}\simeq0.18$, close to the corresponding models of
\cite{sukhbold:14}. For these models, the maximum difference varying
the wind mass loss algorithm is $\Delta \xi_{2.5}^\mathrm{O
  \ depl}\lesssim 0.05$. This is not surprising, because large
variations of $\xi_{2.5}$ are expected because of the transition from
convective to neutrino-cooled and radiative carbon shell burning,
which happens around $M_\mathrm{ZAMS}\simeq20\,M_\odot$
\citep[][]{sukhbold:14}. Therefore, in this mass range, changing the
wind mass loss algorithm can substantially change $\xi_{2.5}^\mathrm{O
  \ depl}$ by shifting the evolutionary track of the star in the
slightest way.

Our $25\,M_\odot$ models have values of $\xi_{2.5}^\mathrm{O \ depl}$
similar to those of the $20\,M_\odot$ models, in agreement with
\cite{sukhbold:14}. However, once again, we obtain a large variation
of $\xi_{2.5}^\mathrm{O \ depl}$ changing the treatment of mass loss
($\Delta \xi_{2.5}^\mathrm{O \ depl}\lesssim 0.05$).

Most of our $30\,M_\odot$ models have $\xi_{2.5}^\mathrm{O
  \ depl}\simeq0.23$, which is significantly larger than the
corresponding value of $\sim 0.16$ found by \cite{sukhbold:14}. The
variations of the compactness parameter with mass loss algorithm
combination and efficiency are smaller for the $30 M_\odot$ models
than for lower $M_\mathrm{ZAMS}$, except for the model computed with
$\eta=1.0$ and the V-dJ algorithm. This model has $\xi_{2.5}^\mathrm{O
  \ depl}=0.179$, which is much closer to the values of
\cite{sukhbold:14}. We expect that the similar V-NJ algorithm
combination with $\eta=1.0$ would produce a structure close to this
model at oxygen depletion, based on the similarities between the
models at the end of the mass loss phase. The relatively low
compactness of this model is therefore likely a consequence of the V hot phase mass loss algorithm. For
$M_\mathrm{ZAMS}\gtrsim 30\,M_\odot$, V with $\eta=1.0$ produces
substantially more mass loss
than K (cf.~\Tabref{tab:hot_end_res}), and
therefore, the subsequent evolution is closer to the path of less
massive stars -- which are also expected to reach oxygen depletion
with a lower compactness.

We speculate that the quantitative differences between our findings
for the compactness parameter at oxygen depletion and those of
\cite{sukhbold:14} are in fact due primarily to their choice of mass
loss algorithm: they employ the NJ algorithm in both the hot and the
cold phase, which results in overall greater early mass loss than the
K algorithm (and the V algorithm until the bistability jump).  This
notion is corroborated by our finding that our $30\,M_\odot$ model closest to theirs is the one that loses
the most mass during the hot phase (using $\eta = 1$ and the V-dJ
combination; cf.~\Tabref{tab:hot_end_res}).

\subsection{Models at the onset of core collapse}
\label{sec:onset_cc}

We select a subset of six of our models at oxygen depletion for
continuation to the onset of core collapse.  Reducing the model set is
necessary to limit the computational cost of this study. We choose two
$15\,M_\odot$ models with efficiency $\eta=1.0$ and mass loss
algorithm combination V-NJ and K-VL, two $20\,M_\odot$ models with
$\eta=0.33$ computed using the V-vL and the K-dJ combination, and two
$30\,M_\odot$ models with $\eta=0.33$ and the V-dJ or the K-NJ
combination.  When restarting our models from oxygen depletion, we
switch from the 45-isotopes nuclear reaction network used so far to a
larger customized network with 203-isotopes
(see~Appendix~\ref{app:MESA_technical}). This is necessary to capture
core deleptonization due to electron capture during and after silicon
burning. Continuing the evolution with a larger nuclear reaction
network also requires reducing the number of computational mesh points
(from $\sim 10^4$ to $\sim10^3$) to run the simulations within the
memory constraints of MESA (see Appendix~\ref{app:MESA_technical} for
details). By the time oxygen depletion is reached, the effect
of wind mass loss on the core structure is already
pronounced, and our
reduced-resolution models still have spatial
resolution that is comparable to that of published models
\citep[e.g.,][]{woosley:02,woosley:07}. Furthermore,
  resolution tests in Appendix~\ref{app:re-meshing} give us
  confidence that the presently unavoidable reduction in resolution
  does not affect our overall results and conclusions.

\begin{figure}[!thbp]
  \centering
  \setlength{\abovecaptionskip}{-0.6cm}
  \setlength{\belowcaptionskip}{-0.2cm}
  \resizebox{\hsize}{!}{\includegraphics{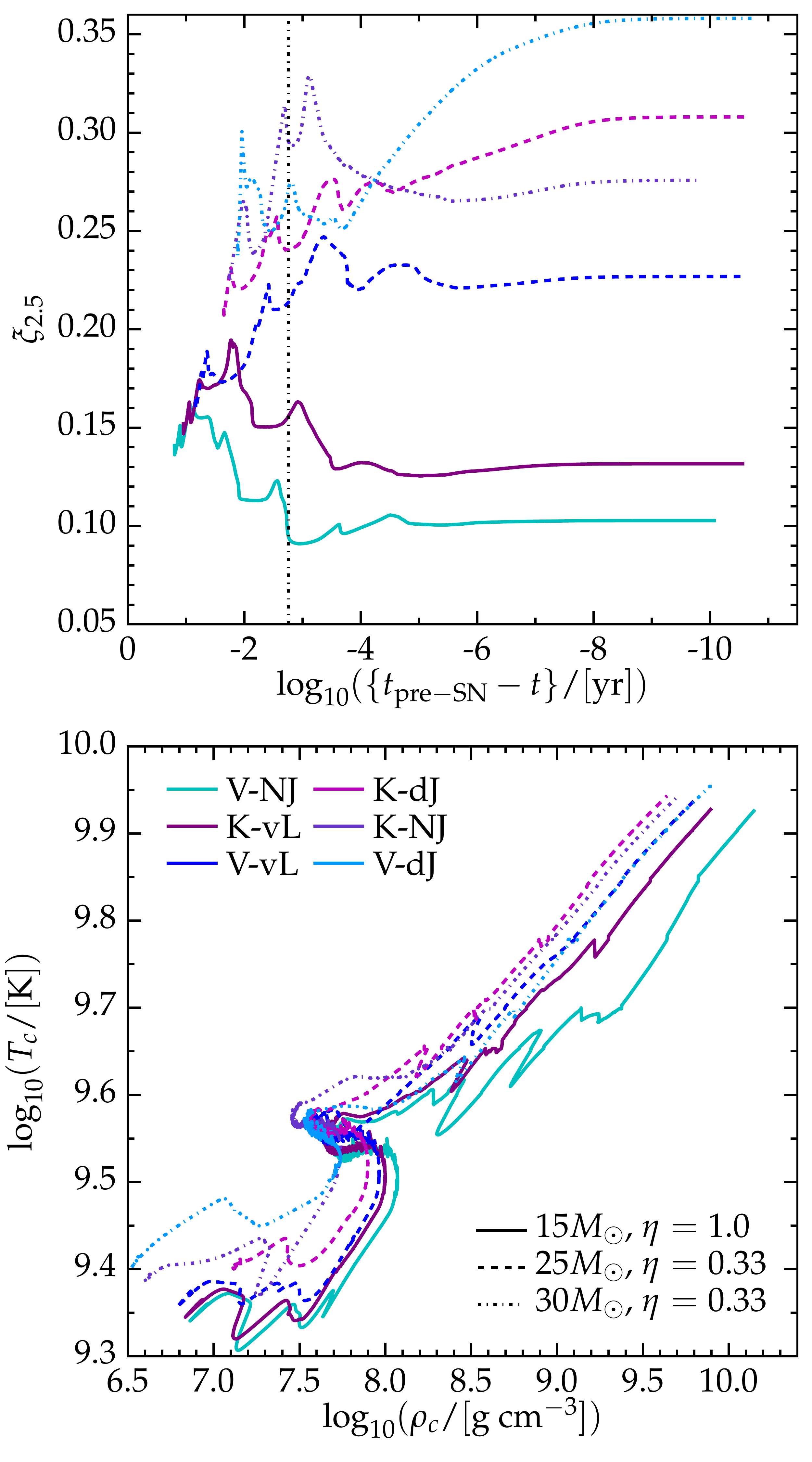}}
  \caption{Time evolution of the compactness
    parameter $\xi_{2.5}$ (top panel) and evolution in $\rho_c -
    T_c$ space (bottom panel) from oxygen depletion to the onset of
    core collapse. Solid curves correspond to $15\,M_\odot$ models
    computed with $\eta=1.0$, dashed and dot-dashed curves correspond
    to $25\,M_\odot$ and $30\,M_\odot$ models, respectively, computed
    with $\eta=0.33$. The vertical dot-dashed line in the top panel
    indicates roughly the time of core silicon
    depletion ($X(^{28}\mathrm{Si})\leq 0.01$).}
  \label{fig:xi_cc_evol}
\end{figure}

Figure~\ref{fig:xi_cc_evol} shows the time evolution of $\xi_{2.5}$
(top panel) and the central temperature--central density evolutionary
tracks (bottom panel) from oxygen depletion to the onset of core
collapse for our pre-SN model set. The top panel shows that
$\xi_{2.5}$ settles onto its final value before the criterion for the
onset of core collapse is reached. The bottom panel clearly shows a
hook at $\log_{10}(T_c/\mathrm{[K]})\simeq 9.55$, where $\rho_c$
decreases at roughly constant temperature, indicating the point of
silicon core ignition. This is also the more ``noisy'' part of these
tracks, indicating that this phase of nuclear burning with very high
and nearly balancing reaction rates is the most challenging to
simulate \citep[][]{hix:96, hix:07}.

\begin{table*}[!htbp]
\centering
\caption{Properties of the subset of models run to the onset of
  core collapse, defined as the time when
  $\mathrm{max}\{ |v| \} \geq 10^3 \ \mathrm{km \ s^{-1}}$. 
  $M_4$ and $\mu_4$ are the parameters used to predict the SN outcome
  of a stellar model in \cite{ertl:16}, see also text.
  $M_{\rho_6} \udef M(\rho = 10^6 \mathrm{g \ cm^{-3})}$ is the mass
  enclosed in the location where the density drops below $10^6
  \ \mathrm{g \ cm^{-3}}$, $M_\mathrm{CO}$ and $M_\mathrm{Fe}$ are the
  carbon-oxygen and iron core masses, respectively. 
   \label{tab:onset_cc}}
 \vspace*{-0.4cm}
\begin{tabular}{c|cl|cccccc}
  \hline
  \hline
  \multicolumn{9}{c}{Onset of core collapse: $\mathrm{max}\{ |v| \} \geq 10^3 \ \mathrm{km \ s^{-1}}$ }\\
  \hline
  \hline
  $M_\mathrm{ZAMS} \ [M_\odot]$ & $\eta$ & \multicolumn{1}{l|}{ ID}  & $ \xi_{2.5}^\mathrm{pre-SN}$ & $M_4 \ [M_\odot] $ & $\mu_4$ & $M_{\rho_6} \ [M_\odot]$ &  $M_\mathrm{CO} \ [M_\odot]$ & $M_\mathrm{Fe} \ [M_\odot]$ \\\hline\hline 
  \multirow{2}{*}{15} & \multirow{2}{*}{1.0} & V-NJ & 0.103 & 1.71 & 0.045 & 1.68 &  2.91 & 1.39\\ 
                      &                      & K-vL & 0.132 & 1.78 & 0.051 & 1.79 &  3.07 & 1.50\\ 
  \hline
  \multirow{2}{*}{25} & \multirow{2}{*}{0.33} & V-vL & 0.227 & 1.73 & 0.084 & 1.84 &  6.38 & 1.51\\ 
                      &                       & K-dJ & 0.308 & 2.05 & 0.100 & 2.19 &  6.40 & 1.63\\ 
  \hline
  \multirow{2}{*}{30} & \multirow{2}{*}{0.33} & V-dJ & 0.358 & 1.60 & 0.163 & 2.21  & 7.98 & 1.56\\ 
                      &                       & K-NJ & 0.276 & 1.82 & 0.100 & 1.98 &  7.90 & 1.58\\ 

\hline
\end{tabular}
\end{table*}

\begin{figure}[!htbp]
  \centering
  \setlength{\abovecaptionskip}{-0.6cm}
  \setlength{\belowcaptionskip}{-0.1cm}
\resizebox{\hsize}{!}{\includegraphics{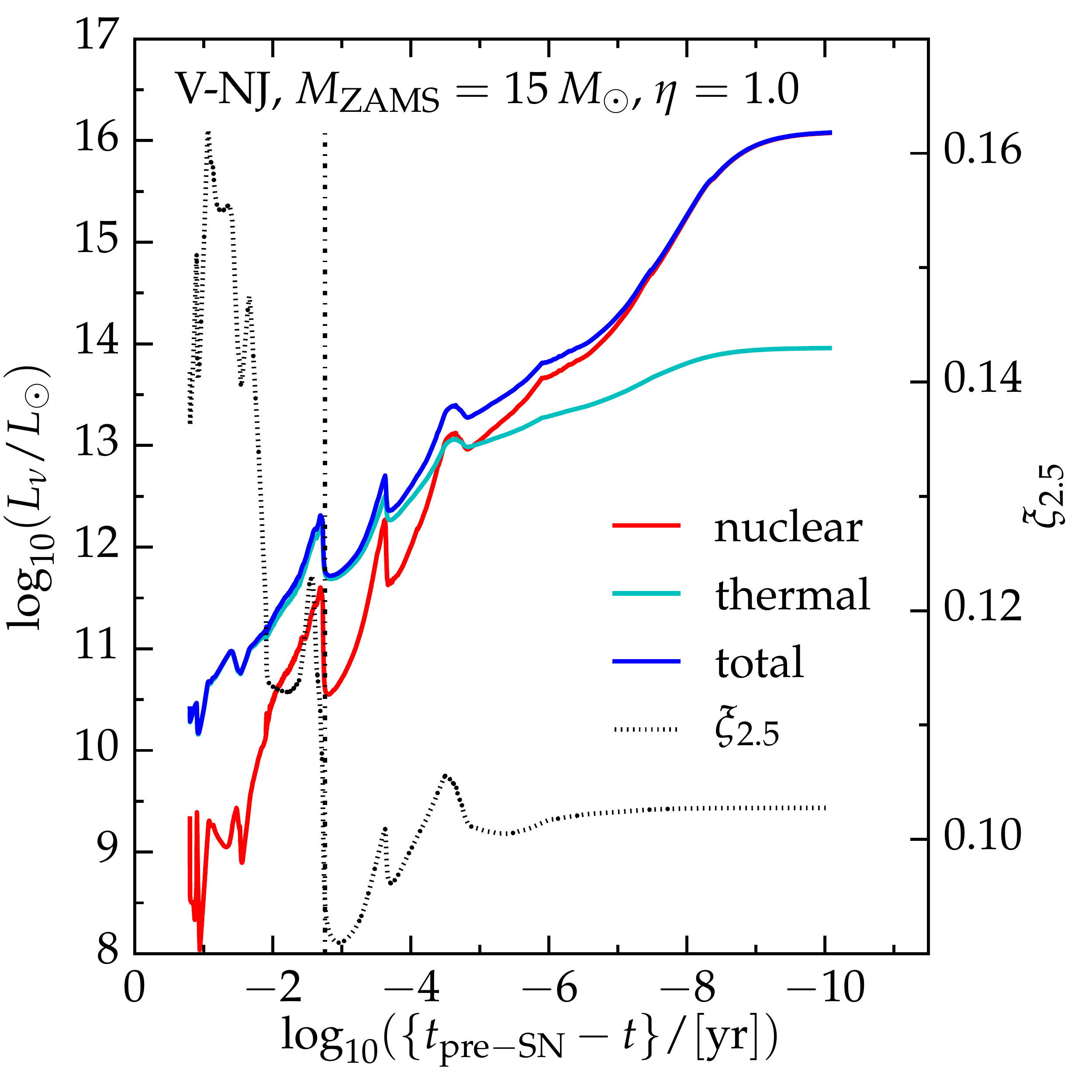}}
  \caption{Neutrino luminosity from
      oxygen depletion to core collapse for the $15\,M_\odot$ model run with the V-NJ
    scheme and $\eta=1.0$. This shows that variations in the nuclear
    burning rate cause the oscillations in $\xi_{2.5}$. The red curve
    shows only neutrinos coming from nuclear reactions, the cyan curve
    shows thermal neutrinos, and the blue curve is the sum of the
    two. The dotted line shows the compactness parameter (right hand
    side vertical axis). The vertical dot-dashed line indicates the
    approximate time of core silicon depletion.  The $\xi_{2.5}$
    oscillations are driven by variations in silicon and oxygen shell
    burning.}
  \label{fig:Lneu}
\end{figure}

The evolution of the compactness parameter shows maxima and minima,
which are related to oxygen shell ignition, around
$\log_{10}(\{t_\mathrm{pre-SN}-t\}/\mathrm{[yr]})\simeq -2$, silicon
core ignition around
$\log_{10}(\{t_\mathrm{pre-SN}-t\}/\mathrm{[yr]})\simeq -3$, and
silicon shell ignition at about
$\log_{10}(\{t_\mathrm{pre-SN}-t\}/\mathrm{[yr]})\simeq
-4.5$. However, note that the ignition times and the durations of
these burning phases are mass-dependent. Figure~\ref{fig:Lneu} shows
the corresponding increase in the neutrino emission from nuclear
reactions for the $15\,M_\odot$ model computed with the V-NJ
combination and $\eta=1.0$. Similar features are present in the
$T_c(\rho_c)$ evolutionary tracks. However, while the $T_c(\rho_c)$
track only probes the innermost part of the stellar core, the
$\xi_{2.5}$ evolution is determined by the interplay between silicon,
oxygen, and carbon burning shells, core contraction, and onset of
electron degeneracy.

Interestingly, the final compactness of the $15\,M_\odot$ models is
lower at the onset of core collapse than at oxygen depletion: for
example, the $15\,M_\odot$ model computed with K-VL and $\eta=1.0$ has
$\xi_{2.5}^\mathrm{pre-SN}=0.132<\xi_{2.5}^\mathrm{O
  \ depl}=0.152$. This is because of the presence of nuclear burning
shells within $\mathcal{M}=2.5\,M_\odot$ whose energy generation tends
to expand the material in the layers above them (the same process
happens during the Hertzsprung gap for hydrogen-shell-burning
stars). The location of the shells can be estimated using the core
masses listed in \Tabref{tab:onset_cc} (and
\Tabref{tab:runs_to_O_depl}): the $15\,M_\odot$ models have two shells
of nuclear burning (Si and O, respectively) within the $2.5\,M_\odot$
mass coordinate, while only one shell exists in this region at oxygen
depletion. Models with $M_\mathrm{ZAMS}>15\,M_\odot$ settle on a
pre-SN compactness that is higher than the corresponding
$\xi_{2.5}^\mathrm{O \ depl}$, because only the silicon burning shell
is within $\mathcal{M}=2.5\,M_\odot$.

Table~\ref{tab:onset_cc} lists the properties of the six models that
we run to the onset of core collapse.  $\xi_{2.5}^\mathrm{pre-SN}$,
and ($M_4$, $\mu_4$), where $M_4\udef M(s=4)$ is the mass location where the
specific entropy is $s=4\,\mathrm{k_b\,baryon^{-1}}$ and
$\mu_4\udef dm/dr|_{s=4}$ is the mass gradient at that location, offer two different ways to estimate how hard it
will be for the SN shock to unbind the stellar mantle and leave a NS
remnant \citep[see][]{ertl:16}. The total mass at density
  higher than $10^6\,\mathrm{g\ cm^{-3}}$ ($M_{\rho6}$), the
  carbon-oxygen core mass ($M_\mathrm{CO}$), and the iron core mass ($M_\mathrm{Fe}$), defined as
the location where $X(^{28}\mathrm{Si})<0.01$) can be used to estimate
the nickel yields of the possible SN explosion and the remnant mass
\citep[][]{fryer:12,sukhbold:16}.

The final $M_\mathrm{CO}$ and $M_\mathrm{Fe}$ depend, although weakly,
on the mass loss algorithm adopted during the hot evolutionary phase
(V or K). The V algorithm yields slightly smaller cores (and total
masses at the end of the hot phase, cf. \Secref{sec:hot}). The
$15\,M_\odot$ models with $\eta=1.0$ reach core collapse with
$M_\mathrm{CO}=2.91\,(3.07)\,M_\odot$, and
$M_\mathrm{Fe}=1.39\,(1.50)\,M_\odot$ when using the combination V-NJ
(K-vL). For $25\,M_\odot$ models with
$\eta=0.33$, we find $M_\mathrm{CO}=6.38\,(6.40)\,M_\odot$, and
$M_\mathrm{Fe}=1.51\,(1.63)\,M_\odot$ for V-vL (K-dJ). Finally, the
$30\,M_\odot$ models with $\eta=0.33$ yield
$M_\mathrm{CO}=7.98\,(7.90)\,M_\odot$, and
$M_\mathrm{Fe}=1.56\,(1.58)\,M_\odot$ for the combination V-dJ (K-NJ).
The differences in $M_\mathrm{CO}$ (and to a lesser extent
$M_\mathrm{Fe}$) decrease with $M_\mathrm{ZAMS}$.  However, note that
the decreasing difference is most likely caused by the lower
wind efficiency $\eta=0.33$ for the $25\,M_\odot$ and $30\,M_\odot$ models,
while our $15M_\odot$ models use full efficiency, i.e.\ $\eta=1.0$.

As anticipated in Sec.~\ref{sec:O_depl_res}, the spread in $\xi_{2.5}$
increases between oxygen depletion and the onset of core collapse: the
final variations are about $\sim 30\%$ for models with the same
initial mass (cf. \Tabref{tab:onset_cc}). The two $15\,M_\odot$ models
have $\Delta \xi_{2.5}^\mathrm{O\ depl}=0.011$, and
$\Delta\xi_{2.5}^\mathrm{pre-SN}=0.029$. For the $25\,M_\odot$ models,
the spread at oxygen depletion is $\Delta
\xi_{2.5}^\mathrm{O\ depl}=0.046$, while it is $\Delta
\xi_{2.5}^\mathrm{pre-SN}=0.081$ at the onset of core collapse. The
two $30\,M_\odot$ models go from $\Delta \xi_{2.5}^\mathrm{O
  \ depl}=0.015$ to $\Delta \xi_{2.5}^\mathrm{pre-SN}=0.082$.

\begin{figure}[!tbp]
  \centering
  \resizebox{\hsize}{!}{\includegraphics{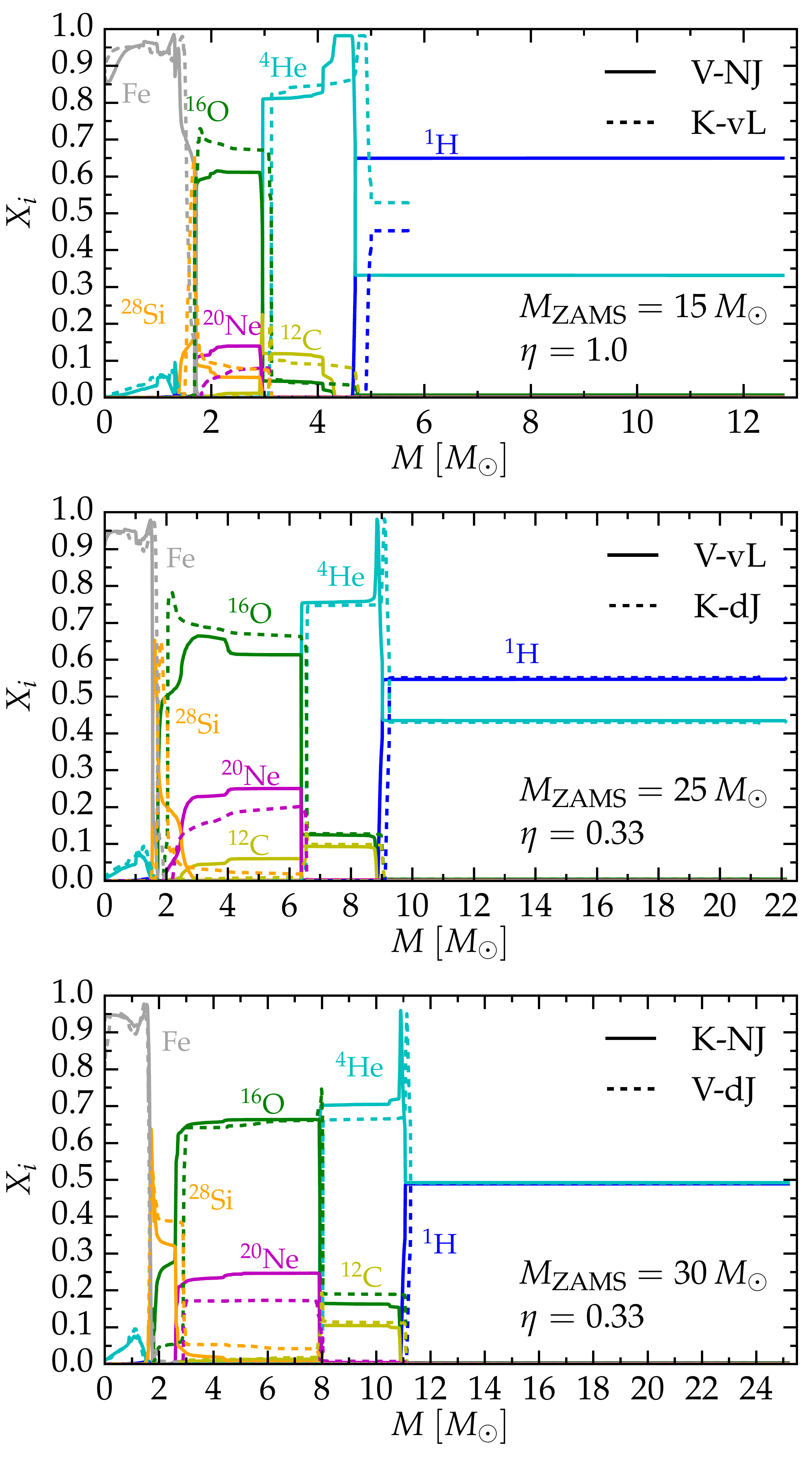}}
  \caption{Chemical composition profiles at the onset
    of core collapse. We show all models computed to this point and listed in \Tabref{tab:onset_cc}. Blue, cyan, yellow, green,
    magenta, orange, and gray curves correspond to the mass fractions of
    $^1\mathrm{H}$, $^{4}\mathrm{He}$,$^{12}\mathrm{C}$,
    $^{16}\mathrm{O}$, $^{20}\mathrm{Ne}$, $^{28}\mathrm{Si}$, and
    iron group elements (i.e.\ with atomic number $50\leq A\leq70$),
    respectively.
    Note the different horizontal scale in each
    panel.}
  \label{fig:chem_profiles}
\end{figure}

The pre-SN chemical abundances in the core are also affected by the
choice of the mass loss algorithm
combination. \Figref{fig:chem_profiles} shows a comparison of the
distribution of the dominant isotopes for our pre-SN models. The
composition of the oxygen rich layer is sensitive to the early (hot
phase) mass loss, with the V scheme producing a lower ratio of
$X(^{16}\mathrm{O})/X(^{20}\mathrm{Ne})$, owing to the higher early
mass loss (cf. \Secref{sec:hot}), and thus lower core temperature
during the late phases (cf. bottom panel of
\Figref{fig:xi_cc_evol}). The distribution of the chemical elements in
mass coordinate is also sensitive to the adopted mass loss algorithm
combination. However, it is likely to depend more strongly on the
treatment of mixing processes (which we do not vary here), mainly
convection and overshooting, which are the only processes fast enough
to have an effect inside the CO core of a star before the onset of
collapse.

\section{Discussion}
\label{sec:discussion}

\paragraph{\bf Sensitivity of wind mass loss to evolving stellar properties.}
For a given $M_\mathrm{ZAMS}$, the luminosity varies little between
models that experience different mass loss rates and the effective
temperature (and radius) varies much more. Once mass-loss induced
differences in $T_\mathrm{eff}$ appear, they feed back on mass loss
to amplify these differences. Hence, the $T_\mathrm{eff}$ evolution
is most important for governing the mass loss of models of a
given $M_\mathrm{ZAMS}$. The dependence of $T_\mathrm{eff}$ is
particularly strong if the dust-driven vL algorithm is included. 

\paragraph{\bf Role of dust in stellar winds.}
The role of dust as a driver of RSG winds is a subject of debate
\citep[e.g.,][]{vanloon:05,ferrarotti:06, bennett:10}. Dust-driven mass
  loss might occur in only a part of RSG evolution, and
  possibly even under only rather specific conditions. Even if dust exists in the
envelopes of cool, evolved, massive stars, uncertainties in the grain
properties result in highly uncertain mass loss rates
\citep[][]{vanloon:05}. The extreme dust-driven vL mass loss seen in
our models is just one example (see \Secref{sec:cool}). In this
context, it is important to mention that our models switch to the
vL algorithm already at $T_\mathrm{eff} \lesssim 15\,000\, \mathrm{K}$
($\lesssim 11\,000\, \mathrm{K}$) when K (V) is used in the hot
phase. These temperatures are clearly far too hot for dust to form.
However, the strong temperature dependence $(\sim T^{-6.3})$ in
combination with the extremely short time spent in the Hertzsprung gap
prevent this from having a substantial consequence for the stellar
mass. Hence, the extreme mass loss we find for
  $M_\mathrm{ZAMS}\lesssim20\,M_\odot$ with vL and $\eta = 1$ is not
an artifact of how we use the vL algorithm. Within its framework the
predicted very low final masses for $15\, M_\odot$ and $20\,M_\odot$
stars, $5.25-5.70\,M_\odot$ and $8.8\,M_\odot$, respectively, are
robust. \emph{It may thus be possible to rule out the extreme case of vL
with $\eta = 1$ using pre-explosion imaging and SN ejecta mass
estimates.}

We also note that in stars with
$M_\mathrm{ZAMS}\gtrsim 20\,M_\odot$ the vL algorithm self-regulates
since its extreme mass loss uncovers hotter layers of the
stars. The steep temperature dependence of the vL algorithm then
results in lower mass loss rates than dJ and NJ.

\paragraph{\bf The role of the efficiency parameter $\eta$.} The wind efficiency
$\eta$ is the parameter that has the greatest impact on the evolution
of initially identical models (cf.~\Figref{fig:mrel_mzams} and
\Tabref{tab:res}). However, it lacks an interpretation from first
principles. Here, we investigate only values $\eta\leq1$, focusing on
reduced wind mass loss motivated by possible inhomogeneities in the wind structure \citep[also
called ``clumpiness'',][and references therein]{smith:14,
  puls:08}. However, starting from first principles, it cannot be
excluded that clumps might actually enhance the mass loss rate
\citep[][]{lucy:80}. This might be the case if the overdense clumps
are efficiently pushed outward by impinging photons.

Furthermore, we assume a constant efficiency factor throughout the
evolution, but in principle the ``clumpiness'' of the wind might
evolve (possibly even in a stochastic way) and may require different
$\eta$ in different evolutionary phases. The mass loss routines in
MESA\footnote{These are available at
  \url{https://stellarcollapse.org/renzo2017}} are already adapted for
varying this parameter in different evolutionary phases. It
  is also possible that during each single evolutionary phase we
  define, the efficiency of mass loss may vary significantly,
  producing enhanced mass loss episodes separated by reduced mass loss
  phases.

Although changing $\eta$ induces substantial changes in the
mass loss rate and final mass, the appearance (i.e., luminosity,
effective temperature) is less affected, making it difficult to use
observed stellar populations to constrain $\eta$ \citep[][]{thesis}.

\paragraph{\bf Metallicity effects.} We do not investigate
the effects of decreasing the metallicity in this study. However, an approximate picture can be drawn by
considering models with
reduced efficiency $\eta$ as proxy for low metallicity models. Most
stellar evolution codes implement the metallicity dependence of the
wind mass loss by just rescaling the mass loss rate at solar
metallicity (cf.~\Eqref{eq:scaling}), which is exactly the purpose of
$\eta$. While the metallicity at the surface of the star can change
throughout the evolution, the main element driving a wind is
iron \citep[][]{vink:01,tramper:16}, and its abundance is unlikely to
change because of upward mixing from the stellar interior. Therefore,
it is not unrealistic to consider a constant metallicity-related
reduction factor for the entire evolution. Nevertheless,
the approach of considering reduced $\eta$ as
a proxy for lower metallicity does not take into account metallicity
effects on stellar radius and nuclear burning.  These could
indirectly affect wind mass loss. 

\paragraph{\bf WR stars.} We emphasize in \Secref{sec:comb} the
shortcomings of the computational definition of WR stars adopted in
stellar evolution codes ($X_s<0.4$).  Although this definition is
artificial, we stress that the mass loss algorithms used during this phase
are derived from the observation of real WR stars. In the mass range
considered here, few WR stars are expected, and indeed our results
show that only $M_\mathrm{ZAMS}=35\,M_\odot$ stars with $\eta=1.0$ can
become sufficiently hydrogen-depleted at their surface to switch to a
WR mass loss algorithm. We compare only two WR mass loss algorithms
(NL and H), and neglect algorithms obtained by fitting either very
luminous (i.e.\ more massive) or hydrogen-free WR stars
\citep[e.g.][]{grafener:08,tramper:16}, since none of our WR models
reach the corresponding regions of the parameter space.  Therefore, in
the framework of single nonrotating stars with wind efficiency
$\eta=1.0$, the minimum ZAMS mass to obtain a WR model is somewhere
between 30 and 35$\,M_\odot$. Although on the high end, this is in
relatively good agreement with the results obtained with other stellar
evolution codes \citep[e.g.][]{woosley:02,limongi:06,eldridge:06,
  georgy:15}. Even with full efficiency of the wind before the WR
phase, our models would likely underestimate the number of observable
WR stars.

Lowering $\eta$ has the obvious consequence of decreasing the mass
loss rate, and thus increasing the minimum ZAMS mass for single
nonrotating WR stars. However, the standard picture of a single
nonrotating star misses pieces of physics of great importance for the
formation of WR stars \citep[see, e.g., ][]{maeder:96,
  meynet:03,eldridge:06}. These include, but are not necessarily
limited to, rotationally-enhanced mass loss, rotational mixing
processes \citep[which can help depleting hydrogen from the
  surface,][]{meynet:03, maeder:96}, and binary interactions. Binarity
can lead to the formation of WR stars via envelope stripping in
RLOF. Alternatively, accretion or merger with a companion could
increase the mass (and luminosity) of the star sufficiently to enhance
the wind mass loss and remove the hydrogen-rich envelope. Also note
that the envelope hydrogen depletion needed to switch to a WR mass
loss algorithm can be reached also because of upward mixing of
thermonuclearly processed material, e.g.,~because of efficient
rotational mixing, and not only because of mass
loss. The choice for the algorithmic representation of mixing
processes (see Appendix~\ref{app:MESA_technical}) influences directly
the surface mass fraction of hydrogen, but also the core mass, and
consequently the luminosity. Indirectly, these effects can change the
mass loss rate and consequently the fate of a model from/to WR.

\paragraph{\bf Nucleosynthetic yields.} The nucleosynthetic yields of
massive stars are mass loss (and angular momentum loss) dependent
\citep[see, e.g.,][]{maeder:92,frischknecht:16}. Processes such as rotational
mixing can bring thermonuclearly processed material upward that is
then lost through winds. This is especially relevant for s-process
elements which are synthesized during the hydrostatic lifetime of
massive stars and the ratio of carbon to oxygen abundance in
  the stellar wind yields. On top of this, the success or failure of the SN
explosion, and the details of the explosive nucleosynthesis, depend on
the interior structure of the exploding star and thus on its mass loss
history \citep[e.g.,][]{sukhbold:16}.

\paragraph{\bf Consequences for SN explosions and compact remnants.}
We find that mass loss affects the core structure and the burning
shells surrounding the core.  Mass loss during the hot phase of the
evolution (i.e.\ the main sequence, roughly speaking) is important for
the core structure, because the core itself re-adjusts
quasi-statically to the wind from the stellar atmosphere. During this
phase, different algorithms produce small variations in the core,
which are then amplified by the subsequent evolution
(cf.~\Secref{sec:hot} and \Figref{fig:xi_t_25}). The general trend is
that a higher mass loss rate during the hot phase produces structures
with lower core compactness (cf.~\Secref{sec:xi_O_depl}). The cool
phase mass loss also impacts the core compactness, but more
indirectly, through its effect on the burning shells. Most of the mass
is lost during the cool phase of the evolution
(cf.~\Secref{sec:cool}), and the vigor, extent and type of mixing in
the burning shells all depend on the amount and the timing of mass
loss.  Since our present understanding of core-collapse SN explosions
strongly depends on the details of the input stellar models
\citep[e.g.,][]{janka:12,couch:15,chatzopoulos:16}, overlooking the
impact of wind mass loss on the core structure might bias detailed
hydrodynamical simulations of stellar explosions.
This, in turn, can have
significant implications for the NS/BH ratio and mass distribution,
and consequently also for the inferred gravitational wave sources. We
provide\footnote{Data are available at
  \url{https://zenodo.org/record/292924\#.WK_eENWi60i} and input
  parameter files at \url{https://stellarcollapse.org/renzo2017}} stellar
models at the end of the main sequence, at the end of the hot phase of
the evolution, at the end of the mass loss phase ($T_c \geq
10^9\,\mathrm{[K]}$), at oxygen depletion, 
 and at the onset of core collapse. These models can
be used as starting points for stellar experiments 
during late evolutionary phases
\citep[see e.g.][]{couch:15,chatzopoulos:16}. 

\begin{figure}[!tbp]
  \centering
  \includegraphics[width=0.5\textwidth]{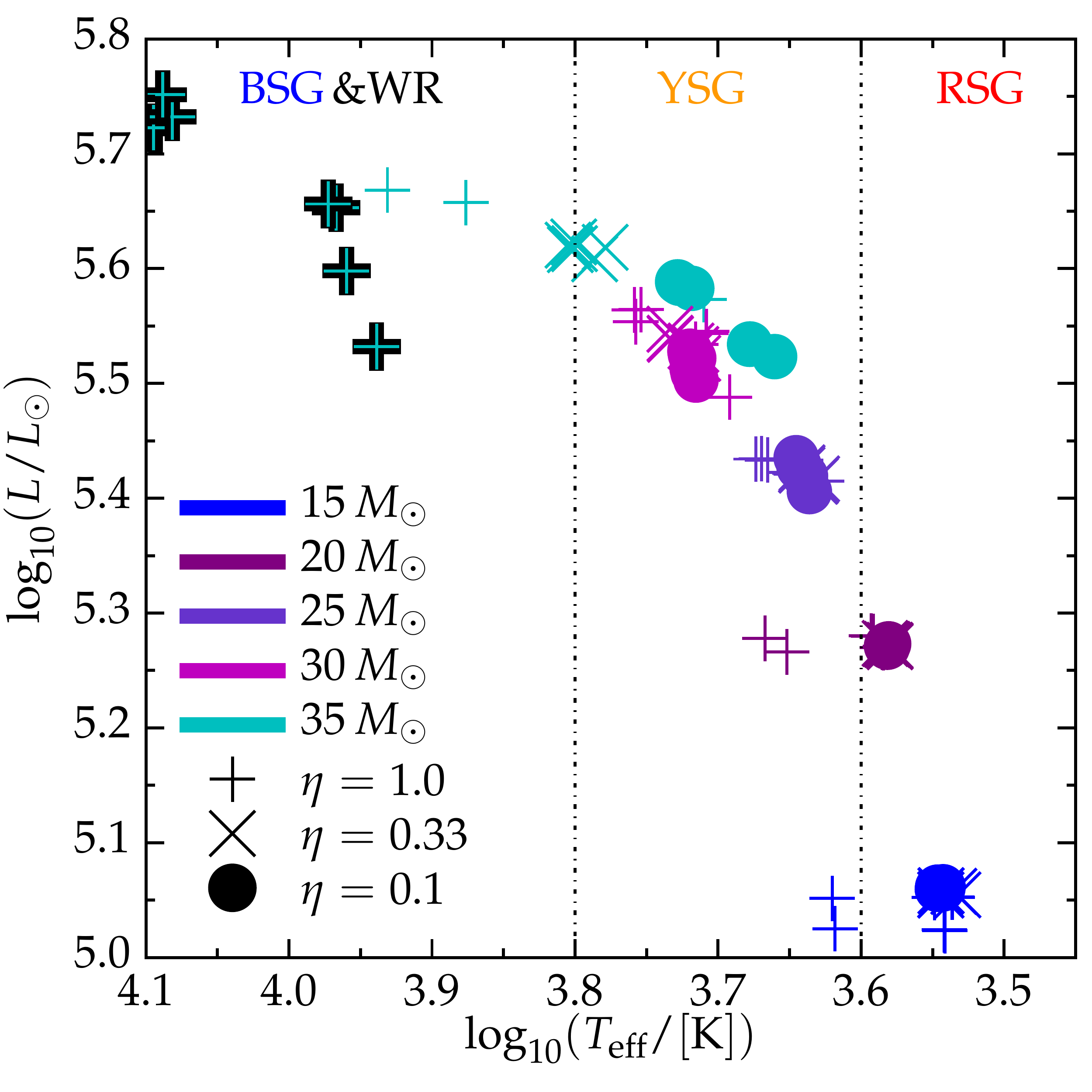}
  \caption{Spread in pre-SN appearance due to the uncertainty in wind
    mass loss. Each color represents an initial mass
    $M_\mathrm{ZAMS}$, each marker corresponds to a different
    efficiency factor $\eta$. The vertical dot-dashed lines indicate
    the BSG-YSG boundary, and YSG-RSG boundary according to
    \cite{georgy:16}. $35\,M_\odot$ models exhibit a larger spread,
    because some models (indicated by a thick black edge on their
    markers) develop into WR stars.
  \vspace*{-0.5cm}}
  \label{fig:HR_pre-SN}
\end{figure}

\paragraph{\bf Observational $M_\mathrm{ZAMS}$
  estimates of SN progenitors.} The large vertical spread in final
mass caused by differences in mass loss for a given $M_\mathrm{ZAMS}$
(\Figref{fig:mrel_mzams}) suggests at first sight that these large
uncertainties may map to equally large uncertainties in
$M_\mathrm{ZAMS}$ estimates from pre-explosion observations.  However,
the pre-SN mass is not a direct observable. Rather, $M_\mathrm{ZAMS}$
estimates typically rely on observational measurements of luminosity
and effective temperature that are then compared with stellar models
(e.g., \citealt{smartt:09review}). In \Figref{fig:HR_pre-SN}, we plot
the final luminosity $L$ and effective temperature $T_\mathrm{eff}$
for our entire model set. Table~\ref{tab:colors} summarizes the
numerical results. These results demonstrate that wind mass
loss variations have very little effect on the final luminosity for
stars with $M_\mathrm{ZAMS} \lesssim 30\,M_\odot$ and luminosity
variations are smaller than typical observational uncertainties (see,
e.g., \citealt{smartt:09review}). The similarity in luminosity of
models of a given $M_\mathrm{ZAMS}$ is a consequence of the rather
small effect that mass loss has on the core mass
(cf.~\Tabref{tab:res}). Interestingly, $T_\mathrm{eff}$ variations are
also small for $M_\mathrm{ZAMS} \lesssim 30\,M_\odot$. The
maximum variation from the average $T_\mathrm{eff}$ of a ZAMS mass is
only $0.08\,\mathrm{dex}$ and comes from the $\eta = 1$ K-vL and V-vL
$20\,M_\odot$ models that are YSGs at the end of their lives. Wind
mass-loss dependent variations in $L$ and $T_\mathrm{eff}$ are much
larger for $35\,M_\odot$ models, some of which die as BSGs and some as
WR stars.

Given the above results, it appears that for massive stars with
$M_\mathrm{ZAMS}\lesssim30\,M_\odot$ wind mass loss uncertainties do
not increase the overall level of uncertainty with which the SN
progenitor $M_\mathrm{ZAMS}$ can be estimated from pre-explosion
observations. However, our results do not say anything about
  the other possibly existing mass loss channels \citep[e.g.,\ binary
    interactions and/or impulsive phenomena, see, for
    example,][]{smith:14,smith:14b,morozova:15,margutti:17} that are
  usually neglected in stellar evolution calculations, or included in
  an very simplified way using enhanced winds \citep[see,
    e.g.,][]{meynet:15}.

\paragraph{\bf The RSG problem.} The observed lack of RSG SN
progenitors with $M_\mathrm{ZAMS}\gtrsim16\,M_\odot$
\citep[][]{smartt:09,smartt:09review} might be explained with the
effects of mass loss on the pre-SN stellar appearance
\citep[][]{georgy:16}. Table~\ref{tab:colors} and
\Figref{fig:HR_pre-SN} show that different wind algorithms can change
the pre-explosion appearance of a massive star. However, this effect
is relatively small in luminosity and $T_\mathrm{eff}$, and the
relative number of YSG to RSG we find (cf.~\Secref{sec:res}) largely
depends on the adopted definition of YSG vs.\ RSG. Therefore, the
systematic uncertainty in the treatment of wind mass loss for
$16\lesssim M_\mathrm{ZAMS}/M_\odot\lesssim30$ does not seem
sufficient to solve the RSG problem. It is likely that either
(\emph{i}) stars with initial mass in the range $\sim16-30\,M_\odot$
do not produce a bright transient at their death, or (\emph{ii}) mass
loss phenomena other than winds, happening at late stages in the
evolution, change the pre-SN appearance of the star.
SN observations can constrain the total
ejected mass and pre-explosion
imaging might also provide constraints (although subject to large
uncertainties) on the total mass lost and on the mass loss timing
\citep[if the surface chemical composition can be inferred,
e.g.,][]{smartt:09,gordon:16}.

\paragraph{\bf Potential observational constraints.}
A range of special systems, events, and phenomena offer alternative
and complimentary ways to constrain mass loss to the more traditional
spectral observations of massive stars.  These special systems include
bow shocks of runaway stars \citep[e.g.,][]{gull:79,meyer:16}, SN
shocks running into the circumstellar material
\citep[e.g.,][]{maeda:15,chakraborti:16,margutti:17},
flash-spectroscopy of material ejected shortly before core collapse
\citep[e.g.,][]{khazov:16}, accretion in wind-fed high mass X-ray
binaries, or binary wind collisions. Constraints from special systems
can be combined with those arising from observed populations of stars
and their compact remnants (including gravitational wave sources).
The different mass loss algorithms that we compare here give a range
of mass loss timing, which suggests that the chemical composition and
dust properties of the circumstellar material may also contain hints
regarding the mass loss of massive stars.

\paragraph{\bf Other known unknowns and uncertainties.} In this study, we have only
varied the mass loss algorithms used throughout the evolution and
their efficiency parameter $\eta$. However, other uncertainties (with
possible degeneracies) are known, for example in the treatment of
rotation, magnetic fields \citep[e.g.,][]{petit:17}, convective mixing, and overshooting
\citep[e.g.,][]{arnett:15b,arnett:15, arnett:16, farmer:16}. The
coupling of these uncertainties with the wind mass loss may modify the
outcomes of our numerical experiments.

\section{Conclusions}
\label{sec:conclusion}

Massive star mass loss is a longstanding issue
in stellar evolution. Despite decades of observational and
theoretical work it remains incompletely understood. Mass loss
influences the lifetime and appearance of massive stars,
their internal structure at the onset of core collapse, and their
total nucleosynthetic yields. Through its effect on the pre-supernova
(pre-SN) structure, wind mass loss can impact the outcome of core
collapse and the nature of the compact remant, with potential
implications for gravitational wave astronomy. 

We studied the impact of a broad range of wind mass loss algorithm
combinations on the evolution and pre-collapse structure of
nonrotating, single, solar-metallicity stars with initial masses of
$15,\,20,\,25,\,30,$ and $35\,M_\odot$. We compared 12 different mass
loss algorithm combinations, drawing from 2 algorithms for the hot phase of the
evolution (corresponding roughly to the main sequence), 3 for the cool
phase of evolution, and 2 for the Wolf-Rayet phase (if it is
reached). We explored the effects of reducing the mass loss rate with
an efficiency scaling factor $\eta=1.0,\,0.33$, or $0.1$ to crudely
account for reduced stellar mass loss that could be caused, for
example, by inhomogeneities in the wind (e.g., clumpiness).  The
resulting differences in stellar structure and total mass at various
stages of the evolution are caused by the different algorithmic
representations of stellar winds.

The different mass loss efficiencies and algorithm
combinations have profound effects on the evolution and the pre-SN
masses of massive stars. On the one hand, this can be expected given
the inherent differences of the various mass loss algorithms and the
various assumptions that enter them. On the other hand, these
algorithms all attempt to describe the same physical process --
steady wind mass loss -- and less sensitivity to the
theoretical/empirical treatment of mass loss would in general be
desirable.

\medskip
We find that the choice of wind efficiency scaling factor $\eta$ has
the greatest impact on our stellar models. It affects their 
total mass loss, their evolutionary path, and their pre-SN
structure. $\eta$ is therefore the main uncertainty and limiting
factor for our present understanding of wind mass loss and its
effects. If the wind efficiency is low, the differences between
various mass loss algorithms are less important.

Considering the full range of wind efficiencies $0.1\leq\eta\leq1.0$,
we find that there is a $\sim 50\%$ uncertainty in the pre-SN mass for
a given $M_\mathrm{ZAMS}$.  For fixed efficiency $\eta=1.0$, the
uncertainty varies with initial mass and is $\sim15-30\%$ in most
cases. Impulsive mass loss events (eruptions, pulsational
instabilities, etc., all neglected here) could only make the
uncertainties in the initial to pre-SN mapping more severe.

Despite the large uncertainty in the pre-SN mass, we find that the key
observables from pre-SN imaging, the luminosity and effective
temperature before explosion, are only mildly affected by varying the
mass loss algorithm combination and wind
efficiency. The uncertainties in $L$ and $T_\mathrm{eff}$ from our
models are within observational limits, suggesting that wind mass-loss
uncertainties do not affect observational estimates of SN progenitor
masses from pre-SN observations for most
massive stars (assuming a single star evolution
scenario). Nevertheless, the impact of stellar winds on the
  internal structure of the star can affect the mass and composition
  of the SN ejecta.

Independent of the employed algorithm during the hot phase, the amount
of mass lost in this phase is only a small
fraction ($\sim$ few percent) of the total mass. However, since the
core can respond directly to it, mass loss during the hot phase
creates seed differences that grow during the subsequent evolution,
leading to changes in the pre-SN structure and composition of the
stellar core. Later, burning shells re-adjust to the mass loss instead
of the core itself, and the effect of mass loss is more indirect.

Most of the mass is lost during the late and short cool phase of the
evolution.  Wind mass loss during this phase is more uncertain. On the
cool side of the Hertzsprung-Russel diagram, two different wind
driving mechanism might exist: line-driving or dust-driving. The
latter, assumed by the \cite{vanloon:05} (vL) algorithm, has a much
stronger temperature dependence and produces much higher mass loss
rates than algorithms describing line-driven mass loss. vL mass loss
can be so strong that it reveals the deep and hot layers of the
star. This results in self-damping of the wind itself for higher mass progenitors and thus a
higher value of the final mass. The vL algorithm produces very
different evolutionary tracks from those obtained with algorithms
assuming line driving, with the vL algorithm driving blueward
displacements on the Hertzsprung-Russel diagram. The more indirect
effect of late mass loss on the pre-SN core structure is
difficult to pinpoint with our limited model grid.

In our model grid, only models with
$M_\mathrm{ZAMS}\geq35\,M_\odot$ and full wind efficiency ($\eta=1.0$)
develop into Wolf-Rayet stars. The absence of Wolf-Rayet models with
reduced wind efficiency suggests that either (\emph{i}) our initial
mass range is too small to produce Wolf-Rayet stars from single,
nonrotating stars, or (\emph{ii}) other formation channels, such as
binarity or impulsive mass loss events, might be dominant.
The standard picture for the evolution of massive stars, the so-called
``Conti scenario'' \citep[e.g.,][]{maeder:94,lamers:13, smith:15b},
predicts the formation of Wolf-Rayet stars
with $M_\mathrm{ZAMS}\lesssim40\,M_\odot$, therefore,
if mass loss occurs in nature with reduced
efficiency ($\eta < 1$), then our results with reduced wind
efficiency disagree with this prediction. 

From our limited sample of six models that we were able to evolve to
the pre-SN stage, we find that changing the wind mass loss algorithm
combination can lead to changes of the compactness parameter
$\xi_{2.5}$ of up to $30\%$ for a given $M_\mathrm{ZAMS}$. Moreover,
the pre-SN models show a spread in terms of composition profiles,
density profiles, and core masses. These uncertainties add to those
arising from the incomplete understanding of mixing processes (mainly
convection and overshooting). They complicate the study of the
core-collapse SN explosion mechanism by adding uncertainty to the
initial conditions from which core-collapse SN simulations
start. \emph{This finding underlines that systematic uncertainty in
  massive star mass loss can have important implications for the
  relative number of BHs and NSs resulting from core collapse events,
  and consequently for gravitational wave sources, and for the
  nucleosynthetic yields from the explosions of massive stars.}

Although the present study provides new insights into the effects of
wind mass loss on the evolution and pre-SN structure and appearance of
massive stars, it suffers from a number of important limitations that
must be addressed by future work.

Our model grid coarsely samples a limited mass range, includes only
nonrotating solar-metalicity models, and we could evolve only six
models to the onset of core collapse.  Extending this grid to finer
mass sampling, higher masses, lower metallicity, and evolving all
models to the pre-SN stage will be important future work needed to
infer more robustly how wind mass loss affects pre-SN
structure. Furthermore, we did not consider time-dependent wind
efficiency, rotation and magnetic fields, and varitations in mixing
processes, which all may have important implications for wind mass
loss and its effects on evolution and pre-SN structure.

As in any computational astrophysics study, numerical resolution is a
major concern in our work. Initially, the effect of mass loss on the
core structure is small, and impossible to resolve with a coarse
spatial mesh. We tested our numerical resolution
(cf.~Appendix~\ref{app:MESA_technical}) and ran our calculations at
unprecedented spatial resolution (between $20\,000$ and $100\,000$
mesh points) until oxygen depletion to capture the delayed effect of
mass loss on the core structure. However, in order to follow the core
deleptonization after oxygen depletion with a large nuclear reaction
network, we were forced to reduce the spatial resolution. We find
that by the end of core oxygen burning, the differences in core
structure due to different mass loss algorithm combinations are
already pronounced. The limited resolution study that we were able to
perform suggests that resolution effects are smaller than the overall
effects of mass loss and do not affect our conclusions. Future work is
needed to more formally demonstrate robustness and numerical
convergence of simulations of the late evolutionary stages.

\begin{acknowledgements}
  We wish to thank B.~Paxton, F.~Timmes, and R.~Farmer for invaluable
  help with MESA. We acknowledge helpful exchanges with W.~D.~Arnett,
  M.~Cantiello, D.~Clausen, A.~de~Koter, L.~Dessart, Y.~G{\"o}tberg,
  A.~L.~Piro, J.~Fuller, N.~Smith, and E.~Zapartas. We are also
  grateful to the referee, G.~Meynet, for the helpful comments and
  suggested improvements. MR thanks the University of Pisa, where this
  study was initiated as part of his Master Thesis in Physics, and
  acknowledges the precious support of the ``Lorentz group''. This
  work is supported in part by NSF under grant numbers AST-1205732,
  AST-1212170, PHY-1151197, and PHY-1125915 and by the Sherman
  Fairchild Foundation.  Some of the computations used resources of
  NSF's XSEDE network under allocation TG-PHY100033. Most of the
  simulation were carried out using the Caltech compute cluster Zwicky
  funded through NSF grant no. PHY-0960291 and the Sherman Fairchild
  Foundation. The runs to the onset of core collapse were carried out
  on the Dutch national e-infrastructure (Cartesius, project number
  15162) with the support of the SURF Cooperative.
\end{acknowledgements}

\bibliographystyle{aa}

\begin{thebibliography}{137}
\expandafter\ifx\csname natexlab\endcsname\relax\def\natexlab#1{#1}\fi

\bibitem[{Abbott {et~al.}(2016)Abbott, Abbott, Abbott, Abernathy, Acernese,
  Ackley, Adams, Adams, Addesso, Adhikari, Adya, Affeldt, Agathos, Agatsuma,
  Aggarwal, Aguiar, Aiello, Ain, Ajith, Allen, Allocca, Altin, Anderson,
  Anderson, Arai, Araya, Arceneaux, Areeda, Arnaud, Arun, Ascenzi, Ashton, Ast,
  Aston, Astone, Aufmuth, Aulbert, Babak, Bacon, Bader, Baker, Baldaccini,
  Ballardin, Ballmer, Barayoga, Barclay, Barish, Barker, Barone, Barr,
  Barsotti, Barsuglia, Barta, Bartlett, Bartos, Bassiri, Basti, Batch, Baune,
  Bavigadda, Bazzan, Behnke, Bejger, Belczynski, Bell, Bell, Berger, Bergman,
  Bergmann, Berry, Bersanetti, Bertolini, Betzwieser, Bhagwat, Bhandare,
  Bilenko, Billingsley, Birch, Birney, Biscans, Bisht, Bitossi, Biwer,
  Bizouard, Blackburn, Blair, Blair, Blair, Bloemen, Bock, Bodiya, Boer,
  Bogaert, Bogan, Bohe, Bojtos, Bond, Bondu, Bonnand, Boom, Bork, Boschi, Bose,
  Bouffanais, Bozzi, Bradaschia, Brady, Braginsky, Branchesi, Brau, Briant,
  Brillet, Brinkmann, Brisson, Brockill, Brooks, Brown, Brown, Brown, Buchanan,
  Buikema, Bulik, Bulten, Buonanno, Buskulic, Buy, Byer, Cadonati, Cagnoli,
  Cahillane, Bustillo, Callister, Calloni, Camp, Cannon, Cao, Capano, Capocasa,
  Carbognani, Caride, Diaz, Casentini, Caudill, Cavaglià, Cavalier, Cavalieri,
  Cella, Cepeda, Baiardi, Cerretani, Cesarini, Chakraborty, Chalermsongsak,
  Chamberlin, Chan, Chao, Charlton, Chassande-Mottin, Chen, Chen, Cheng,
  Chincarini, Chiummo, Cho, Cho, Chow, Christensen, Chu, Chua, Chung, Ciani,
  Clara, Clark, Cleva, Coccia, Cohadon, Colla, Collette, Cominsky, Jr., Conte,
  Conti, Cook, Corbitt, Cornish, Corsi, Cortese, Costa, Coughlin, Coughlin,
  Coulon, Countryman, Couvares, Cowan, Coward, Cowart, Coyne, Coyne, Craig,
  Creighton, Cripe, Crowder, Cumming, Cunningham, Cuoco, Canton, Danilishin,
  D’Antonio, Danzmann, Darman, Dattilo, Dave, Daveloza, Davier, Davies, Daw,
  Day, DeBra, Debreczeni, Degallaix, Laurentis, Deléglise, Pozzo, Denker,
  Dent, Dereli, Dergachev, DeRosa, DeRosa, DeSalvo, Dhurandhar, Díaz, Fiore,
  Giovanni, Lieto, Pace, Palma, Virgilio, Dojcinoski, Dolique, Donovan, Dooley,
  Doravari, Douglas, Downes, Drago, Drever, Driggers, Du, Ducrot, Dwyer, Edo,
  Edwards, Effler, Eggenstein, Ehrens, Eichholz, Eikenberry, Engels, Essick,
  Etzel, Evans, Evans, Everett, Factourovich, Fafone, Fair, Fairhurst, Fan,
  Fang, Farinon, Farr, Farr, Favata, Fays, Fehrmann, Fejer, Ferrante, Ferreira,
  Ferrini, Fidecaro, Fiori, Fiorucci, Fisher, Flaminio, Fletcher, Fournier,
  Franco, Frasca, Frasconi, Frei, Freise, Frey, Frey, Fricke, Fritschel,
  Frolov, Fulda, Fyffe, Gabbard, Gair, Gammaitoni, Gaonkar, Garufi, Gatto,
  Gaur, Gehrels, Gemme, Gendre, Genin, Gennai, George, Gergely, Germain, Ghosh,
  Ghosh, Giaime, Giardina, Giazotto, Gill, Glaefke, Goetz, Goetz, Gondan,
  González, Castro, Gopakumar, Gordon, Gorodetsky, Gossan, Gosselin, Gouaty,
  Graef, Graff, Granata, Grant, Gras, Gray, Greco, Green, Groot, Grote,
  Grunewald, Guidi, Guo, Gupta, Gupta, Gushwa, Gustafson, Gustafson, Hacker,
  Hall, Hall, Hammond, Haney, Hanke, Hanks, Hanna, Hannam, Hanson, Hardwick,
  Harms, Harry, Harry, Hart, Hartman, Haster, Haughian, Heidmann, Heintze,
  Heitmann, Hello, Hemming, Hendry, Heng, Hennig, Heptonstall, Heurs, Hild,
  Hoak, Hodge, Hofman, Hollitt, Holt, Holz, Hopkins, Hosken, Hough, Houston,
  Howell, Hu, Huang, Huerta, Huet, Hughey, Husa, Huttner, Huynh-Dinh, Idrisy,
  Indik, Ingram, Inta, Isa, Isac, Isi, Islas, Isogai, Iyer, Izumi, Jacqmin,
  Jang, Jani, Jaranowski, Jawahar, Jiménez-Forteza, Johnson, Jones, Jones,
  Jonker, Ju, K, Kalaghatgi, Kalogera, Kandhasamy, Kang, Kanner, Karki,
  Kasprzack, Katsavounidis, Katzman, Kaufer, Kaur, Kawabe, Kawazoe, Kéfélian,
  Kehl, Keitel, Kelley, Kells, Kennedy, Key, Khalaidovski, Khalili, Khan, Khan,
  Khan, Khazanov, Kijbunchoo, Kim, Kim, Kim, Kim, Kim, Kim, King, King, Kinzel,
  Kissel, Kleybolte, Klimenko, Koehlenbeck, Kokeyama, Koley, Kondrashov,
  Kontos, Korobko, Korth, Kowalska, Kozak, Kringel, Krishnan, Królak, Krueger,
  Kuehn, Kumar, Kuo, Kutynia, Lackey, Landry, Lange, Lantz, Lasky, Lazzarini,
  Lazzaro, Leaci, Leavey, Lebigot, Lee, Lee, Lee, Lee, Lenon, Leonardi, Leong,
  Leroy, Letendre, Levin, Levine, Li, Libson, Littenberg, Lockerbie, Logue,
  Lombardi, Lord, Lorenzini, Loriette, Lormand, Losurdo, Lough, Lück,
  Lundgren, Luo, Lynch, Ma, MacDonald, Machenschalk, MacInnis, Macleod,
  Magaña-Sandoval, Magee, Mageswaran, Majorana, Maksimovic, Malvezzi, Man,
  Mandel, Mandic, Mangano, Mansell, Manske, Mantovani, Marchesoni, Marion,
  Márka, Márka, Markosyan, Maros, Martelli, Martellini, Martin, Martin,
  Martynov, Marx, Mason, Masserot, Massinger, Masso-Reid, Matichard, Matone,
  Mavalvala, Mazumder, Mazzolo, McCarthy, McClelland, McCormick, McGuire,
  McIntyre, McIver, McManus, McWilliams, Meacher, Meadors, Meidam, Melatos,
  Mendell, Mendoza-Gandara, Mercer, Merilh, Merzougui, Meshkov, Messenger,
  Messick, Meyers, Mezzani, Miao, Michel, Middleton, Mikhailov, Milano, Miller,
  Millhouse, Minenkov, Ming, Mirshekari, Mishra, Mitra, Mitrofanov,
  Mitselmakher, Mittleman, Moggi, Mohan, Mohapatra, Montani, Moore, Moore,
  Moraru, Moreno, Morriss, Mossavi, Mours, Mow-Lowry, Mueller, Mueller, Muir,
  Mukherjee, Mukherjee, Mukherjee, Mukund, Mullavey, Munch, Murphy, Murray,
  Mytidis, Nardecchia, Naticchioni, Nayak, Necula, Nedkova, Nelemans, Neri,
  Neunzert, Newton, Nguyen, Nielsen, Nissanke, Nitz, Nocera, Nolting,
  Normandin, Nuttall, Oberling, Ochsner, O’Dell, Oelker, Ogin, Oh, Oh, Ohme,
  Oliver, Oppermann, Oram, O’Reilly, O’Shaughnessy, Ott, Ottaway, Ottens,
  Overmier, Owen, Pai, Pai, Palamos, Palashov, Palomba, Pal-Singh, Pan, Pankow,
  Pannarale, Pant, Paoletti, Paoli, Papa, Paris, Parker, Pascucci, Pasqualetti,
  Passaquieti, Passuello, Patricelli, Patrick, Pearlstone, Pedraza, Pedurand,
  Pekowsky, Pele, Penn, Perreca, Phelps, Piccinni, Pichot, Piergiovanni,
  Pierro, Pillant, Pinard, Pinto, Pitkin, Poggiani, Popolizio, Post, Powell,
  Prasad, Predoi, Premachandra, Prestegard, Price, Prijatelj, Principe,
  Privitera, Prix, Prodi, Prokhorov, Puncken, Punturo, Puppo, Pürrer, Qi, Qin,
  Quetschke, Quintero, Quitzow-James, Raab, Rabeling, Radkins, Raffai, Raja,
  Rakhmanov, Rapagnani, Raymond, Razzano, Re, Read, Reed, Regimbau, Rei, Reid,
  Reitze, Rew, Reyes, Ricci, Riles, Robertson, Robie, Robinet, Rocchi, Rolland,
  Rollins, Roma, Romano, Romano, Romanov, Romie, Rosińska, Rowan, Rüdiger,
  Ruggi, Ryan, Sachdev, Sadecki, Sadeghian, Salconi, Saleem, Salemi, Samajdar,
  Sammut, Sanchez, Sandberg, Sandeen, Sanders, Sassolas, Sathyaprakash,
  Saulson, Sauter, Savage, Sawadsky, Schale, Schilling, Schmidt, Schmidt,
  Schnabel, Schofield, Schönbeck, Schreiber, Schuette, Schutz, Scott, Scott,
  Sellers, Sentenac, Sequino, Sergeev, Serna, Setyawati, Sevigny, Shaddock,
  Shah, Shahriar, Shaltev, Shao, Shapiro, Shawhan, Sheperd, Shoemaker,
  Shoemaker, Siellez, Siemens, Sigg, Silva, Simakov, Singer, Singer, Singh,
  Singh, Singhal, Sintes, Slagmolen, Smith, Smith, Smith, Son, Sorazu,
  Sorrentino, Souradeep, Srivastava, Staley, Steinke, Steinlechner,
  Steinlechner, Steinmeyer, Stephens, Stevenson, Stone, Strain, Straniero,
  Stratta, Strauss, Strigin, Sturani, Stuver, Summerscales, Sun, Sutton,
  Swinkels, Szczepańczyk, Tacca, Talukder, Tanner, Tápai, Tarabrin,
  Taracchini, Taylor, Theeg, Thirugnanasambandam, Thomas, Thomas, Thomas,
  Thorne, Thorne, Thrane, Tiwari, Tiwari, Tokmakov, Tomlinson, Tonelli, Torres,
  Torrie, Töyrä, Travasso, Traylor, Trifirò, Tringali, Trozzo, Tse, Turconi,
  Tuyenbayev, Ugolini, Unnikrishnan, Urban, Usman, Vahlbruch, Vajente, Valdes,
  van Bakel, van Beuzekom, van~den Brand, van~den Broeck, Vander-Hyde, van~der
  Schaaf, van Heijningen, van Veggel, Vardaro, Vass, Vasúth, Vaulin, Vecchio,
  Vedovato, Veitch, Veitch, Venkateswara, Verkindt, Vetrano, Viceré,
  Vinciguerra, Vine, Vinet, Vitale, Vo, Vocca, Vorvick, Voss, Vousden,
  Vyatchanin, Wade, Wade, Wade, Walker, Wallace, Walsh, Wang, Wang, Wang, Wang,
  Wang, Ward, Warner, Was, Weaver, Wei, Weinert, Weinstein, Weiss, Welborn,
  Wen, Weßels, Westphal, Wette, Whelan, White, Whiting, Williams, Williamson,
  Willis, Willke, Wimmer, Winkler, Wipf, Wittel, Woan, Worden, Wright, Wu,
  Yablon, Yam, Yamamoto, Yancey, Yap, Yu, Yvert, Zadrożny, Zangrando, Zanolin,
  Zendri, Zevin, Zhang, Zhang, Zhang, Zhang, Zhao, Zhou, Zhou, Zhu, Zucker,
  Zuraw, Zweizig, Collaboration, \& Collaboration}]{LVC:16b}
Abbott, B.~P., Abbott, R., Abbott, T.~D., {et~al.} 2016, \apjl, 818, L22

\bibitem[{{Abbott}(1982)}]{abbott:82}
{Abbott}, D.~C. 1982, \apj, 259, 282

\bibitem[{{Almeida} {et~al.}(2016){Almeida}, {Sana}, {Taylor}, {Barb{\'a}},
  {Bonanos}, {Crowther}, {Damineli}, {de Koter}, {de Mink}, {Evans}, {Gieles},
  {Grin}, {H{\'e}nault-Brunet}, {Langer}, {Lennon}, {Lockwood}, {Ma{\'{\i}}z
  Apell{\'a}niz}, {Moffat}, {Neijssel}, {Norman}, {Ram{\'{\i}}rez-Agudelo},
  {Richardson}, {Schootemeijer}, {Shenar}, {Soszy{\'n}ski}, {Tramper}, \&
  {Vink}}]{almeida:16}
{Almeida}, L.~A., {Sana}, H., {Taylor}, W., {et~al.} 2016, accepted for
  publication in A\&A, arXiv:1610.03500

\bibitem[{{Arnett}(2015)}]{arnett:15}
{Arnett}, W.~D. 2015, in IAU Symposium, Vol. 307, New Windows on Massive Stars,
  ed. G.~{Meynet}, C.~{Georgy}, J.~{Groh}, \& P.~{Stee}, 459--469

\bibitem[{{Arnett} \& {Meakin}(2016)}]{arnett:16}
{Arnett}, W.~D. \& {Meakin}, C. 2016, Reports on Progress in Physics, 79,
  102901

\bibitem[{{Arnett} {et~al.}(2015){Arnett}, {Meakin}, {Viallet}, {Campbell},
  {Lattanzio}, \& {Moc{\'a}k}}]{arnett:15b}
{Arnett}, W.~D., {Meakin}, C., {Viallet}, M., {et~al.} 2015, \apj, 809, 30

\bibitem[{{Belczynski} {et~al.}(2010){Belczynski}, {Bulik}, {Fryer}, {Ruiter},
  {Valsecchi}, {Vink}, \& {Hurley}}]{belczynski:10}
{Belczynski}, K., {Bulik}, T., {Fryer}, C.~L., {et~al.} 2010, \apj, 714, 1217

\bibitem[{{Bennett}(2010)}]{bennett:10}
{Bennett}, P.~D. 2010, in Astronomical Society of the Pacific Conference
  Series, Vol. 425, Hot and Cool: Bridging Gaps in Massive Star Evolution, ed.
  C.~{Leitherer}, P.~D. {Bennett}, P.~W. {Morris}, \& J.~T. {Van Loon}, 181

\bibitem[{{Bestenlehner} {et~al.}(2011){Bestenlehner}, {Vink}, {Gr{\"a}fener},
  {Najarro}, {Evans}, {Bastian}, {Bonanos}, {Bressert}, {Crowther}, {Doran},
  {Friedrich}, {H{\'e}nault-Brunet}, {Herrero}, {de Koter}, {Langer}, {Lennon},
  {Ma{\'{\i}}z Apell{\'a}niz}, {Sana}, {Soszynski}, \&
  {Taylor}}]{bestenlehner:11}
{Bestenlehner}, J.~M., {Vink}, J.~S., {Gr{\"a}fener}, G., {et~al.} 2011, \aap,
  530, L14

\bibitem[{{Blaauw}(1961)}]{blaauw:61}
{Blaauw}, A. 1961, \bain, 15, 265

\bibitem[{{Boersma}(1961)}]{boersma:61}
{Boersma}, J. 1961, \bain, 15, 291

\bibitem[{{B{\"o}hm-Vitense}(1958)}]{bohmvitense:58}
{B{\"o}hm-Vitense}, E. 1958, \zap, 46, 108

\bibitem[{{Bouret} {et~al.}(2005){Bouret}, {Lanz}, \& {Hillier}}]{bouret:05}
{Bouret}, J., {Lanz}, T., \& {Hillier}, D.~J. 2005, \aap, 438, 301

\bibitem[{{Castor} {et~al.}(1975){Castor}, {Abbott}, \& {Klein}}]{castor:75}
{Castor}, J.~I., {Abbott}, D.~C., \& {Klein}, R.~I. 1975, \apj, 195, 157

\bibitem[{{Chakraborti} {et~al.}(2016){Chakraborti}, {Ray}, {Smith},
  {Margutti}, {Pooley}, {Bose}, {Sutaria}, {Chandra}, {Dwarkadas}, {Ryder}, \&
  {Maeda}}]{chakraborti:16}
{Chakraborti}, S., {Ray}, A., {Smith}, R., {et~al.} 2016, \apj, 817, 22

\bibitem[{{Chatzopoulos} {et~al.}(2016){Chatzopoulos}, {Couch}, {Arnett}, \&
  {Timmes}}]{chatzopoulos:16}
{Chatzopoulos}, E., {Couch}, S.~M., {Arnett}, W.~D., \& {Timmes}, F.~X. 2016,
  \apj, 822, 61

\bibitem[{{Chini} {et~al.}(2012){Chini}, {Hoffmeister}, {Nasseri}, {Stahl}, \&
  {Zinnecker}}]{chini:12}
{Chini}, R., {Hoffmeister}, V.~H., {Nasseri}, A., {Stahl}, O., \& {Zinnecker},
  H. 2012, \mnras, 424, 1925

\bibitem[{{Clausen} {et~al.}(2015){Clausen}, {Piro}, \& {Ott}}]{clausen:15}
{Clausen}, D., {Piro}, A.~L., \& {Ott}, C.~D. 2015, \apj, 799, 190

\bibitem[{{Conti}(1975)}]{conti:75}
{Conti}, P.~S. 1975, Memoires of the Societe Royale des Sciences de Liege, 9,
  193

\bibitem[{{Couch} \& {Ott}(2015)}]{couch:15}
{Couch}, S.~M. \& {Ott}, C.~D. 2015, \apj, 799, 5

\bibitem[{{Crowther}(2007)}]{crowther:07}
{Crowther}, P.~A. 2007, \araa, 45, 177

\bibitem[{{Crowther} {et~al.}(2002){Crowther}, {Hillier}, {Evans}, {Fullerton},
  {De Marco}, \& {Willis}}]{crowther:02}
{Crowther}, P.~A., {Hillier}, D.~J., {Evans}, C.~J., {et~al.} 2002, \apj, 579,
  774

\bibitem[{{Crowther} {et~al.}(2010){Crowther}, {Schnurr}, {Hirschi}, {Yusof},
  {Parker}, {Goodwin}, \& {Kassim}}]{crowther:10}
{Crowther}, P.~A., {Schnurr}, O., {Hirschi}, R., {et~al.} 2010, \mnras, 408,
  731

\bibitem[{{de Jager} {et~al.}(1988){de Jager}, {Nieuwenhuijzen}, \& {van der
  Hucht}}]{dejager:88}
{de Jager}, C., {Nieuwenhuijzen}, H., \& {van der Hucht}, K.~A. 1988, \aaps,
  72, 259

\bibitem[{{de Koter} {et~al.}(1997){de Koter}, {Heap}, \&
  {Hubeny}}]{dekoter:97}
{de Koter}, A., {Heap}, S.~R., \& {Hubeny}, I. 1997, \apj, 477, 792

\bibitem[{{Dessart} {et~al.}(2013){Dessart}, {Hillier}, {Waldman}, \&
  {Livne}}]{dessart:13}
{Dessart}, L., {Hillier}, D.~J., {Waldman}, R., \& {Livne}, E. 2013, \mnras,
  433, 1745

\bibitem[{{Dessart} \& {Owocki}(2005)}]{dessart:05}
{Dessart}, L. \& {Owocki}, S.~P. 2005, \aap, 437, 657

\bibitem[{Eggenberger {et~al.}(2007)Eggenberger, Meynet, Maeder, Hirschi,
  Charbonnel, Talon, \& Ekstr\"{o}m}]{eggenberger:07}
Eggenberger, P., Meynet, G., Maeder, a., {et~al.} 2007, Astrophysics and Space
  Science, 316, 43

\bibitem[{{Ekstr{\"o}m} {et~al.}(2008){Ekstr{\"o}m}, {Meynet}, \&
  {Maeder}}]{ekstrom:08}
{Ekstr{\"o}m}, S., {Meynet}, G., \& {Maeder}, A. 2008, in American Institute of
  Physics Conference Series, Vol. 990, First Stars III, ed. B.~W. {O'Shea} \&
  A.~{Heger}, 220--224

\bibitem[{{Eldridge} \& {Tout}(2004)}]{eldridge:04}
{Eldridge}, J.~J. \& {Tout}, C.~A. 2004, \mnras, 353, 87

\bibitem[{Eldridge \& Vink(2006)}]{eldridge:06}
Eldridge, J.~J. \& Vink, J.~S. 2006, A\&A, 452, 295

\bibitem[{{Ertl} {et~al.}(2016){Ertl}, {Janka}, {Woosley}, {Sukhbold}, \&
  {Ugliano}}]{ertl:16}
{Ertl}, T., {Janka}, H.-T., {Woosley}, S.~E., {Sukhbold}, T., \& {Ugliano}, M.
  2016, \apj, 818, 124

\bibitem[{{Evans} {et~al.}(2004){Evans}, {Crowther}, {Fullerton}, \&
  {Hillier}}]{evans:04}
{Evans}, C.~J., {Crowther}, P.~A., {Fullerton}, A.~W., \& {Hillier}, D.~J.
  2004, \apj, 610, 1021

\bibitem[{{Farmer} {et~al.}(2016){Farmer}, {Fields}, {Petermann}, {Dessart},
  {Cantiello}, {Paxton}, \& {Timmes}}]{farmer:16}
{Farmer}, R., {Fields}, C.~E., {Petermann}, I., {et~al.} 2016, \apjs, 227, 22

\bibitem[{{Feldmeier}(1995)}]{feldmeier:95}
{Feldmeier}, A. 1995, \aap, 299, 523

\bibitem[{{Ferrarotti} \& {Gail}(2006)}]{ferrarotti:06}
{Ferrarotti}, A.~S. \& {Gail}, H.-P. 2006, \aap, 447, 553

\bibitem[{{Fibonacci}(1202)}]{Fibonacci}
{Fibonacci}, L. 1202, {Liber Abbaci}

\bibitem[{{Frischknecht} {et~al.}(2016){Frischknecht}, {Hirschi}, {Pignatari},
  {Maeder}, {Meynet}, {Chiappini}, {Thielemann}, {Rauscher}, {Georgy}, \&
  {Ekstr{\"o}m}}]{frischknecht:16}
{Frischknecht}, U., {Hirschi}, R., {Pignatari}, M., {et~al.} 2016, \mnras, 456,
  1803

\bibitem[{{Fryer} {et~al.}(2012){Fryer}, {Belczynski}, {Wiktorowicz},
  {Dominik}, {Kalogera}, \& {Holz}}]{fryer:12}
{Fryer}, C.~L., {Belczynski}, K., {Wiktorowicz}, G., {et~al.} 2012, \apj, 749,
  91

\bibitem[{{Fullerton} {et~al.}(2006){Fullerton}, {Massa}, \&
  {Prinja}}]{fullerton:06}
{Fullerton}, A.~W., {Massa}, D.~L., \& {Prinja}, R.~K. 2006, \apj, 637, 1025

\bibitem[{{Georgy}(2012)}]{georgy:12}
{Georgy}, C. 2012, \aap, 538, L8

\bibitem[{{Georgy} {et~al.}(2015){Georgy}, {Ekstr{\"o}m}, {Hirschi}, {Meynet},
  {Groh}, \& {Eggenberger}}]{georgy:15}
{Georgy}, C., {Ekstr{\"o}m}, S., {Hirschi}, R., {et~al.} 2015, in Wolf-Rayet
  Stars: Proceedings of an International Workshop held in Potsdam, Germany, 1-5
  June 2015. Edited by Wolf-Rainer Hamann, Andreas Sander, Helge Todt.
  Universit{\"a}tsverlag Potsdam, 2015., p.229-232, ed. W.-R. {Hamann},
  A.~{Sander}, \& H.~{Todt}, 229--232

\bibitem[{{Georgy} {et~al.}(2016){Georgy}, {Saio}, {Ekstr{\"o}m}, \&
  {Meynet}}]{georgy:16}
{Georgy}, C., {Saio}, H., {Ekstr{\"o}m}, S., \& {Meynet}, G. 2016, Proceedings
  of the conference 'The B[e] Phenomenom: Forty Years of Studies',
  arxiv:1610.07332

\bibitem[{{Glebbeek} {et~al.}(2009){Glebbeek}, {Gaburov}, {de Mink}, {Pols}, \&
  {Portegies Zwart}}]{glebbeek:09}
{Glebbeek}, E., {Gaburov}, E., {de Mink}, S.~E., {Pols}, O.~R., \& {Portegies
  Zwart}, S.~F. 2009, \aap, 497, 255

\bibitem[{{Gordon} {et~al.}(2016){Gordon}, {Humphreys}, \& {Jones}}]{gordon:16}
{Gordon}, M.~S., {Humphreys}, R.~M., \& {Jones}, T.~J. 2016, \apj, 825, 50

\bibitem[{{Gr{\"a}fener} \& {Hamann}(2008)}]{grafener:08}
{Gr{\"a}fener}, G. \& {Hamann}, W.-R. 2008, \aap, 482, 945

\bibitem[{{Groh} {et~al.}(2013){Groh}, {Meynet}, \& {Ekstr{\"o}m}}]{groh:13}
{Groh}, J.~H., {Meynet}, G., \& {Ekstr{\"o}m}, S. 2013, \aap, 550, L7

\bibitem[{{Groh} {et~al.}(2014){Groh}, {Meynet}, {Ekstr{\"o}m}, \&
  {Georgy}}]{groh:14}
{Groh}, J.~H., {Meynet}, G., {Ekstr{\"o}m}, S., \& {Georgy}, C. 2014, \aap,
  564, A30

\bibitem[{{Gull} \& {Sofia}(1979)}]{gull:79}
{Gull}, T.~R. \& {Sofia}, S. 1979, \apj, 230, 782

\bibitem[{{Hamann} \& {Koesterke}(1998)}]{hamann:98}
{Hamann}, W.~R. \& {Koesterke}, L. 1998, \aap, 335, 1003

\bibitem[{{Hamann} {et~al.}(1995){Hamann}, {Koesterke}, \&
  {Wessolowski}}]{hamann:95}
{Hamann}, W.-R., {Koesterke}, L., \& {Wessolowski}, U. 1995, \aap, 299, 151

\bibitem[{{Hamann} {et~al.}(1982){Hamann}, {Schoenberner}, \&
  {Heber}}]{hamann:82}
{Hamann}, W.-R., {Schoenberner}, D., \& {Heber}, U. 1982, \aap, 116, 273

\bibitem[{{Heger} {et~al.}(2005){Heger}, {Woosley}, \& {Spruit}}]{heger:05}
{Heger}, A., {Woosley}, S.~E., \& {Spruit}, H.~C. 2005, \apj, 626, 350

\bibitem[{{Hillier} {et~al.}(2003){Hillier}, {Lanz}, {Heap}, {Hubeny}, {Smith},
  {Evans}, {Lennon}, \& {Bouret}}]{hillier:03}
{Hillier}, D.~J., {Lanz}, T., {Heap}, S.~R., {et~al.} 2003, \apj, 588, 1039

\bibitem[{{Hix} {et~al.}(2007){Hix}, {Parete-Koon}, {Freiburghaus}, \&
  {Thielemann}}]{hix:07}
{Hix}, W.~R., {Parete-Koon}, S.~T., {Freiburghaus}, C., \& {Thielemann}, F.-K.
  2007, \apj, 667, 476

\bibitem[{{Hix} \& {Thielemann}(1996)}]{hix:96}
{Hix}, W.~R. \& {Thielemann}, F.-K. 1996, \apj, 460, 869

\bibitem[{{Janka}(2013)}]{janka:13}
{Janka}, H.-T. 2013, \mnras, 434, 1355

\bibitem[{{Janka}(2016)}]{janka:16}
{Janka}, H.-T. 2016, submitted to ApJ, arXiv:1611.07562

\bibitem[{{Janka} {et~al.}(2012){Janka}, {Hanke}, {H{\"u}depohl}, {Marek},
  {M{\"u}ller}, \& {Obergaulinger}}]{janka:12}
{Janka}, H.-T., {Hanke}, F., {H{\"u}depohl}, L., {et~al.} 2012, Progress of
  Theoretical and Experimental Physics, 2012, 01A309

\bibitem[{{Joss} {et~al.}(1973){Joss}, {Salpeter}, \& {Ostriker}}]{joss:73}
{Joss}, P.~C., {Salpeter}, E.~E., \& {Ostriker}, J.~P. 1973, \apj, 181, 429

\bibitem[{{Khazov} {et~al.}(2016){Khazov}, {Yaron}, {Gal-Yam}, {Manulis},
  {Rubin}, {Kulkarni}, {Arcavi}, {Kasliwal}, {Ofek}, {Cao}, {Perley},
  {Sollerman}, {Horesh}, {Sullivan}, {Filippenko}, {Nugent}, {Howell}, {Cenko},
  {Silverman}, {Ebeling}, {Taddia}, {Johansson}, {Laher}, {Surace},
  {Rebbapragada}, {Wozniak}, \& {Matheson}}]{khazov:16}
{Khazov}, D., {Yaron}, O., {Gal-Yam}, A., {et~al.} 2016, \apj, 818, 3

\bibitem[{{Kiminki} \& {Kobulnicky}(2012)}]{kiminki:12}
{Kiminki}, D.~C. \& {Kobulnicky}, H.~A. 2012, \apj, 751, 4

\bibitem[{{Kippenhahn} {et~al.}(1980){Kippenhahn}, {Ruschenplatt}, \&
  {Thomas}}]{kippenhahn:80}
{Kippenhahn}, R., {Ruschenplatt}, G., \& {Thomas}, H.-C. 1980, \aap, 91, 175

\bibitem[{{Kobulnicky} {et~al.}(2014){Kobulnicky}, {Kiminki}, {Lundquist},
  {Burke}, {Chapman}, {Keller}, {Lester}, {Rolen}, {Topel}, {Bhattacharjee},
  {Smullen}, {Vargas {\'A}lvarez}, {Runnoe}, {Dale}, \&
  {Brotherton}}]{kobulnicky:14}
{Kobulnicky}, H.~A., {Kiminki}, D.~C., {Lundquist}, M.~J., {et~al.} 2014,
  \apjs, 213, 34

\bibitem[{{Kudritzki} {et~al.}(1989){Kudritzki}, {Pauldrach}, {Puls}, \&
  {Abbott}}]{kudritzki:89}
{Kudritzki}, R.~P., {Pauldrach}, A., {Puls}, J., \& {Abbott}, D.~C. 1989, \aap,
  219, 205

\bibitem[{{Lamers}(2013)}]{lamers:13}
{Lamers}, H.~J.~G.~L.~M. 2013, in Astronomical Society of the Pacific
  Conference Series, Vol. 470, 370 Years of Astronomy in Utrecht, ed.
  G.~{Pugliese}, A.~{de Koter}, \& M.~{Wijburg}, 97

\bibitem[{{Lamers} \& {Cassinelli}(1999)}]{lamers:99}
{Lamers}, H.~J.~G.~L.~M. \& {Cassinelli}, J.~P. 1999, {Introduction to Stellar
  Winds} (Cambridge University Press, Cambridge, UK)

\bibitem[{{Langer}(1997)}]{langer:97}
{Langer}, N. 1997, in Astronomical Society of the Pacific Conference Series,
  Vol. 120, Luminous Blue Variables: Massive Stars in Transition, ed. A.~{Nota}
  \& H.~{Lamers}, 83

\bibitem[{{Langer}(2012)}]{langer:12}
{Langer}, N. 2012, \araa, 50, 107

\bibitem[{{Ledoux}(1947)}]{ledoux:47}
{Ledoux}, P. 1947, \apj, 105, 305

\bibitem[{{Limongi} \& {Chieffi}(2006)}]{limongi:06}
{Limongi}, M. \& {Chieffi}, A. 2006, \apj, 647, 483

\bibitem[{{Lovegrove} \& {Woosley}(2013)}]{lovegrove:13}
{Lovegrove}, E. \& {Woosley}, S.~E. 2013, \apj, 769, 109

\bibitem[{{Lucy} \& {Solomon}(1970)}]{lucy:70}
{Lucy}, L.~B. \& {Solomon}, P.~M. 1970, \apj, 159, 879

\bibitem[{{Lucy} \& {White}(1980)}]{lucy:80}
{Lucy}, L.~B. \& {White}, R.~L. 1980, \apj, 241, 300

\bibitem[{{Maeda} {et~al.}(2015){Maeda}, {Hattori}, {Milisavljevic},
  {Folatelli}, {Drout}, {Kuncarayakti}, {Margutti}, {Kamble}, {Soderberg},
  {Tanaka}, {Kawabata}, {Kawabata}, {Yamanaka}, {Nomoto}, {Kim}, {Simon},
  {Phillips}, {Parrent}, {Nakaoka}, {Moriya}, {Suzuki}, {Takaki}, {Ishigaki},
  {Sakon}, {Tajitsu}, \& {Iye}}]{maeda:15}
{Maeda}, K., {Hattori}, T., {Milisavljevic}, D., {et~al.} 2015, \apj, 807, 35

\bibitem[{{Maeder}(1992)}]{maeder:92}
{Maeder}, A. 1992, \aap, 264, 105

\bibitem[{{Maeder}(1996)}]{maeder:96}
{Maeder}, A. 1996, in Liege International Astrophysical Colloquia, Vol.~33,
  Liege International Astrophysical Colloquia, ed. J.~M. {Vreux}, A.~{Detal},
  D.~{Fraipont-Caro}, E.~{Gosset}, \& G.~{Rauw}, 39

\bibitem[{{Maeder} \& {Conti}(1994)}]{maeder:94}
{Maeder}, A. \& {Conti}, P.~S. 1994, \araa, 32, 227

\bibitem[{{Maeder} \& {Meynet}(1988)}]{maeder:88}
{Maeder}, A. \& {Meynet}, G. 1988, \aaps, 76, 411

\bibitem[{{Maeder} \& {Meynet}(1989)}]{maeder:89}
{Maeder}, A. \& {Meynet}, G. 1989, \aap, 210, 155

\bibitem[{{Marchenko} {et~al.}(2010){Marchenko}, {Moffat}, \&
  {Crowther}}]{marchenko:10}
{Marchenko}, S.~V., {Moffat}, A.~F.~J., \& {Crowther}, P.~A. 2010, \apjl, 724,
  L90

\bibitem[{{Margutti} {et~al.}(2017){Margutti}, {Kamble}, {Milisavljevic},
  {Zapartas}, {de Mink}, {Drout}, {Chornock}, {Risaliti}, {Zauderer},
  {Bietenholz}, {Cantiello}, {Chakraborti}, {Chomiuk}, {Fong}, {Grefenstette},
  {Guidorzi}, {Kirshner}, {Parrent}, {Patnaude}, {Soderberg}, {Gehrels}, \&
  {Harrison}}]{margutti:17}
{Margutti}, R., {Kamble}, A., {Milisavljevic}, D., {et~al.} 2017, \apj, 835,
  140

\bibitem[{{Mason} {et~al.}(2009){Mason}, {Hartkopf}, {Gies}, {Henry}, \&
  {Helsel}}]{mason:09}
{Mason}, B.~D., {Hartkopf}, W.~I., {Gies}, D.~R., {Henry}, T.~J., \& {Helsel},
  J.~W. 2009, \aj, 137, 3358

\bibitem[{{Mauron} \& {Josselin}(2011)}]{mauron:11}
{Mauron}, N. \& {Josselin}, E. 2011, \aap, 526, A156

\bibitem[{{Meyer} {et~al.}(2016){Meyer}, {van Marle}, {Kuiper}, \&
  {Kley}}]{meyer:16}
{Meyer}, D.~M.-A., {van Marle}, A.-J., {Kuiper}, R., \& {Kley}, W. 2016,
  \mnras, 459, 1146

\bibitem[{{Meynet} {et~al.}(2015){Meynet}, {Chomienne}, {Ekstr{\"o}m},
  {Georgy}, {Granada}, {Groh}, {Maeder}, {Eggenberger}, {Levesque}, \&
  {Massey}}]{meynet:15}
{Meynet}, G., {Chomienne}, V., {Ekstr{\"o}m}, S., {et~al.} 2015, \aap, 575, A60

\bibitem[{{Meynet} \& {Maeder}(2003)}]{meynet:03}
{Meynet}, G. \& {Maeder}, A. 2003, \aap, 404, 975

\bibitem[{{Meynet} {et~al.}(1994){Meynet}, {Maeder}, {Schaller}, {Schaerer}, \&
  {Charbonnel}}]{meynet:94}
{Meynet}, G., {Maeder}, A., {Schaller}, G., {Schaerer}, D., \& {Charbonnel}, C.
  1994, \aaps, 103

\bibitem[{{Moravveji} {et~al.}(2016){Moravveji}, {Townsend}, {Aerts}, \&
  {Mathis}}]{moravveji:16}
{Moravveji}, E., {Townsend}, R.~H.~D., {Aerts}, C., \& {Mathis}, S. 2016, \apj,
  823, 130

\bibitem[{{Morozova} {et~al.}(2015){Morozova}, {Piro}, {Renzo}, {Ott},
  {Clausen}, {Couch}, {Ellis}, \& {Roberts}}]{morozova:15}
{Morozova}, V., {Piro}, A.~L., {Renzo}, M., {et~al.} 2015, \apj, 814, 63

\bibitem[{{Nieuwenhuijzen} \& {de Jager}(1990)}]{nieuwenhuijzen:90}
{Nieuwenhuijzen}, H. \& {de Jager}, C. 1990, \aap, 231, 134

\bibitem[{{Nugis} \& {Lamers}(2000)}]{nugis:00}
{Nugis}, T. \& {Lamers}, H.~J.~G.~L.~M. 2000, \aap, 360, 227

\bibitem[{{O'Connor} \& {Ott}(2011)}]{oconnor:11}
{O'Connor}, E. \& {Ott}, C.~D. 2011, \apj, 730, 70

\bibitem[{{O'Connor} \& {Ott}(2013)}]{oconnor:13}
{O'Connor}, E. \& {Ott}, C.~D. 2013, \apj, 762, 126

\bibitem[{{Owocki} {et~al.}(1988){Owocki}, {Castor}, \& {Rybicki}}]{owocki:88}
{Owocki}, S.~P., {Castor}, J.~I., \& {Rybicki}, G.~B. 1988, \apj, 335, 914

\bibitem[{{Owocki} \& {Puls}(1999)}]{owocki:99}
{Owocki}, S.~P. \& {Puls}, J. 1999, \apj, 510, 355

\bibitem[{{Owocki} \& {Rybicki}(1984)}]{owocki:84}
{Owocki}, S.~P. \& {Rybicki}, G.~B. 1984, \apj, 284, 337

\bibitem[{{Pauldrach} {et~al.}(1986){Pauldrach}, {Puls}, \&
  {Kudritzki}}]{pauldrach:86}
{Pauldrach}, A., {Puls}, J., \& {Kudritzki}, R.~P. 1986, \aap, 164, 86

\bibitem[{{Pauldrach} {et~al.}(1994){Pauldrach}, {Kudritzki}, {Puls}, {Butler},
  \& {Hunsinger}}]{pauldrach:94}
{Pauldrach}, A.~W.~A., {Kudritzki}, R.~P., {Puls}, J., {Butler}, K., \&
  {Hunsinger}, J. 1994, \aap, 283, 525

\bibitem[{{Paxton} {et~al.}(2011){Paxton}, {Bildsten}, {Dotter}, {Herwig},
  {Lesaffre}, \& {Timmes}}]{paxton:11}
{Paxton}, B., {Bildsten}, L., {Dotter}, A., {et~al.} 2011, \apjs, 192, 3

\bibitem[{{Paxton} {et~al.}(2013){Paxton}, {Cantiello}, {Arras}, {Bildsten},
  {Brown}, {Dotter}, {Mankovich}, {Montgomery}, {Stello}, {Timmes}, \&
  {Townsend}}]{paxton:13}
{Paxton}, B., {Cantiello}, M., {Arras}, P., {et~al.} 2013, \apjs, 208, 4

\bibitem[{{Paxton} {et~al.}(2015){Paxton}, {Marchant}, {Schwab}, {Bauer},
  {Bildsten}, {Cantiello}, {Dessart}, {Farmer}, {Hu}, {Langer}, {Townsend},
  {Townsley}, \& {Timmes}}]{paxton:15}
{Paxton}, B., {Marchant}, P., {Schwab}, J., {et~al.} 2015, \apjs, 220, 15

\bibitem[{{Petit} {et~al.}(2017){Petit}, {Keszthelyi}, {MacInnis}, {Cohen},
  {Townsend}, {Wade}, {Thomas}, {Owocki}, {Puls}, \& {ud-Doula}}]{petit:17}
{Petit}, V., {Keszthelyi}, Z., {MacInnis}, R., {et~al.} 2017, \mnras, 466, 1052

\bibitem[{{Puls} {et~al.}(2015){Puls}, {Sundqvist}, \& {Markova}}]{puls:15}
{Puls}, J., {Sundqvist}, J.~O., \& {Markova}, N. 2015, in IAU Symposium, Vol.
  307, New Windows on Massive Stars, ed. G.~{Meynet}, C.~{Georgy}, J.~{Groh},
  \& P.~{Stee}, 25--36

\bibitem[{{Puls} {et~al.}(2008){Puls}, {Vink}, \& {Najarro}}]{puls:08}
{Puls}, J., {Vink}, J.~S., \& {Najarro}, F. 2008, \aap, 16, 209

\bibitem[{{Quataert} {et~al.}(2016){Quataert}, {Fern{\'a}ndez}, {Kasen},
  {Klion}, \& {Paxton}}]{quataert:16}
{Quataert}, E., {Fern{\'a}ndez}, R., {Kasen}, D., {Klion}, H., \& {Paxton}, B.
  2016, \mnras, 458, 1214

\bibitem[{{Quataert} \& {Shiode}(2012)}]{quataert:12}
{Quataert}, E. \& {Shiode}, J. 2012, \mnras, 423, L92

\bibitem[{{Renzo}(2015)}]{thesis}
{Renzo}, M. 2015, Master's thesis, Universit\`a di Pisa, Italy,
  \url{https://etd.adm.unipi.it/t/etd-05062015-125630/}

\bibitem[{{Sana} {et~al.}(2012){Sana}, {de Mink}, {de Koter}, {Langer},
  {Evans}, {Gieles}, {Gosset}, {Izzard}, {Le Bouquin}, \&
  {Schneider}}]{sana:12}
{Sana}, H., {de Mink}, S.~E., {de Koter}, A., {et~al.} 2012, Science, 337, 444

\bibitem[{{Sana} \& {Evans}(2011)}]{sana:11}
{Sana}, H. \& {Evans}, C.~J. 2011, in IAU Symposium, Vol. 272, Active OB Stars:
  Structure, Evolution, Mass Loss, and Critical Limits, ed. C.~{Neiner},
  G.~{Wade}, G.~{Meynet}, \& G.~{Peters}, 474--485

\bibitem[{{Schmutz} \& {Drissen}(1999)}]{schmutz:99}
{Schmutz}, W. \& {Drissen}, L. 1999, in Revista Mexicana de Astronomia y
  Astrofisica, vol. 27, Vol.~8, Revista Mexicana de Astronomia y Astrofisica
  Conference Series, ed. N.~I. {Morrell}, V.~S. {Niemela}, \& R.~H.
  {Barb{\'a}}, 41--48

\bibitem[{{Shara} {et~al.}(2017){Shara}, {Crawford}, {Vanbeveren}, {Moffat},
  {Zurek}, \& {Crause}}]{shara:17}
{Shara}, M.~M., {Crawford}, S.~M., {Vanbeveren}, D., {et~al.} 2017, \mnras,
  464, 2066

\bibitem[{{Shiode} \& {Quataert}(2014)}]{shiode:14}
{Shiode}, J.~H. \& {Quataert}, E. 2014, \apj, 780, 96

\bibitem[{{Smartt}(2009)}]{smartt:09review}
{Smartt}, S.~J. 2009, \araa, 47, 63

\bibitem[{{Smartt} {et~al.}(2009){Smartt}, {Eldridge}, {Crockett}, \&
  {Maund}}]{smartt:09}
{Smartt}, S.~J., {Eldridge}, J.~J., {Crockett}, R.~M., \& {Maund}, J.~R. 2009,
  \mnras, 395, 1409

\bibitem[{{Smith}(2014)}]{smith:14}
{Smith}, N. 2014, \araa, 52, 487

\bibitem[{{Smith} \& {Arnett}(2014)}]{smith:14b}
{Smith}, N. \& {Arnett}, W.~D. 2014, \apj, 785, 82

\bibitem[{{Smith} {et~al.}(2011){Smith}, {Li}, {Filippenko}, \&
  {Chornock}}]{smith:11}
{Smith}, N., {Li}, W., {Filippenko}, A.~V., \& {Chornock}, R. 2011, \mnras,
  412, 1522

\bibitem[{{Smith} \& {Owocki}(2006)}]{smith:06}
{Smith}, N. \& {Owocki}, S.~P. 2006, \apjl, 645, L45

\bibitem[{{Smith} \& {Tombleson}(2015)}]{smith:15b}
{Smith}, N. \& {Tombleson}, R. 2015, \mnras, 447, 598

\bibitem[{{Sukhbold} {et~al.}(2016){Sukhbold}, {Ertl}, {Woosley}, {Brown}, \&
  {Janka}}]{sukhbold:16}
{Sukhbold}, T., {Ertl}, T., {Woosley}, S.~E., {Brown}, J.~M., \& {Janka}, H.-T.
  2016, \apj, 821, 38

\bibitem[{{Sukhbold} \& {Woosley}(2014)}]{sukhbold:14}
{Sukhbold}, T. \& {Woosley}, S.~E. 2014, \apj, 783, 10

\bibitem[{{Tramper} {et~al.}(2016){Tramper}, {Sana}, \& {de
  Koter}}]{tramper:16}
{Tramper}, F., {Sana}, H., \& {de Koter}, A. 2016, \apj, 833, 133

\bibitem[{{Ugliano} {et~al.}(2012){Ugliano}, {Janka}, {Marek}, \&
  {Arcones}}]{ugliano:12}
{Ugliano}, M., {Janka}, H.-T., {Marek}, A., \& {Arcones}, A. 2012, \apj, 757,
  69

\bibitem[{{van der Hucht}(2001)}]{vanderhutch:01}
{van der Hucht}, K.~A. 2001, \nar, 45, 135

\bibitem[{{van Loon} {et~al.}(2005){van Loon}, {Cioni}, {Zijlstra}, \&
  {Loup}}]{vanloon:05}
{van Loon}, J.~T., {Cioni}, M.-R.~L., {Zijlstra}, A.~A., \& {Loup}, C. 2005,
  \aap, 438, 273

\bibitem[{{Vink}(2015)}]{vink:14}
{Vink}, J.~S. 2015, {Mass-Loss Rates of Very Massive Stars}, In: Very Massive
  Stars in the Local Universe, ed. J.~S.~Vink (Springer Verlag, Heidelberg,
  Germany), 77

\bibitem[{{Vink} {et~al.}(2000){Vink}, {de Koter}, \& {Lamers}}]{vink:00}
{Vink}, J.~S., {de Koter}, A., \& {Lamers}, H.~J.~G.~L.~M. 2000, \aap, 362, 295

\bibitem[{{Vink} {et~al.}(2001){Vink}, {de Koter}, \& {Lamers}}]{vink:01}
{Vink}, J.~S., {de Koter}, A., \& {Lamers}, H.~J.~G.~L.~M. 2001, \aap, 369, 574

\bibitem[{{Wachter} {et~al.}(2002){Wachter}, {Schr{\"o}der}, {Winters},
  {Arndt}, \& {Sedlmayr}}]{wachter:02}
{Wachter}, A., {Schr{\"o}der}, K.-P., {Winters}, J.~M., {Arndt}, T.~U., \&
  {Sedlmayr}, E. 2002, \aap, 384, 452

\bibitem[{{Wellstein} \& {Langer}(1999)}]{wellstein:99}
{Wellstein}, S. \& {Langer}, N. 1999, \aap, 350, 148

\bibitem[{{Woosley}(2016)}]{woosley:16b}
{Woosley}, S.~E. 2016, accepted for publication in ApJ, arXiv:1608.08939

\bibitem[{{Woosley} \& {Heger}(2007)}]{woosley:07}
{Woosley}, S.~E. \& {Heger}, A. 2007, \physrep, 442, 269

\bibitem[{{Woosley} {et~al.}(2002){Woosley}, {Heger}, \& {Weaver}}]{woosley:02}
{Woosley}, S.~E., {Heger}, A., \& {Weaver}, T.~A. 2002, Rev. Mod. Phys., 74,
  1015

\bibitem[{{Woosley} {et~al.}(1995){Woosley}, {Langer}, \&
  {Weaver}}]{woosley:95}
{Woosley}, S.~E., {Langer}, N., \& {Weaver}, T.~A. 1995, \apj, 448, 315

\bibitem[{{Yoon} {et~al.}(2010){Yoon}, {Woosley}, \& {Langer}}]{yoon:10b}
{Yoon}, S.-C., {Woosley}, S.~E., \& {Langer}, N. 2010, \apj, 725, 940

\bibitem[{{Zwicky}(1957)}]{zwicky:57}
{Zwicky}, F. 1957, \zap, 44, 64

\end{thebibliography}

\appendix
\section{Parametric Wind  Algorithms}
\label{app:schemes_review}

Here, we give a brief review of the physical assumptions entering in
each mass loss algorithm and the corresponding prescribed rate. We
summarize in \Tabref{tab:scaling} the scaling of the mass loss rate
with physical parameters of the stellar model for each mass loss algorithm
considered in this study.

\begin{table*}[!ht]
\setlength{\abovecaptionskip}{-0.1cm}
\setlength{\belowcaptionskip}{-0.2cm}
\begin{center}
\caption{Functional dependence of the mass loss rate $\dot{M}$ on
  stellar and wind parameters for each algorithm combination and mass
  loss phase.  $L$ is the luminosity, $M$ is the mass,
  $v_\mathrm{esc}$ is the escape velocity, $v_\infty$ is the final
  velocity of the wind, $v_\mathrm{th}$ is the thermal velocity,
  $T_\mathrm{eff}$ is the effective temperature, $R$ the radius,
  $\Gamma_\mathrm{E} = L/L_\mathrm{Edd}$ the Eddington ratio, and
  $X_s$, $Y_s$, and $Z_s$ are the surface hydrogen abundance, helium
  abundance, and metallicity, respectively.  The scalings are obtained
  from the algorithms described in Appendix~\ref{app:schemes_review},
  and errors and overall multiplying factors are omitted for the sake
  of brevity. See \Tabref{tab:comb} for the naming convention for mass
  loss algorithm combinations. \label{tab:scaling} }

\newcommand{\mc}[3]{\multicolumn{#1}{#2}{#3}}
\renewcommand{\arraystretch}{2.5}
\begin{tabular}{c|c|c|c|c|c|c}
\hline\hline
ID & V-dJ-NL & V-NJ-NL & V-vL-NL & V-dJ-H & V-NJ-H & V-vL-H \\\hline
\multirow{2}{*}{Hot} &
\mc{6}{l}{$L^{2.210}M^{-1.339}\left(\frac{v_\infty}{2v_\mathrm{esc}}\right)^{-1.601}T_\mathrm{eff}^{1.07}Z^{0.85}$\phantom{aaaaaaaaaaaaaaaa}
  \hfill if $T_\mathrm{eff} < 22.5 \ \mathrm{kK}$}  \\
 & \mc{6}{l}{$L^{2.194}M^{-1.313}\left(\frac{v_\infty}{2v_\mathrm{esc}}\right)^{-1.226}Z^{0.85}T_\mathrm{eff}^{(0.933-10.92\log_{10}(T_\mathrm{eff}/
     40 \mathrm{kK}))}$ \hfill if $T_\mathrm{eff} > 27.5 \ \mathrm{kK}$} \\\hline
Cool & $L^{1.769}T_\mathrm{eff}^{-1.676}$ & $L^{1.24}M^{0.16}R^{0.81}$
& $L^{1.05}T_\mathrm{eff}^{-6.3}$& $L^{1.769}T_\mathrm{eff}^{-1.676}$ & $L^{1.24}M^{0.16}R^{0.81}$
& $L^{1.05}T_\mathrm{eff}^{-6.3}$ \\\hline
 \multirow{2}{*}{WR}& \mc{3}{l}{\multirow{2}{*}{$L^{1.29}Y_s^{1.73}Z_s^{0.47}$}}\hfill &
 \mc{3}{|l}{$L^{1.5}10^{-2.85X_s}$ \hfill if $\log_{10}(L/L_\odot)>4.5$} \\
 & \mc{3}{l}{} & \mc{3}{|l}{$L^{6.8}$\phantom{aaaaaaaa} \hfill if $\log_{10}(L/L_\odot)\leq4.5
   $}\\\hline\hline
ID & K-dJ-NL & K-NJ-NL & K-vL-NL & K-dJ-H & K-NJ-H & K-vL-H \\\hline
Hot  &
\mc{6}{l}{$L^{1.779}v_\mathrm{th}^{-1.169}[GM(1-\Gamma_\mathrm{E})]^{0.610}$}\hfill\\\hline
Cool & $L^{1.769}T_\mathrm{eff}^{-1.676}$ & $L^{1.24}M^{0.16}R^{0.81}$
& $L^{1.05}T_\mathrm{eff}^{-6.3}$& $L^{1.769}T_\mathrm{eff}^{-1.676}$ & $L^{1.24}M^{0.16}R^{0.81}$
& $L^{1.05}T_\mathrm{eff}^{-6.3}$ \\\hline
 \multirow{2}{*}{WR}&  \mc{3}{l}{\multirow{2}{*}{$L^{1.29}Y_s^{1.73}Z_s^{0.47}$}}\hfill &
 \mc{3}{|l}{$L^{1.5}10^{-2.85X_s}$ \hfill if $\log_{10}(L/L_\odot)>4.5$}\\
 & \mc{3}{l}{} & \mc{3}{|l}{$L^{6.8}$\phantom{aaaaaaaa} \hfill if $\log_{10}(L/L_\odot)\leq4.5$}\\
\end{tabular}
\end{center}
\end{table*}

\subsection{Vink \emph{et al.} (V)}
\label{sec:vink}
The wind mass loss algorithm proposed by \cite{vink:00,vink:01} is
based on Monte Carlo simulations of the photon transport in the
stellar atmosphere to evaluate the radiative acceleration. Their
algorithm is explicitly metallicity dependent and is supposed to be
applied only to single OB stars
with metallicity $1/30\leq Z/Z_\odot\leq 3$ and effective temperature
$12\,500 \ \mathrm{K} < T_\mathrm{eff} < 50\,000 \ \mathrm{K}$, so during
the hot, blue evolutionary phase.  In this temperature range, a
non-monotonic behavior of $\dot{M}$ as a function of $T_\mathrm{eff}$
is expected (the so-called ``bi-stability jumps''): normally the lower
the temperature, the lower the mass loss rate (because the radiation
pressure is proportional to $T^4$). However, when the temperature
drops below $\sim 25\,000 \ \mathrm{K}$, the recombination of
$\mathrm{Fe \ IV} \rightarrow \mathrm{Fe \ III}$ provides a new ion
with an increased number of lines to drive the wind, resulting in an
increased mass loss rate at lower temperatures around
$T_\mathrm{eff}\sim 25\,000 \ \mathrm{K}$. This happens with different
iron ion recombinations also at
$T_\mathrm{eff}\sim 12\,000 \ \mathrm{K}$.

\cite{vink:00,vink:01} provide two different formulae for above and
below $T_\mathrm{eff} \sim 25\,000 \ \mathrm{K}$,
\begin{equation}
  \label{eq:vink_hot}
  \begin{aligned}
    \log_{10}(-\dot{M}) =
    -6.697(61)+2.194(21)\log_{10}\left(\frac{L}{10^5L_\odot}\right)+ \\
    -1.313(46)\log_{10}\left(\frac{M}{30M_\odot}\right)-1.226(37)\log_{10}\left(\frac{v_\infty}{2v_\mathrm{esc}}\right)+\\
    +0.933(64)\log_{10}\left(\frac{T_\mathrm{eff}}{40\,000 \
        \mathrm{K}}\right)-10.92(90)\log_{10}^2\left(\frac{T_\mathrm{eff}}{40\,000 \ \mathrm{K}}\right)+\\
    +0.85(10)\log_{10}\left(\frac{Z}{Z_\odot}\right)  \ \ ,
  \end{aligned}
\end{equation}
for $27\,500 \ \mathrm{K}< T_\mathrm{eff}\leq 50\,000 \ \mathrm{K}$, and
\begin{equation}
  \label{eq:vink_cool}
  \begin{aligned}
    \log_{10}(-\dot{M}) =
    -6.668(80)+2.210(31)\log_{10}\left(\frac{L}{10^5L_\odot}\right)+\\
    -1.339(68)\log_{10}\left(\frac{M}{30M_\odot}\right)-1.601(55)\log_{10}\left(\frac{v_\infty}{2v_\mathrm{esc}}\right)+ \\
    +1.07(10)\log_{10}\left(\frac{T_\mathrm{eff}}{40\,000 \ \mathrm{K}}\right) +0.85(10)\log_{10}\left(\frac{Z}{Z_\odot}\right)  \ \ ,
  \end{aligned}
\end{equation}
for $12 500 \ \mathrm{K}< T_\mathrm{eff} \leq 22500 \ \mathrm{K}$. In
between these two temperature ranges, it is common practice to simply
interpolate between the two formulae
(cf.~Appendix~\ref{app:MESA_technical}). The numbers in parenthesis
are the estimates of the error on the last digit reported, according
to \cite{vink:01}. These error estimates are usually neglected in
stellar evolutionary calculations.

\subsection{Kudritzki \emph{et al.} (K)}
\label{sec:kudr}
The mass loss algorithm proposed by \cite{kudritzki:89} is based on an
analytic solution of the equations for a stationary, isothermal,
spherically-symmetric, ideal (neither viscosity nor heat conduction)
gas flow with no magnetic fields and no rotation.  They include the
radiative acceleration $g_\mathrm{ph}$ in the momentum equation using
the standard parametrization of \cite{castor:75} \citep[see also
  Eq.~16 in][]{pauldrach:86},
\begin{equation}\label{eq:rad_acc}
  g_\mathrm{ph} = g_\mathrm{rad}^\mathrm{Th}
  \left(1+k\left(\frac{\sigma_\mathrm{Th}
        c_s}{dv/dr}\right)^{-\alpha}\left(\frac{2
        n_e}{\sqrt{1-(1-(R/r)^2}}\right)^\delta
    \mathrm{CF}\left(r,v,\frac{dv}{dr}, \alpha \right)\right) \ \ ,
\end{equation}
where $g_\mathrm{rad}^\mathrm{Th}$ is the radiative acceleration due
to Thomson scattering, and the second term in the brackets is the
so-called ``force multiplier'', i.e.\ the line acceleration in units
of $g_\mathrm{rad}^\mathrm{Th}$. It depends on the Thomson cross
section $\sigma_\mathrm{Th}$, the speed of sound $c_s$, the electron
number density $n_e$, and three free parameters $k$, $\alpha$, and
$\delta$. $k$ can be interpreted roughly as the number of lines strong
enough to have an effect, and $\alpha$ as the slope of the
distribution of the number of lines as a function of their strength.
The parameter $\delta$ and the correction factor $\mathrm{CF}$, which
is the ratio between the opacity as a function of the incoming angle
and the opacity in the radial direction, are used to include the
``finite cone-angle effect'' to account for photons travelling in
non-radial directions \citep[e.g.,][]{castor:75}. If $\delta$ and
$\mathrm{CF}$ were both equal to one, the parametrization would be
  valid only in the ``radial streaming'' limit
  \citep{abbott:82,pauldrach:86}, i.e.\ considering only incoming
    photons from the radial direction.  While this is a good
    approximation in the outer portion of the wind, where $r \gg R$
    and $R$ is the stellar radius, it is quite poor in the inner
    portion, where the mass loss rate is determined and photons
    traveling in non-radial directions can have a relevant effect. For
    completeness, we report the expression for the
    correction factor $\mathrm{CF}$ as a function of the dimensionless
    radial coordinate $x\udef r/R$ and $h \udef
    d\log_{10}(x)/d\log_{10}(v)$, cf. Eq.~(4) in \cite{kudritzki:89},
\begin{equation}
  \label{eq:CF}
  \mathrm{CF}\left(r,v,\frac{dv}{dr}, \alpha \right) = \frac{1}{\alpha+1}\frac{x^2}{1-h}\left( 1 -
    \left(1-\frac{1}{x^2}+\frac{h}{x^2} \right)^{\alpha+1}\right) \ \ .\
\end{equation}

We refer the reader to \cite{castor:75, pauldrach:86, vink:14} and
references therein for more details on this parametrization.

To find an analytic solution to their model, further assumptions are
needed. \cite{kudritzki:89} impose a ``$\beta$-law'' velocity field (a
common assumption found to be close to numerical solutions, see
\citealt{lamers:13}),

\begin{equation}
  \label{eq:beta_law}
  v(r) = v_\infty \left( 1- \frac{R}{r}\right)^\beta \ \ ,
\end{equation}
where $\beta$ is a free parameter assumed to be $\beta=1$, $v_\infty$
is the asymptotic velocity of the wind and $R$ is the stellar
radius. Note that this assumption means \cite{kudritzki:89} do not
solve for the dynamics of the system, but rather assume the velocity structure and
solve self-consistently for the density and the acceleration. In the
limit where $v \gg c_s$ (reasonable in the outer portion of the wind),
\cite{kudritzki:89} find a solution of the form

\begin{equation}
  \label{eq:kudritzki}
  \begin{aligned}
    \dot{M}\equiv
    \dot{M}(\alpha,\delta, k , M, L, v_{th}) =
    \tilde{D}(\alpha,\delta,v)\left(\frac{\sigma_\mathrm{Th}kL}{4 \pi c}\right)^{1/(\alpha-\delta)}\times \\
    \times \left(\frac{4\pi\alpha}{\sigma_\mathrm{Th}v_\mathrm{th}}\right)^{\alpha/(\alpha-\delta)}\left(\frac{1-\alpha}{GM(1-\Gamma)}\right)^{(1-\alpha)/(\alpha-\delta)} \ \ ,
  \end{aligned}
\end{equation}

where $v_\mathrm{th}$ is the thermal velocity of protons, $k$,
$\alpha$, and $\delta$ are the free parameters, $L$ is the luminosity,
$v_\mathrm{th}$ is the thermal velocity, $\Gamma = L/ L_\mathrm{Edd}$
is the Eddington ratio, and $\tilde{D}$ is a function of the free
parameters and the velocity \citep[cf.~Eqs.~47,\,62,\,65
  in][]{kudritzki:89}.

The main limitation of
\Eqref{eq:kudritzki} is that the parameters $k$, $\alpha$ and $\delta$
are not constants, but rather depend on the optical depth. The
numerical values commonly adopted are $\alpha=0.657$, $\beta=1$,
$\delta=0.095$, and $k=0.085$ and they are calibrated on $\zeta$
Puppis by \cite{pauldrach:94}. These should be interpreted as values
averaged over the optical depth. 

\subsection{de Jager \emph{et al.} (dJ)}
\label{sec:dejager}
The wind algorithm proposed by
\cite{dejager:88} is an empirical relationship of the form $\dot{M}
\equiv \dot{M}(T_\mathrm{eff}, L)$. This choice of variables uses only
observable quantities, making it easy to track the mass loss rate
while the star moves on the HR diagram.  This allows a better
understanding of how mass loss changes during stellar evolution. The
drawback is that no information about the physical mechanism driving
the wind is considered.

To formulate a reliable mass loss algorithm, \cite{dejager:88} collect
from the literature mass loss rates observed with different techniques
for a sample of galactic stars with spectral types from O to M. They
determine the ``average'' measured mass loss rate for stars observed
with multiple techniques and the deviation from this average for each
available measurement and for each star. Then, they define the
``average intrinsic error per determination'' as the one-sigma value
of the distribution of these deviations. They fit their entire data
sample with a sum of Chebychev polynomials of the first kind $T_n(x)=
\cos(n\ \mathrm{arccos}(x))$:

\begin{equation}
  \label{eq:complete_fit}
  \log_{10}(-\dot{M}) = \sum_{n=0}^N\sum_{\stackrel{i=0}{j=n-i}}^{i=n} a_{ij}T_i(\log_{10}(T_\mathrm{eff}))\cdot T_j(\log_{10}(L)) \ \ .
\end{equation}

What is commonly used in stellar evolution codes is the first-order
approximation to \Eqref{eq:complete_fit},

\begin{equation}
  \label{eq:dejager}
  \log_{10}(-\dot{M}) = 1.769\log_{10}(L/L_\odot)-1.676\log_{10}(T_\mathrm{eff}/[\mathrm{K}])-8.158 \ \ .
\end{equation}

To asses the quality of this fit, they derive the distribution of the
differences between the observed values of $\log_{10}(-\dot{M})$ with
the result of \Eqref{eq:dejager}. The standard deviation of this
distribution is $\sim 0.45$, slightly larger than the "averaged
intrinsic error per determination" previously determined. This
indicates that quantities other than $T_\mathrm{eff}$ and $L$ must
physically enter in the determination of $\dot{M}$, and the
parametrization in \Eqref{eq:dejager} is incomplete.  The main
limitation of the dJ algorithm is that it is representative of the
``averaged statistical behavior'' of stellar winds in the entire HR
diagram, which might average over different physical regimes. It is
important to note that WR and Be stars are intentionally excluded from
the data sample used to derive \Eqref{eq:dejager}.

\subsection{Nieuwenhuijzen \& de Jager (NJ)} 
\label{sec:nieuw}

The algorithm of \cite{nieuwenhuijzen:90} is intended as an
improvement over the \cite{dejager:88} algorithm of \Eqref{eq:dejager}, since
it is derived from the same data sample with a similar method but
includes the dependence of mass loss on the total stellar
mass. The goal of this is to capture one of
the missing stellar parameters entering in the mass loss
determination indicated by the large standard deviation of
\Eqref{eq:dejager}. The \cite{nieuwenhuijzen:90} algorithm
also translates the temperature dependence into a radius
dependence.

Since the total mass is not a directly observable quantity for single
stars, the authors' mass determination is based on stellar model
calculations.  The theoretical models used are from \cite{maeder:88,
  maeder:89}.  However, different stellar evolution codes consider a
large variety of physical processes (e.g.\ for mixing, mass loss,
etc.), or just use different implementations of them. Hence, there is
spread in the stellar masses found at the same point of an
evolutionary track. This implies that the \cite{nieuwenhuijzen:90}
mass loss algorithm depends on the set of stellar models used to
derive it.  This drawback applies to all mass loss algorithms
involving a functional dependence of the form
$\dot{M}\equiv\dot{M}(M)$ derived from stellar evolution
calculations. However, \cite{mauron:11} suggests that the dependence
on the total mass of the \cite{nieuwenhuijzen:90} mass loss rate is so
weak (see Eq.~\ref{eq:nieuwenhuijzen}) that it can be averaged (by
substituting $M$, which changes during the evolution, with a constant
value) without dramatic consequences on the
evolved stellar model.

Stars with different masses pass through the same point on the HR
diagram at different stages of their evolution. Therefore, in order to
include the total mass as a variable for the mass loss rate,
\cite{nieuwenhuijzen:90} determine an ``average expected mass''
$\bar{M}$ of a star at a given ($T_\mathrm{eff}, L$) point.  This
value is derived as follows. The authors define a \emph{dwell time}
representing the time for a star to travel over a unit length track on
the HR diagram, i.e.\ ,
\begin{equation}
  \label{eq:dwell_time}
  t^{(d)} \udef \frac{\delta t}{\sqrt{[\delta
      \log_{10}(T_\mathrm{eff}/ \mathrm{[K]})]^2+[\delta \log_{10}(L/L_\odot)]^2}} \ \ ,
\end{equation}
where $\delta t$ is the time spent to travel over the $(\delta
\log_{10}(T_\mathrm{eff}/\mathrm{[K]}),\delta \log_{10}(L/L_\odot))$ distance.

For every point on the HR diagram, there are N left-ward or
right-ward subtracks of the stellar evolutionary tracks crossing it.
Let $t^{(d)}_n$ be the dwell time for the n-th subtrack. The average
expected mass $\bar{M}$ is obtained as:
\begin{equation}\label{eq:avgM}
  \bar{M}=\frac{\sum_{n=1}^N \Psi(M_n)\frac{d M_n}{d \log_{10}(L/L_\odot)}t^{(d)}_nM_n}{\sum_{n=1}^N \Psi(M_n)\frac{d M_n}{d \log_{10}(L/L_\odot)}t^{(d)}_n} \ \ ,
\end{equation}
where $\Psi$ is the initial mass function for stars on the subtrack
considered, and $d M_n / d \log_{10}(L/L_\odot)$ is the density of tracks over a
unit $\log_{10}(L/L_\odot)$ interval.

The authors perform a fit of the data set\footnote{Which excludes WR and Be stars.}  used in \cite{dejager:88},
adding the value of $\bar{M}$ from \Eqref{eq:avgM} to the set, and
they find the interpolation formula\footnote{The formula
  in the abstract of \cite{nieuwenhuijzen:90} has a typo, see their Eq.~2.} (where $\bar{M}$ is substitued by
the total mass $M$ to use the formula in a stellar evolution simulation)
\begin{equation}
  \label{eq:nieuwenhuijzen}
  \begin{aligned}
    \log_{10}(-\dot{M}) = -14.02 + 1.24\log_{10}(L/L_\odot)+\\ +0.16\log_{10}(M/M_\odot)
    +\, 0.81\log_{10}(R/R_\odot) \ \ .
  \end{aligned}
\end{equation}
With the inclusion of $M$ among the parameters determining $\dot{M}$,
\cite{nieuwenhuijzen:90} obtain a standard deviation of $\sim 0.37$
for the distribution of the differences between the prediction of
\Eqref{eq:nieuwenhuijzen} and the observed values, comparable to the
standard deviation of the distribution of the differences between
  individual mass loss determination used as input data. The main
  limitations of this algorithm are its dependence on the input
  stellar models, and its statistical-average nature \citep[in the
    same sense as][]{dejager:88}.

\subsection{Van Loon \emph{et al.} (vL)}
\label{sec:van_loon}

The mass loss rate of \cite{vanloon:05} is empirically determined on
the basis of observations of a sample of oxygen-rich AGB and RSG stars
in the Large Magellanic Cloud (LMC).  The \cite{vanloon:05} analysis
is based on a dust-driven wind model: AGB and RSG stars have very
extended and cool envelopes where dust grains might form through
sublimation. The photons from the radiation field transfer momentum to
these grains, pushing them away. The dust grains drag the gas with
them through collisional coupling. To obtain their mass loss
algorithm, \cite{vanloon:05} fit the observed IR spectra to synthetic
spectra obtained with a simple model of the gas/dust mixture
(identical grains and dust-to-gas ratio set to the value observed at
$Z_\odot$ rescaled to $Z_\mathrm{LMC}$), using $T_\mathrm{eff}$ and
$L$ as variables. They obtain the relationship
\begin{equation}
  \begin{aligned}
    \label{eq:van_loon}
    \log_{10}(-\dot{M}) = -5.65(15) + 1.05(14)\log_{10}(L/10^4 L_\odot) + \\
    -6.3(1.2)\log_{10}(T_\mathrm{eff}/3\,500\,\mathrm{K}) \ \ ,
  \end{aligned}
\end{equation}
where the numbers in parentheses indicate the estimated errors on the
last digits. These errors are typically neglected in stellar evolution
codes. The main limitation of this algorithm is the high uncertainty
of the dust grain properties (mass fraction, opacity, when they form,
etc.). Since it gives
such a high mass loss rate, the \cite{vanloon:05} algorithm is
sometimes combined with a high efficiency factor to \emph{ad hoc}
mimic non-wind mass loss phenomena \citep[e.g.,][]{meynet:15}.

\subsection{Nugis \& Lamers (NL)}
\label{sec:nl}

The mass loss algorithm derived by \cite{nugis:00} applies only to WR
stars. The wind mass loss rate of these stars depends strongly on
their chemical composition: not only the metallicity has an important
role, but also the helium mass fraction $Y$. This is because the
amount of helium in the stellar atmosphere influences its temperature
and therefore the ionization fraction and the level populations of all
other atoms and ions.  \cite{nugis:00} derive a mass loss rate
algorithm as a function of the luminosity and the chemical composition
starting from a relevant sample of observed galactic WR stars. Their
sample is made of two subsets of stars: one in which both mass and
distance (i.e.\ luminosity) are known, thanks to binarity and
membership association in open clusters; and another subset for which
the intrinsic luminosity is not known.  They use stars from the first
subset to derive an empirical bolometric correction\footnote{The
  bolometric correction is the difference between the bolometric
  magnitude and the observed (visual) magnitude, influenced by the
  instrumental intrinsic band pass, $\mathrm{BC}\udef
  M-M_\mathrm{obs}$}.  They then use a theoretical mass-luminosity
relation to infer the luminosity of stars in the second subset, and
correct it with the previously derived bolometric correction.  Note
that the mass-luminosity relation is determined using as input the age
of the star and its spectral type, not its luminosity, therefore this
relation can be consistently used to estimate the luminosity of stars
in the second subset \citep{nugis:00}.

The mass loss rate observed for the stars in the sample are then
fitted as follows. The authors make two independent fits for stars of
different composition and then merge them together in a single
formula, valid for all WR stars:
\begin{equation}\label{nugis_lamers}
  \begin{aligned}
    \log_{10}(-\dot{M}) = -11.0
    +1.29(14)\log_{10}(L/L_\odot)+\\
    1.73(42)\log_{10}(Y)+0.47(09)\log_{10}(Z) \ \ ,
  \end{aligned}
\end{equation}
where the numbers in parentheses are the estimates of the error on the
last digits reported. According to \cite{nugis:00}, the mass loss
algorithm for WR stars cannot be expressed as a function of
$T_\mathrm{eff}$ and/or the radius of the star $R$. This is because WR
winds are so strong (i.e.\ dense) that they are optically thick, and
thus the observed radius is a function of the wavelength: the black
body relation $L=4\pi R^2 \sigma T_\mathrm{eff}^4$, where $\sigma$ is
the Stefan-Boltzmann constant, loses its meaning.

\subsection{Hamann \emph{et al.} (H)}
\label{sec:hamann}
This wind scheme, which applies only to WR stars, is a combination of
the algorithms from \cite{hamann:82}, \cite{hamann:95}, and
\cite{hamann:98}.  It is derived from a spherically-symmetric,
homogeneous, and stationary (but not static, i.e.\ $\partial_t = 0$
but $v \neq 0$) expanding WR atmosphere model.  The authors assume an
\emph{ad-hoc} velocity structure $v\equiv v(r)$ as follows. For the
supersonic part of the wind, they assume a $\beta$-law of the form of
\Eqref{eq:beta_law} with $\beta=1$, while for the subsonic part, $v(r)$
is chosen in such way that the density approaches the hydrostatic
limit. These assumptions are in reasonable agreement with the
numerical solutions. Note that since the velocity field is imposed,
the acceleration is not computed. This allows the authors to adopt a
very simple chemical composition, since they do not need to evaluate
the line-driven acceleration and do not need to keep track of all
atomic/ionic species and their level populations. The authors include
only ions of H and He, and the radiation field is considered only to
determine non-LTE populations of these species.  The temperature
stratification is derived with the assumption of a gray LTE model,
assuming a value of $T_\mathrm{eff}$ at the base of the atmosphere
determined by the stellar luminosity $L$ and the radius $R$ via a
black body relation $L=4\pi R^2\sigma T_\mathrm{eff}^4$, in contrast
to the suggestion of \cite{nugis:00}.  A synthetic spectrum is derived
from the simulations and a best fit to the observed line profiles for
the ions of H and He is obtained via variation of the stellar
parameters \citep[i.e.\ the radius of the inner boundary of the
  atmosphere $R$ and the luminosity $L$, the surface hydrogen mass
  fraction $X_s$ and $\dot{M}$,][]{hamann:95}. Once the stellar
parameters are known from this fit, a mass loss formula is derived for
high luminosity WR stars, i.e.\ $\log_{10}(L/L_\odot) > 4.5$.  The
algorithm for the low luminosity WR stars,
i.e.\ $\log_{10}(L/L_\odot)<4.5$, is derived with a similar technique
in \cite{hamann:82}, but the spectra fitted are from a small sample of
Helium stars (i.e.\ stars undergoing He shell burning with most of the
mass in a CO core, without H lines in their spectra). The resulting
formula is
\begin{equation}
  \label{eq:hamman}
  \log_{10}(-\dot{M}) = \begin{cases} -12.25+1.5\log_{10}\left(\frac{L}{L_\odot}\right)-2.85X_s & \mbox{if } \mbox{$L > 4.5 L_\odot$}, \\ -35.8+6.8\log_{10}\left(\frac{L}{L_\odot}\right) & \mbox{if } \mbox{$L \leq 4.5 L_\odot$}. \end{cases}
\end{equation}
Improvements that take into account inhomogeneities in the wind are
suggested in \cite{hamann:98}. Specifically, the authors suggest to
reduce the wind efficiency by a factor between 2 and 3 to account for
the wind clumpiness, which strongly affects the
fitted spectral lines.

\section{MESA input physics, customization, and resolution study}
\label{app:MESA_technical}

Here, we provide and discuss the MESA parameters not directly related
to the physics of mass loss that we omit in \Secref{sec:methods} for
brevity.

We employ the Ledoux criterion \citep{ledoux:47} to determine
convective stability. This means that we consider the effects of the
temperature and chemical composition gradients on the stability of a
region, and allow for semiconvective mixing driven by compositional
gradients. We follow the suggestions of \cite{sukhbold:14} for the
treatment of convective mixing \citep[relying on mixing length
  theory;][]{bohmvitense:58} and overshooting, but we also include
thermohaline mixing using the default MESA treatment of this process
\citep[][]{kippenhahn:80}. Our specific MESA parameter choices are as
follows: mixing length parameter $\alpha_\mathrm{mlt} = 2$,
exponential overshooting/undershooting parameters $f_\mathrm{ov} =
0.025$ and $f_0=0.05$ for all regions (see
\citealt{moravveji:16,paxton:11}), semiconvection efficiency of $0.2$,
and thermohaline mixing with coefficient $2$ according to
\cite{kippenhahn:80}.

We employ the 45-isotope nuclear reaction network
\texttt{mesa\_45.net} (in
\texttt{\$MESA\_DIR/}\texttt{data/}\texttt{net\_data/}\texttt{nets/})
for all models until oxygen depletion. We switch to a customized
203-isotope network for models evolved further, and initialize to zero
the abundances of all the new isotopes introduced
(i.e.\ \texttt{adjust\_abundances\_for\_new\_isos = .false.}).  We
obtain the 203-isotope network by taking the union of the isotope sets
of \texttt{mesa\_45.net} and \texttt{mesa\_201.net}. The list of
isotopes included in the network is available at
\href{https://stellarcollapse.org/renzo2017}{\url{https://stellarcollapse.org/renzo2017}}.

Stars more massive than $\sim 20\,M_\odot$ can develop a
radiation-dominated, near super-Eddington convective envelope (e.g.,
because of the iron opacity bump) in which the convective flux is not
sufficient to transport energy outward. This can lead to numerical
(and possibly physical, e.g., \citealt{langer:97}) density and
pressure inversions and consequently dynamical instabilities
\citep[see, e.g.,][and references therein]{joss:73, paxton:13}.  To
handle these envelope issues, MESA introduces the so-called MLT++
scheme \citep[see][]{paxton:13}.  MLT++ arbitrarily decreases the
superadiabaticity in near-to-Eddington ($L\geq 0.5 L_\mathrm{Edd}$)
radiation dominated ($P_\mathrm{gas}/P_\mathrm{tot} \leq 0.35$)
convective envelopes in order to prevent numerical issues. We note
that MLT++ can modify the radius and luminosity of the star by
artificially enhancing the energy flux carried by convection, and thus
indirectly, it can also change the mass loss rate. Furthermore, we
find that MLT++ combined with the \texttt{simple\_atmosphere} option
for the outer boundary condition causes an unphysical oscillation of
the mass loss rate and surface characteristics \citep[e.g., luminosity
  $L$, effective temperature $T_\mathrm{eff}$, and radius
  $R$;][]{thesis}.  To avoid this issue, we employ the more realistic
atmosphere boundary condition provided by the \texttt{Eddington\_grey}
option in all our models. This uses the Eddington gray $T(\tau)$
relation,
\begin{equation}
  \label{eq:Edd_T_tau}
  T^4 = \frac{3}{4}T_\mathrm{eff}^4\left(\tau +\frac{2}{3}\right) \ \ ,
\end{equation}
to find the boundary pressure on the photosphere, instead of guessing
the approximate photosphere location. 

\begin{figure}[!tp]
  \resizebox{\hsize}{!}{\includegraphics{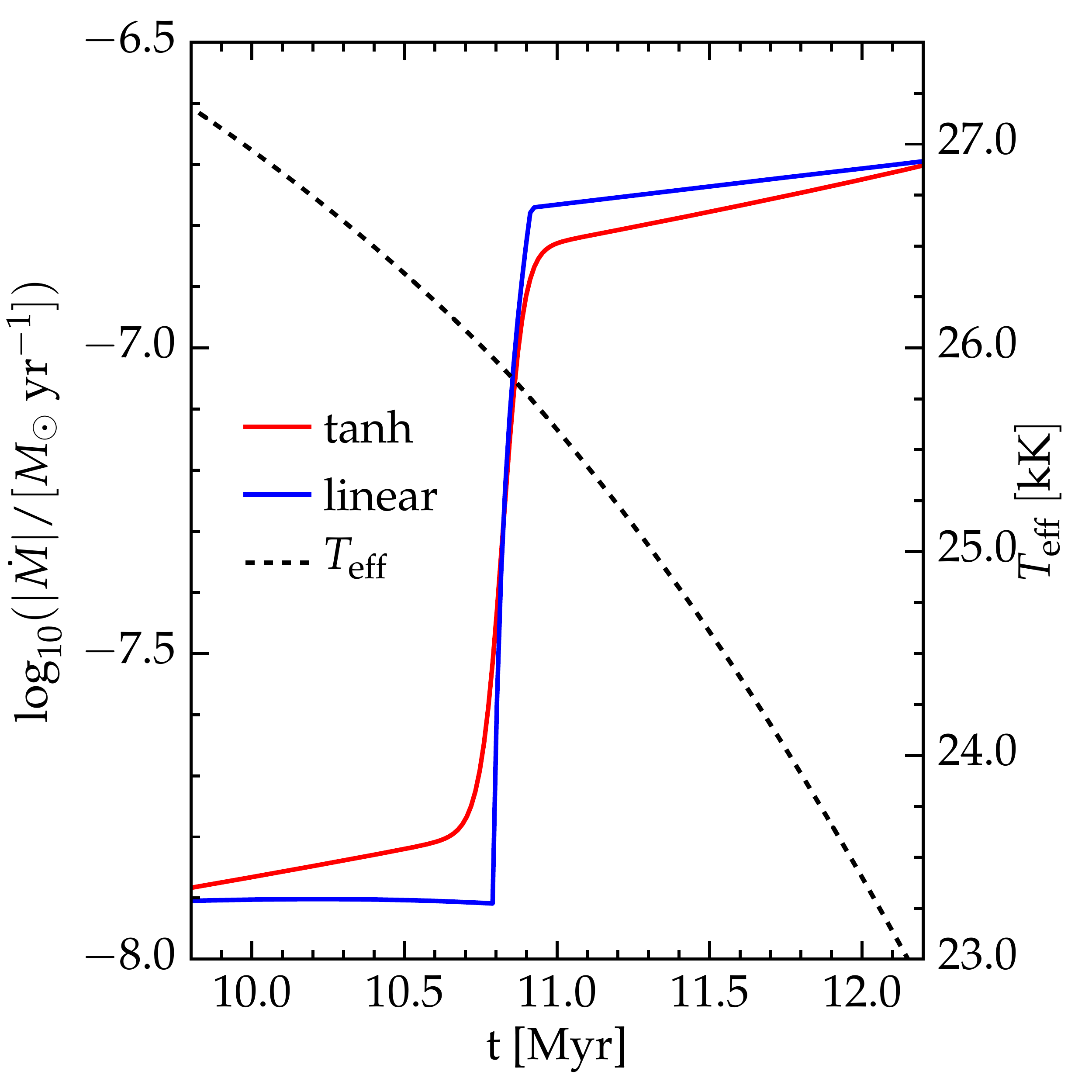}}
  \caption{Time evolution of the mass loss
    rate around the bistability jump of the \cite{vink:00,vink:01} (V) mass loss rate (\Secref{sec:vink}) in a $M=15M_\odot$, $\eta=1.0$ model at $Z=Z_\odot$. The black dashed line is the effective temperature
    $T_\mathrm{eff}$, reached near the terminal age main sequence
    (TAMS, $X_c<0.01$) in these simulations. The blue curve
    corresponds to the mass loss rate from the default MESA routine,
    which uses a linear interpolation between \Eqref{eq:vink_hot} and
    \Eqref{eq:vink_cool}. The red curve is the mass loss rate from our
    customized routine, where we interpolate using a hyperbolic
    tangent. \label{fig:vink_intrp}}
\end{figure}

At the beginning of each evolutionary step, MESA evaluates the mass
loss rate $\dot{M}$ according to the algorithm specified by the user,
and then removes from the surface the amount of mass
$\dot{M}\times\Delta t$, where $\Delta t$ is the timestep.  The mass
loss algorithm to evaluate $\dot{M}$ can be chosen from the many
built-in algorithms, or it can be implemented by the user using the
\texttt{run\_star\_extras.f} hooks to override the default MESA
routines\footnote{See also
  \url{http://mesa.sourceforge.net/run_star_extras.html}}. We use the
latter option to combine three different algorithms (one for the hot
phase of the evolution, one for the cool phase, and one for the WR
phase, see \Secref{sec:comb}). When possible, we call the built-in
MESA mass loss routines from our \texttt{run\_star\_extras.f}, except
for the V algorithm. The default implementation of the V mass loss
algorithm in MESA uses a linear interpolation between
\Eqref{eq:vink_hot} and \Eqref{eq:vink_cool}. This results in a jump
of $\dot{M}$ (as a function of time $t$) with discontinuous derivative
(i.e.\ $M(t)$ is not $\mathcal{C}^2$) as $T_\mathrm{eff}$ drops below
$25\,000 \ \mathrm{K}$ during the evolution. For a physically more
realistic, smoother time dependence of $\dot{M}$, we implement a
hyperbolic tangent interpolation between the two formulae. This is
depicted in \Figref{fig:vink_intrp}.

To save MESA ``\texttt{photos}''\footnote{These are binary files that
  allow one to restart a run and obtain bit-to-bit identical results
  (provided that the parameter set does not change). See also
  \url{http://mesa.sourceforge.net/}.} and to call the built-in
implementation of the K algorithm from
 \texttt{run\_star\_extras.f}, we
copy the Fortran modules
\texttt{write\_model.mod} and
\texttt{kuma.mod} from
\texttt{\$MESA\_DIR/star/make} into \texttt{\$MESA\_DIR/include} in
our standard MESA installation. This is
  necessary since these two modules do not have a ``public''
interface \citep[][]{paxton:11} in MESA
release version
7624.

We describe the settings for spatial and
temporal resultion in the next \Secref{sec:res_study}.
We use the default MESA settings for massive stars
(see \texttt{\$MESA\_DIR/star/inlist\_massive\_defaults}) for anything
else not explicitly mentioned.

\subsection{Resolution Dependence}
\label{sec:res_study}

Any computational study in astrophysics must carefully assess the
effects of numerical resolution (in both space and time) on its
results. To study the sensitivity of our results to variations in
temporal and spatial resolution, we carry out a resolution study using
a 30$M_\odot$ star since stars of this $M_\mathrm{ZAMS}$ appeared to
be the most sensitive to the resolution in our preliminary
calculations using the default MESA parameters.  The
post core carbon depletion evolutionary
tracks produced by the default MESA parameters show large amplitude
oscillations (e.g., in the central temperature--central density plane)
when varying the spatial discretization. It is unlikely that nature
would do this. Thus, such oscillations are most likely artificial, and
they are generally worse at higher initial mass. The aim of this
section is to find a set of parameters that reduces and possibly
eliminates these oscillations.

For each set of resolution parameters, we run our test model to oxygen
depletion with two different mesh refinement parameters
(\texttt{mesh\_delta\_coeff}=1.0 and \texttt{mesh\_delta\_coeff}=0.5),
and two different wind mass loss algorithm combinations (V-dJ and
K-NJ; both with $\eta=1.0$). The use of two different mass loss
algorithm combinations allows us to check that our settings are not
cherry-picked for a particular model in our grid. However, the results
of our study indicate that the resolution dependence is insensitive to
the mass loss algorithm combination.

To obtain numerically converged results, we use both the available MESA controls
(specified in the \texttt{inlist}s) and customized routines in
\texttt{run\_star\_extras.f}. Both are available at
\href{https://stellarcollapse.org/renzo2017}{\url{https://stellarcollapse.org/renzo2017}}.

\begin{figure}[!t]
  \centering
   \resizebox{\hsize}{!}{\includegraphics{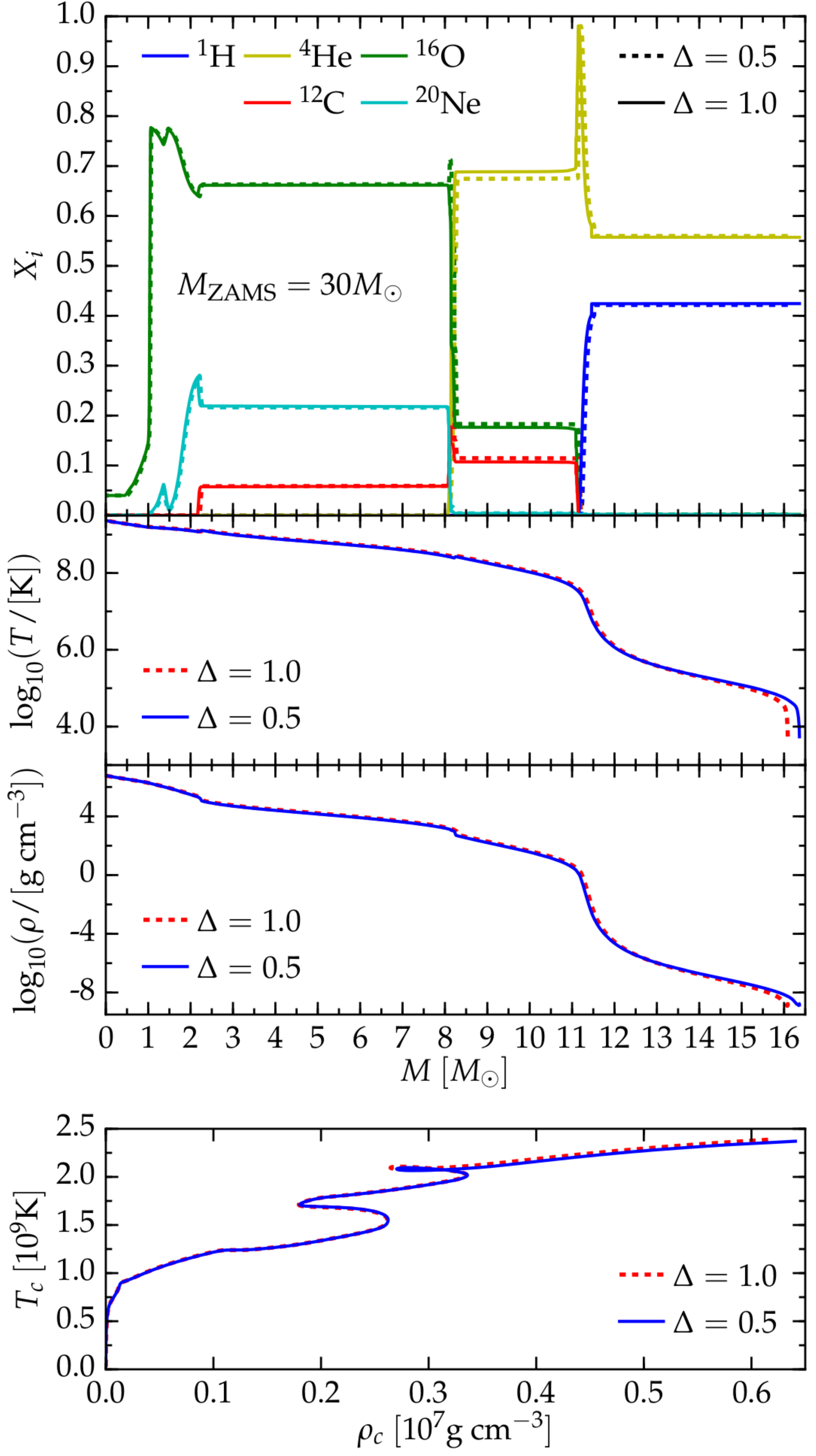}}
  \caption{Mass fractions, temperature, and density profiles (top
    three panels) at oxygen depletion and central temperature --
    central density evolutionary track (bottom panel) from ZAMS to
    core oxygen depletion for the $M_\mathrm{ZAMS}=30M_\odot$ test case with the V-dJ algorithm and $\eta=1.0$. We run the test case at the standard spatial resolution ($\Delta\equiv$ \texttt{mesh\_delta\_coeff}=1.0, dashed
    lines) and twice that resolution ($\Delta\equiv$ \texttt{mesh\_delta\_coeff}=0.5, solid lines). \label{fig:res_improvements}}
\end{figure}

\subsubsection{Timestep selection throughout the evolution}
\label{app:timestep}
To avoid overstepping relevant physical
processes\footnote{Preliminary calculations with low
    resolution showed that in some rare cases, e.g.~during the
    Herzsprung gap, MESA would try to overstep the Kelvin-Helmholtz timescale.}, we enforce a
customized timestep control for the evolution of our models (see the
routine \texttt{extras\_finish\_step} in our
\texttt{run\_star\_extras.f}). In particular, we tighten the default
MESA (release version 7624) timestep controls
by manually enforcing
\begin{equation}
  \label{eq:timestep_ctrl}
  \Delta t_{n+1} \leq \min \{ t_\mathrm{KH}, t_{\dot{M}} \}
  \ \ ,
\end{equation}
where $\Delta t_{n+1}$ is the timestep proposed
at the end of the $n-\mathrm{th}$ step for the next
($(n+1)-\mathrm{th}$) step, and all the timescales on the right hand
side refer to the $n-\mathrm{th}$ step.  On the right hand side of
condition \ref{eq:timestep_ctrl},
\begin{equation}
  t_\mathrm{KH} \udef \frac{3}{4}\frac{GM^2}{RL} \ \ \ \ , \ \ \ \  t_{\dot{M}} \udef \frac{|\dot{M}|}{M} \ \ ,
\end{equation}
are the Kelvin-Helmholtz timescale and the mass change timescale,
respectively.

Note that the right-hand side of 
condition \ref{eq:timestep_ctrl} is evaluated with the quantities of
the $n-\mathrm{th}$ step, but it limits the $(n+1)-\mathrm{th}$
step. Moreover, until oxygen depletion, we limit the timestep by
setting \texttt{varcontrol\_target = 1.0d-4} as the maximum relative
variation for quantities in each computational
cell. We also limit the amount
of matter nuclearly processed in one single timestep by setting the
maximum variation of the mass fraction of each element
due to nuclear buring in each time step
to \texttt{dX\_nuc\_drop\_limit = 1.0d-4}.

\subsubsection{Spatial meshing until oxygen depletion}
Until oxygen depletion, we impose a maximum value for the fraction of
the total mass in each computational
cell by setting \texttt{max\_dq = 0.5d-4} in the \texttt{inlist}. This means
that each of our models has at least
$1/\texttt{max\_dq}=2\times10^4$ computational cells. Moreover, we
refine the mesh around some specific regions of the
star to focus resolution there:

\begin{itemize}
\item[$\bullet$] {\bf Regions with steep temperature gradients:}
  to better resolve the deep
  interior of the star, we limit the maximum variation of
  $d\log_{10}(T)/dm$ across adjacent cells,
  where $T$ is the temperature and $m$ is
  the mass coordinate. Specifically, we impose a maximum variation of
  \texttt{mesh\_delta\_coeff}$/10$ for $\log_{10}\left(
  d\log_{10}(T)/dm+1 \right)$. This is is achieved using the
  \texttt{other\_mesh\_fcns\_data} routine in
  \texttt{run\_star\_extras.f};

\item[$\bullet$] {\bf Stellar surface}: to properly resolve the amount of mass lost at each timestep,
  we impose that the outermost $0.5\,M_\odot$ of the star is sampled
  by at least $500 \times (\texttt{mesh\_delta\_coeff})^{-1}$ cells
  (see \texttt{other\_mesh\_fcns\_data} routine in
  \texttt{run\_star\_extras.f} for the implementation). Note that for
  each timestep $\Delta t$, $\dot{M}\times \Delta t\ll 0.5 M_\odot$;

\item[$\bullet$] {\bf Edges of burning regions}: to resolve the edges
  of burning regions, we constrain the spatial variation of $d
  \log_{10}(\varepsilon_\mathrm{nuc})/d \log_{10}(P)$ by multiplying
  its maximum allowed variation (regulated by
    \texttt{mesh\_delta\_coeff}) by 0.015 \citep[as in][private
    communication]{dessart:13}. This is done separately for each
  nuclear burning process, see
  \texttt{mesh\_dlog\_}*\texttt{\_dlogP\_extra} in the
  \texttt{inlist}s;

\item[$\bullet$] {\bf Edges of the
  cores of different composition}: to resolve sharp variations in
  the chemical composition, we impose \texttt{mesh\_delta\_coeff}$/20$
  as the maximum spatial variation allowed
  for the mass fractions of several isotopes ($^{1}\mathrm{H}$,
  $^{4}\mathrm{He}$, $^{12}\mathrm{C}$, $^{16}\mathrm{O}$,
  $^{20}\mathrm{Ne}$, $^{28}\mathrm{Si}$,$^{24}\mathrm{Mg}$,
  $^{32}\mathrm{S}$, $^{54}\mathrm{Fe}$,
  $^{56}\mathrm{Fe}$). These are specified via
  \texttt{xa\_function\_}* in the \texttt{inlists}.
\end{itemize}

Our models with these settings and \texttt{mesh\_delta\_coeff} = 1.0
have typically between $\sim 50\,000$ and $\sim 100\,000$ mesh points.
Figure~\ref{fig:res_improvements} illustrates for the model
using the V-dJ-NL mass loss algorithm combination and $\eta=1.0$ that our
setup does not produce any appreciable variations when changing
\texttt{mesh\_delta\_coeff} by a factor of 2 (corresponding to an
increase of the number of spatial mesh points from 54\,814 to
77\,400 at oxygen depletion). Note the linear scale for the central
temperature -- central density evolutionary tracks.

\subsubsection{Re-meshing after Oxygen Depletion}
\label{app:re-meshing}
After oxygen depletion, we switch from a 45-isotope to a 203-isotope
nuclear reaction network (cf.~\Secref{app:MESA_technical}). This
switch forces us to decrease the number of spatial mesh points
for the following reason: MESA solves the fully coupled set of equations
for stellar structure and evolution \cite{paxton:11}. This results in
the use of a work array of length $\mathcal{L}$ which scales as
\begin{equation}
  \label{eq:mtx_size}
  \mathcal{L} \sim
  \left((N_\mathrm{iso}+5)\cdot N_\mathrm{z}\right)\cdot(3N_\mathrm{iso}+9)
  \ \ , 
\end{equation}
where $N_\mathrm{iso}$ is number of isotopes in the nuclear reaction
network and $N_\mathrm{z}$ is the number of mesh points (see the
routine \texttt{get\_newton\_work\_sizes} in
\texttt{\$MESA\_DIR/star/private/star\_newton.f90}).  $\mathcal{L}$ is
stored as a 4-byte integer in MESA, which sets the maximum adressable
memory to $\sim17$ GB (R.~Farmer, private communication). Therefore,
for a nuclear reaction network including $\sim200$ isotopes, the
maximum number of mesh points cannot exceed
$\sim17\,000$. Changing the relevant variables
to an 8-byte integer data type would require substantial changes
throughout MESA, which we have opted to defer to future work.

Another significant limitation on the resolution that can be achieved
in the very late evolutionary phases is the stability of stellar
evolution codes: many highly uncertain and/or poorly understood
physical phenomena take place in the cores of evolved massive stars,
and the evolutionary timescale gets progressively closer to
the (neutrino-)thermal timescale and, finally,
the dynamical timescale.

Experiments with the highest achievable resolution resulted in
frequent failures of the code to find solutions to the stellar
structure equations. However, note that by the time core oxygen burning is
complete, the variations in core structure caused by different mass
loss algorithm combinations are already largely developed and will be
amplified by the subsequent evolution. 

We reduce the number of mesh points in our models down to a few
thousand (the precise value varies from model to model). We do this by
restarting from oxygen depletion and running MESA's re-meshing
algorithm for 100 timesteps of $\Delta t < 10^{-9}\,\mathrm{s}$ with
nuclear burning and neutrino cooling turned off (see
\texttt{inlist\_remesh}). At the same time, we shut down thermohaline
mixing, which is a secular process that does not have time to produce
any physically relevant change in the star after core oxygen
burning. We also drop the spatial mesh refinement criteria described
above for the remaining evolution. After the 100 re-meshing timesteps
during which the star is \emph{de facto} frozen in its state, we
resume the evolution by again turning on nuclear burning and neutrino
cooling and not imposing a maximum timestep,
  but we do not re-enable thermohaline mixing.
We compare pre-re-meshing and post-re-meshing
core structure, thermodynamics, and compactness parameter and ensure
that the post-re-meshing core is still very well resolved with the
reduced resolution as we proceed in the evolution toward core collapse.

\begin{figure}[tbp]
  \centering
  \resizebox{\hsize}{!}{\includegraphics{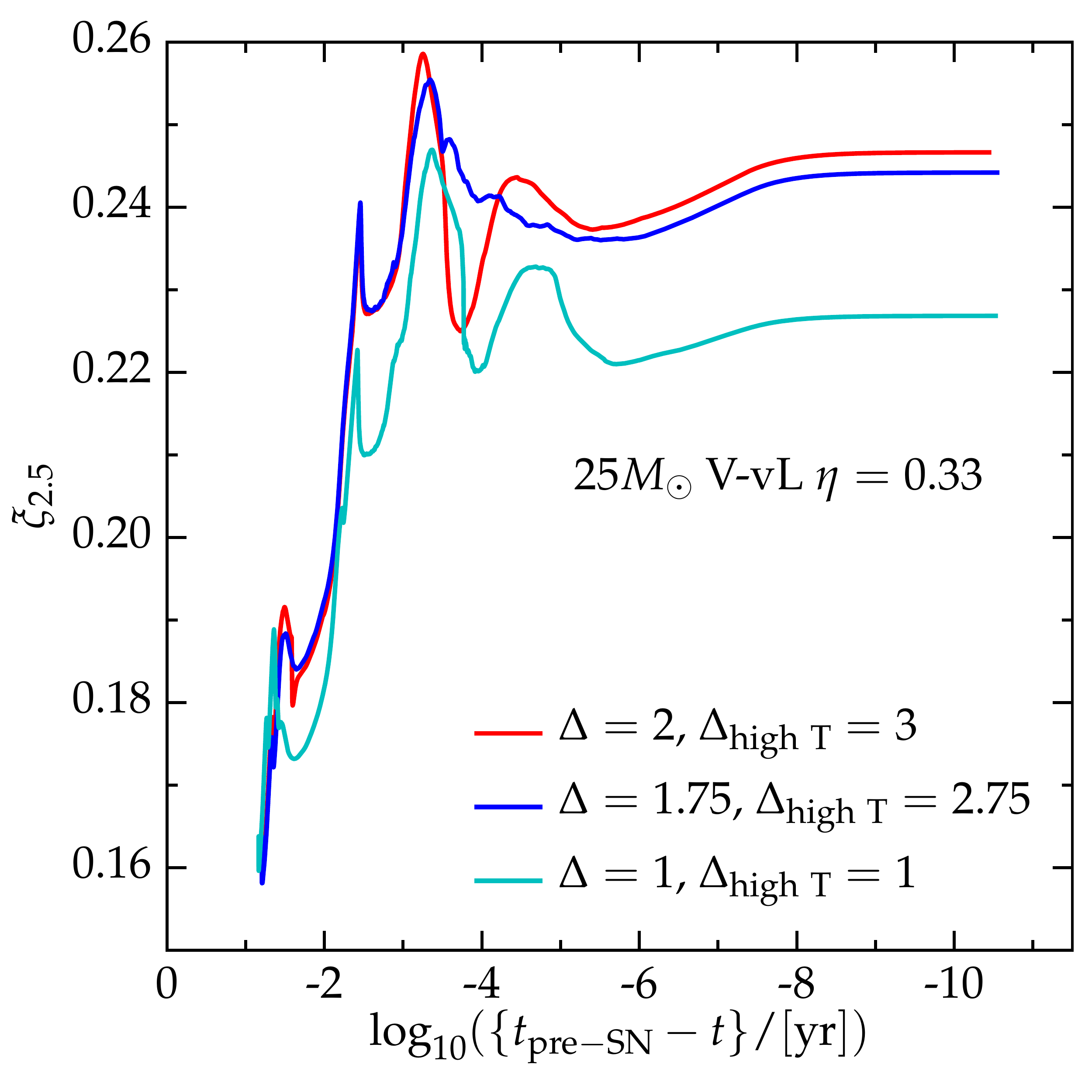}}
  \caption{Variations in the late-time evolution of the compactness
    parameter with spatial resolution. The
    magnitude of these
    variations is smaller than the variations caused by uncertainties
    in the wind mass loss. The curves start at oxygen depletion (after
    the re-meshing described in Appendix~\ref{app:re-meshing}) and end
    at the onset of core collapse. They correspond to $25\,M_\odot$
    models computed with a 203-isotope nuclear reaction network, using
    the V-vL algorithm and $\eta=0.33$ and varying $\Delta \equiv
    \texttt{mesh\_delta\_coeff}$, and $\Delta_\mathrm{high\ T} \equiv
    \texttt{mesh\_delta\_coeff\_for\_highT}$ (higher numbers
    correspond to lower resolution, see text). The highest resolution
    model (cyan curve) is the model described in
    \Secref{sec:onset_cc}.}
  \label{fig:conv_test_cc}
\end{figure}

As the temperature in the core increases, the nuclear burning rate
accelerates. When the central temperature rises above $3 \times
10^9\,\mathrm{K}$, we progressively relax the constraints on the
timestep from nuclear burning. We do this by modifying, at the
beginning of each timestep, the parameters loaded in our
\texttt{inlist} using the routine \texttt{extras\_startup} of our
\texttt{run\_star\_extras.f}. Specifically, for $3.0\leq
T_c/\mathrm{[10^9K]} \leq 3.5$ we impose \texttt{dX\_nuc\_drop =
  5d-3}, and for even higher $T_c$ we only require
\texttt{dX\_nuc\_drop = 5d-2} (cf.~\Secref{app:timestep}).  After
silicon core depletion ($X_c(^{28}\mathrm{Si}) \leq 0.001$), we also
decrease the spatial resolution of the innermost infalling core by
increasing \texttt{mesh\_delta\_coeff\_for\_highT=3.0} (used where
$\log_{10}(T/\mathrm{[K]}) \geq 9.3 $, the value used during the
previous evolution is 1.0). 

Finally, an important question to address is
the sensitivity to numerical resolution of the subsequent evolution
toward core collapse and of the final pre-collapse structure and
compactness parameter. Due to the memory limitations and
computational difficulties described in the above, we are unable to
carry out a rigorous convergence test with MESA at this
time. However, in order to gain some insight into the effects of
resolution, we choose the $25\,M_\odot$ V-vL $\eta = 0.33$ model and
carry out two additional simulations at \emph{reduced} resolution
from oxygen depletion to the onset of core collapse using our
203-isotope nuclear reaction network. Specifically,
we choose (\texttt{mesh\_delta\_coeff},
\texttt{mesh\_delta\_coeff\_for\_highT}) = (1.75, 2.75) and (2, 3),
whereas the standard setting for our models discussed in
\Secref{sec:onset_cc} is (1, 1). Figure~\ref{fig:conv_test_cc} shows the
evolution of the compactness parameter from oxygen depletion
(after re-meshing) to the onset of core collapse for these models. We
find that these lower-resolution models evolve qualitatively very
similar to our standard model, but produce variations in the pre-SN
compactness parameter of about $\sim$ 9\%, which is smaller than the
variations of $\sim 30\%$ caused by different wind mass loss algorithm combinations.

\end{document}